\documentclass[12pt]{article}
\pdfoutput=1
\usepackage[colorlinks,linkcolor=blue,citecolor=blue,bookmarks,bookmarksnumbered]{hyperref}
\usepackage[scaled=0.85]{helvet}
\usepackage{amsmath,amssymb,accents,mathrsfs,XoohmE}
\usepackage{graphicx,color}
\usepackage{booktabs}
\usepackage{multirow}
\usepackage{placeins}
\usepackage{amsmath}

\usepackage{enumitem}

\usepackage[aligntableaux=center,boxsize=1.2em]{ytableau}

\usepackage{XoohmE}

\definecolor{Green}  {rgb}{0.10,0.70,0.10} %  1
\definecolor{Orange} {rgb}{1.00,0.50,0.15} %  2
\definecolor{Red}    {rgb}{0.90,0.00,0.12} %  3
\definecolor{Purple} {rgb}{0.50,0.25,0.55} %  4
\definecolor{Turque} {rgb}{0.00,0.65,0.85} %  5
\definecolor{Blue}   {rgb}{0.00,0.00,1.00} %  6
\definecolor{Magenta}{rgb}{1.00,0.00,1.00} %  7
\definecolor{Gold}   {rgb}{1.00,0.75,0.25} %  8
\definecolor{Seaweed}{rgb}{0.01,0.24,0.09} %  9
\definecolor{Brown}  {rgb}{0.43,0.26,0.32} % 10
\definecolor{grey1}  {rgb}{0.20,0.20,0.20} % 11
\definecolor{grey2}  {rgb}{0.40,0.40,0.40} % 12
\definecolor{grey3}  {rgb}{0.60,0.60,0.60} % 13
\definecolor{grey4}  {rgb}{0.80,0.80,0.80} % 14
\definecolor{grey5}  {rgb}{0.90,0.90,0.90} % 15
\def\C#1#2{{\ifcase#1\or%Greg's color scheme
             \color{Green}\or \color{Orange}\or \color{Red}\or
              \color{Purple}\or \color{Turque}\or \color{Blue}\or
               \color{Magenta}\or \color{Gold}\or \color{Seaweed}\or
                \color{Brown}\or\color{grey1}\or\color{grey2}\or
                 \color{grey3}\else\color{grey4}\fi#2}}

\definecolor{Slate} {rgb}{0.00,0.45,0.55}

\usepackage[enableskew,vcentermath]{youngtab}
\let\TC=\textcolor
\definecolor{Hey}{rgb}{.9,.05,.4}
\definecolor{orange}{rgb}{1,.5,0}
\definecolor{plum}{rgb}{.4,0,.6}
\definecolor{R}{rgb}{1,0,0}
\definecolor{G}{rgb}{0,1,0}
\definecolor{B}{rgb}{0,0,1}

\long\def\CMTgrn#1{\leavevmode\TC{green}{\sf#1}}
\long\def\CMTred#1{\leavevmode\TC{red}{\sf#1}}
\long\def\CMTorg#1{\leavevmode\TC{orange}{\sf#1}}
\long\def\CMTR#1{\leavevmode\TC{R}{\sf#1}}
\long\def\CMTG#1{\leavevmode\TC{G}{\sf#1}}
\long\def\CMTB#1{\leavevmode\TC{B}{\sf#1}}
\usepackage{lipsum}
\usepackage{listings}
\definecolor{MyDarkGreen}{rgb}{0.0,0.4,0.0} % This is the color used for comments
\def\rD{{\rm D}}

\def\fracm#1#2{\hbox{\large{${\frac{{#1}}{{#2}}}$}}}

\def\be{\begin{equation}}
\def\ee{\end{equation}}
\newcommand{\bea}{\begin{eqnarray}}
\newcommand{\eea}{\end{eqnarray}}
\newcommand{\ena}{\end{eqnarray}}

\def\pp{{\mathchoice
          {%displaystyle
              \kern 1pt%
              \raise 1pt
              \vbox{\hrule width5pt height0.4pt depth0pt
                    \kern -2pt
                    \hbox{\kern 2.3pt
                          \vrule width0.4pt height6pt depth0pt
                          }
                    \kern -2pt
                    \hrule width5pt height0.4pt depth0pt}%
                    \kern 1pt
           }
            {%textstyle
              \kern 1pt%
              \raise 1pt
              \vbox{\hrule width4.3pt height0.4pt depth0pt
                    \kern -1.8pt
                    \hbox{\kern 1.95pt
                          \vrule width0.4pt height5.4pt depth0pt
                          }
                    \kern -1.8pt
                    \hrule width4.3pt height0.4pt depth0pt}%
                    \kern 1pt
            }
            {%scriptstyle
              \kern 0.5pt%
              \raise 1pt
              \vbox{\hrule width4.0pt height0.3pt depth0pt
                    \kern -1.9pt  %[e]=0.15pt
                    \hbox{\kern 1.85pt
                          \vrule width0.3pt height5.7pt depth0pt
                          }
                    \kern -1.9pt
                    \hrule width4.0pt height0.3pt depth0pt}%
                    \kern 0.5pt
            }
            {%scriptscriptstyle
              \kern 0.5pt%
              \raise 1pt
              \vbox{\hrule width3.6pt height0.3pt depth0pt
                    \kern -1.5pt
                    \hbox{\kern 1.65pt
                          \vrule width0.3pt height4.5pt depth0pt
                          }
                    \kern -1.5pt
                    \hrule width3.6pt height0.3pt depth0pt}%
                    \kern 0.5pt%}
            }
        }}

\def\mm{{\mathchoice
                       {%displaystyle
                             \kern 1pt
               \raise 1pt    \vbox{\hrule width5pt height0.4pt depth0pt
                                  \kern 2pt
                                  \hrule width5pt height0.4pt depth0pt}
                             \kern 1pt}
                       {%textstyle
                            \kern 1pt
               \raise 1pt \vbox{\hrule width4.3pt height0.4pt depth0pt
                                  \kern 1.8pt
                                  \hrule width4.3pt height0.4pt depth0pt}
                             \kern 1pt}
                       {%scriptstyle
                            \kern 0.5pt
               \raise 1pt
                            \vbox{\hrule width4.0pt height0.3pt depth0pt
                                  \kern 1.9pt
                                  \hrule width4.0pt height0.3pt depth0pt}
                            \kern 1pt}
                       {%scriptscriptstyle
                           \kern 0.5pt
             \raise 1pt  \vbox{\hrule width3.6pt height0.3pt depth0pt
                                  \kern 1.5pt
                                  \hrule width3.6pt height0.3pt depth0pt}
                           \kern 0.5pt}
                       }}

\def\ad{{\kern0.5pt
                   \alpha \kern-5.05pt \raise5.8pt\hbox{$\textstyle.$}\kern
0.5pt}}

\def\bd{{\kern0.5pt
                   \beta \kern-5.05pt \raise5.8pt\hbox{$\textstyle.$}\kern
0.5pt}}

\def\qd{{\kern0.5pt
                   q \kern-5.05pt \raise5.8pt\hbox{$\textstyle.$}\kern
0.5pt}}
\def\Dot#1{{\kern0.5pt
     {#1} \kern-5.05pt \raise5.8pt\hbox{$\textstyle.$}\kern
0.5pt}}

\catcode`@=11
\def\un#1{\relax\ifmmode\@@underline#1\else
        $\@@underline{\hbox{#1}}$\relax\fi}
\catcode`@=12

\def\a{\alpha}
\def\b{\beta}
\def\c{\chi}
\def\d{\delta}
\def\e{\epsilon}

\def\g{\gamma}

\def\i{\iota}
\def\j{\psi}
\def\k{\kappa}
\def\l{\lambda}

\def\o{\omega}

\def\r{\rho}
\def\s{\sigma}
\def\t{\tau}

\def\L{\Lambda}
\def\O{\Omega}
\def\P{\Pi}

\def\S{\Sigma}

\def\ca{{\cal A}}

\def\cf{{\cal F}}

\def\dslash{\not{\hbox{\kern-2pt $\partial$}}}
\def\Dslash{\not{\hbox{\kern-4pt $D$}}}
\def\pslash{\not{\hbox{\kern-2.3pt $p$}}}
 \newtoks\slashfraction
 \slashfraction={.13}
 \def\slash#1{\setbox0\hbox{$ #1 $}
 \setbox0\hbox to \the\slashfraction\wd0{\hss \box0}/\box0 }
 
\def\kcr{{\hbox{\ro \char'170}}}                % right-handed rope
\def\ktl{{\hbox{\ro \char'170}}}        % top end for left-handed rope
\def\ktr{{\hbox{\ro \char'170}}}        % " right
\def\kbl{{\hbox{\ro \char'170}}}        % " bottom left
\def\kbr{{\hbox{\ro \char'170}}}        % " right
\def\plpl{\raise-2pt\hbox{$\raise3pt\hbox{$_+$}\hskip-6.67pt\raise0.0pt
\hbox{$^+$}\hskip 0.01pt$}}
\def\mimi{\raise-2pt\hbox{$\raise3pt\hbox{$_-$}\hskip-6.67pt\raise0.0pt
\hbox{$^-$}\hskip 0.01pt$}}

\def\bo{{\raise.15ex\hbox{\large$\Box$}}}               % D'Alembertian
\def\TH{{\raise.2ex\hbox{$\displaystyle \bigodot$}\mskip-4.7mu \llap H \;}}
\def\face{{\raise.2ex\hbox{$\displaystyle \bigodot$}\mskip-2.2mu \llap {$\ddot
        \smile$}}}                                      % happy face
\def\dt#1{\on{\hbox{\bf .}}{#1}}                % (big) dot over
\def\Dot#1{\dt{#1}}

\def\leftrightarrowfill{$\mathsurround=0pt \mathord\leftarrow \mkern-6mu
        \cleaders\hbox{$\mkern-2mu \mathord- \mkern-2mu$}\hfill
        \mkern-6mu \mathord\rightarrow$}
\def\dvec#1{\vbox{\ialign{##\crcr
        \leftrightarrowfill\crcr\noalign{\kern-1pt\nointerlineskip}
        $\hfil\displaystyle{#1}\hfil$\crcr}}}           % <--> accent
\def\dt#1{{\buildrel {\hbox{\LARGE .}} \over {#1}}}     % dot-over for sp/sb
\def\fracm#1#2{\hbox{\large{${\frac{{#1}}{{#2}}}$}}}
\def\sfrac#1#2{{\vphantom1\smash{\lower.5ex\hbox{\small$#1$}}\over
        \vphantom1\smash{\raise.4ex\hbox{\small$#2$}}}} % alternate fraction
\def\bfrac#1#2{{\vphantom1\smash{\lower.5ex\hbox{$#1$}}\over
        \vphantom1\smash{\raise.3ex\hbox{$#2$}}}}       % "
\def\afrac#1#2{{\vphantom1\smash{\lower.5ex\hbox{$#1$}}\over#2}}    % "
\def\ad{{\Dot{\alpha}}}
\def\bd{{\Dot{\beta}}}

 \font\rOpe=cmsy10                        % Ersatz for the non-standard rope font
 \def\ktl{{\hbox{\rOpe\char'170}}}        % top end for left-handed rope
 \def\kbl{{\hbox{\rOpe\char'170}}}        % bottom end for left-handed rope
 \def\kcr{{\reflectbox{\rOpe\char'170}}}        % right-handed rope
 \def\ktr{{\reflectbox{\rOpe\char'170}}}        % top end for right-handed rope
 \def\kbr{{\reflectbox{\rOpe\char'170}}}        % bottom end for right-handed rope
 \def\Border{\vbox{\hsize0pt% braided border
        \setlength{\unitlength}{1mm}
        \newcount\xco
        \newcount\yco
        \xco=-21
        \yco=12
        \begin{picture}(0,0)(-7.5,0)
        \put(\xco,\yco){$\ktl$}
        \advance\yco by-1
        {\loop
        \put(\xco,\yco){$\kcr$}
        \advance\yco by-2
        \ifnum\yco>-240
        \repeat
        \put(\xco,\yco){$\kbl$}}
        \xco=170
        \yco=12
        \put(\xco,\yco){$\ktr$}
        \advance\yco by-1
        {\loop
        \put(\xco,\yco){$\kcr$}
        \advance\yco by-2
        \ifnum\yco>-240
        \repeat
        \put(\xco,\yco){$\kbr$}}
        \put(-19.5,13){\scalebox{.6065}{%
         University of Maryland Center for String and Particle  Theory \&\ Physics Department%
        |University of Maryland Center for String and Particle  Theory \&\ Physics Department}}
        \put(-19.5,-241.5){\scalebox{.5835}{%
         ****University of Maryland * Center for String and
         Particle  Theory* Physics Department****University of Maryland *Center
        for String and Particle  Theory* Physics Department}}
        \end{picture}
        \par\vskip-8mm}}
\definecolor{UMred}{rgb}{.9,.05,.2}
\definecolor{HUblue}{rgb}{.0,.3,.7}

\definecolor{skyblue}{rgb}{0.12, 0.46, 1.00}
\definecolor{brightpink}{rgb}{1.0, 0.0, 0.5}
\definecolor{darkgreen}{rgb}{0.10, 0.75, 0.24}

\definecolor{Red}    {rgb}{0.90,0.00,0.12} %  1
\definecolor{Blue}   {rgb}{0.00,0.00,1.00} %  2
\definecolor{Green}  {rgb}{0.10,0.70,0.10} %  3
\definecolor{Turque} {rgb}{0.00,0.65,0.85} %  4
\definecolor{Orange} {rgb}{1.00,0.50,0.15} %  5
\definecolor{Magenta}{rgb}{1.00,0.00,1.00} %  6
\definecolor{Gold}   {rgb}{1.00,0.75,0.25} %  7
\definecolor{Seaweed}{rgb}{0.01,0.24,0.09} %  8
\definecolor{Purple} {rgb}{0.50,0.25,0.55} %  9
\definecolor{Brown}  {rgb}{0.43,0.26,0.32} % 10
\definecolor{grey1}  {rgb}{0.20,0.20,0.20} % 11
\definecolor{grey2}  {rgb}{0.40,0.40,0.40} % 12
\definecolor{grey3}  {rgb}{0.60,0.60,0.60} % 13
\definecolor{grey4}  {rgb}{0.80,0.80,0.80} % 14
\definecolor{grey5}  {rgb}{0.90,0.90,0.90} % 15
\def\C#1#2{{\ifcase#1\or%TH color scheme
             \color{Red}\or \color{Green}\or \color{Blue}\or\
              \color{Turque}\or \color{Orange}\or \color{Magenta}\or
               \color{Gold}\or \color{Seaweed}\or \color{Purple}\or
                \color{Brown}\or\color{grey1}\or\color{grey2}\or
                 \color{grey3}\else\color{grey4}\fi#2}}

\definecolor{Slate} {rgb}{0.00,0.45,0.55}

\newdimen\parshift\parshift=\parindent
\catcode`@=11
 \long\def\@footnotetext#1{\insert\footins{\reset@font\footnotesize
           \interlinepenalty\interfootnotelinepenalty\splittopskip%
            \footnotesep\splitmaxdepth\dp\strutbox\floatingpenalty\@MM%
             \hsize\columnwidth\addtolength{\hsize}{-2\parindent}
              \@parboxrestore\protected@edef\@currentlabel%
              {\csname p@footnote\endcsname\@thefnmark}%
                \color@begingroup%
                 \@makefntext{\rule\z@\footnotesep\ignorespaces#1%
                  \@finalstrut\strutbox}%
                \color@endgroup}}
 \long\def\@makefntext#1{\hglue\parshift%
           \vbox{\noindent\baselineskip=11pt plus.5pt minus.5pt\hb@xt@0em{\hss\@makefnmark\kern1pt}#1}}
\catcode`@=12

\newskip\humongous \humongous=0pt plus 1000pt minus 1000pt

\newif\ifdtup

\makeatletter
\def\section{\@startsection{section}{1}{\z@}
        {3ex plus-1ex minus-.2ex}{1pt plus1pt}{\large\sf\bfseries\boldmath}}
\def\subsection{\@startsection{subsection}{2}{\z@}
         {1.5ex plus-1ex minus-.2ex}{0.01pt plus1pt}{\sf\slshape}}
\def\subsubsection{\@startsection{subsubsection}{3}{\z@}
          {1.5ex plus-1ex minus-.2ex}{0.01pt plus0.2pt}{\sf\boldmath}}
\def\paragraph{\@startsection{paragraph}{4}{\z@}
           {.75ex \@plus.5ex \@minus.2ex}{-2mm}{\sf\bfseries\boldmath}}
\makeatother

 \allowdisplaybreaks
 \seceq

\newcommand{\aone}{{\un a}_1}
\newcommand{\bone}{{\un b}_1}
\newcommand{\cone}{{\un c}_1}
\newcommand{\done}{{\un d}_1}
\newcommand{\eone}{{\un e}_1}

\newcommand{\atwo}{{\un a}_2}

\newcommand{\athree}{{\un a}_3}

\newcommand{\ap}{{\un a}_p}
\newcommand{\bq}{{\un b}_q}
\newcommand{\cunq}{{\un c}_q}
\newcommand{\fone}{{\un f}_1}
\newcommand{\gone}{{\un g}_1}
\newcommand{\hone}{{\un h}_1}
\newcommand{\ione}{{\un i}_1}
\newcommand{\jone}{{\un j}_1}

\newcommand{\dr}{{\un d}_r}
\newcommand{\er}{{\un e}_r}
\newcommand{\fr}{{\un f}_r}
\newcommand{\gs}{{\un g}_s}
\newcommand{\hs}{{\un h}_s}
\newcommand{\is}{{\un i}_s}
\newcommand{\js}{{\un j}_s}

\newcommand{\tinysixteen}{{\raisebox{-0.03em}{\scriptsize $16$}}}

\newcommand{\tinyeights}{{\raisebox{-0.03em}{\scriptsize $8_s$}}}
\newcommand{\tinyeightc}{{\raisebox{-0.03em}{\scriptsize $8_c$}}}
\newcommand{\tinyeight}{{\raisebox{-0.03em}{\scriptsize $8$}}}
\newcommand{\tinyfour}{{\raisebox{-0.03em}{\scriptsize $4$}}}
\newcommand{\tinyfourbar}{{\raisebox{-0.03em}{\scriptsize $\overline{4}$}}}
\definecolor{skyblue}{rgb}{0.12, 0.46, 1.00}
\definecolor{brightpink}{rgb}{1.0, 0.0, 0.5}
\definecolor{darkgreen}{rgb}{0.10, 0.75, 0.24}

\begin{document}

\thispagestyle{empty}
\noindent{\small
\hfill{$~~$}  \\ % un-comment-out and specify when done}  
{}
}
\begin{center}
{\large \bf
Component Decompositions and Adynkra Libraries for \vskip0.02in
Supermultiplets in Lower Dimensional Superspaces
}   \\   [8mm]
{\large {
S.\ James Gates, Jr.\footnote{sylvester$_-$gates@brown.edu}${}^{,a, b}$,
%%%%%%%%%%%%%%%%%%%%%%%%%%%%%%%%%%%%%%%
Yangrui Hu\footnote{yangrui$_-$hu@brown.edu}${}^{,a,b}$, and
%%%%%%%%%%%%%%%%%%%%%%%%%%%%%%%%%%%%%%%
S.-N. Hazel Mak\footnote{sze$_-$ning$_-$mak@brown.edu}${}^{,a,b}$
%%%%%%%%%%%%%%%%%%%%%%%%%%%%%%%%%%%%%%%
}}
\\*[6mm]
\emph{
\centering
$^{a}$Brown Theoretical Physics Center,
\\[1pt]
Box S, 340 Brook Street, Barus Hall,
Providence, RI 02912, USA 
\\[10pt]
$^{b}$Department of Physics, Brown University,
\\[1pt]
Box 1843, 182 Hope Street, Barus \& Holley,
Providence, RI 02912, USA 
}
 \\*[85mm]
{ ABSTRACT}\\[5mm]
\parbox{142mm}{\parindent=2pc\indent\baselineskip=14pt plus1pt
We present Adynkra Libraries that can be used to explore the embedding of
multiplets of component field (whether on-shell or partial on-shell) within
Salam-Strathdee superfields for theories in dimension nine through four.}
 \end{center}
\vfill
\vfill
\clearpage

%%%%%%%%%%%%%%%%%%%%%%%%%%%%%%%%%%%%%%%%%%%%
%%%%%%%%%%%%%%%%%%%%%%%%%%%%%%%%%%%%%%%%%%%%
%%%%%%%%%%%%%%%%%%%%%%%%%%%%%%%%%%%%%%%%%%%%
%%%%%%%%%%%%%%%%%%%%%%%%%%%%%%%%%%%%%%%%%%%%

\newpage
{\hypersetup{linkcolor=black}
\tableofcontents
}

%%%%%%%%%%%%%%%%%%%%%%%%%%%%%%%%%%%%%%%%%%%%
%%%%%%%%%%%%%%%%%%%%%%%%%%%%%%%%%%%%%%%%%%%%
%%%%%%%%%%%%%%%%%%%%%%%%%%%%%%%%%%%%%%%%%%%%
%%%%%%%%%%%%%%%%%%%%%%%%%%%%%%%%%%%%%%%%%%%%
\newpage
\section{Introduction}

In a continuing series of works \cite{CNT10d,CNT11d,nDx}, we have been 
developing a new approach to supersymmetry and supermultiplets that possesses
the clarity of traditional component field approaches without their incompleteness.  
Simultaneously this new approach also retains the completeness of superfield 
approaches without their opaqueness.  Our approach, likely applicable in all 
dimensions\footnote{The realm in which our explicit constructions and discussions 
have been presented are for 10D and 11D supersymmetrical systems.}, has led to 
the construction of ``adynkrafields'' \cite{nDx}.  Adynkrafields are based on 
adinkra graphs  \cite{Adnk1} and Dynkin Labels \cite{SLnk,GrGi}.

An adynkrafield \cite{nDx} may be obtained by starting with a traditional 
Salam-Strathdee superfield \cite{SandS} and replacing the familiar Grassmann
coordinate expansions required to derive component results by two 
distinct types of special configurations of Young Tableaux (YT's) to 
derive component results.  Both sets of quantities ($\theta$-monomials
versus specialized YT combinations) for expansions, respectively, are
distinct sets of ``basis vectors'' for the description of component 
fields within off-shell supermultiplets.  

However, it is important to realize that taking a starting point that solely begins
with the considerations of Dynkin Labels, YT's and their properties is {\it {logically}} 
{\it {viable}} and independent of the concept of the Salam-Strathdee superfield.
This means that as far as the representation theory of supersymmetry and
supermultiplets go, the latter can be entirely replaced by ordinary concepts that
exist in Lie Algebra theory without reference to the superfield concept.

The algorithmically superiority of YT-expansions over $\theta$-expansions 
can be understood by an analogy.  It is well known in physics, when a 
problem has spherical symmetry (take for example the quantum mechanical 
hydrogen atom) the superior calculational direction is to employ spherical 
($r, \, \theta, \, \phi$) coordinates\footnote{In this triple of coordinates,
of course, $\theta$ denotes the polar angle, not the Grassmann coordinate.}, 
not rectilinear ($x, \, y, \, z$) ones.  Taking the path of using the triplet 
($r, \, \theta, \, \phi$) coordinates leads to the discovery of solutions in terms of 
``hydrogenic wavefunctions.'' If one is engaged in finding numerical solutions for 
atomic physics problems, efficient algorithms begin with the use of ``hydrogenic 
wavefunctions.''  Thus, we assert if one is engaged in finding component results 
for supersymmetric problems, efficient algorithms begin with the use of adynkrafields, 
not superfields.  The most powerful demonstration of this was given in the work seen 
in reference \cite{nDx}.

Continuing to use this analogy, when one examines the eigenfunctions of the 
Laplace operator in three spatial dimensions, the spherical harmonics naturally 
emerge as the angular portion of it solutions.  On the other hand, for the Laplace 
operator in two spatial dimensions the angular portion of the solution is dominated 
by complex exponentials.   Of course, complex exponentials are a part of the 
spherical harmonics.  This phenomenon of the emergence of a lower dimensional 
set of orthogonal function under a dimensional reduction is the main point of this 
current work.

In this approach, all component fields are reduced to two pieces of data,
the corresponding Dynkin Label, and the corresponding engineering height. 
The first piece of data encodes the Lorentz representation of the field and the
second controls how a component field can appear in an action.  In the
construction of adynkras, these two pieces of data are quite simple to track.
Thus when one is examining the question of whether there exists a superfield
into which any boson-fermion pair can be embedded, on a conceptual basis, 
the problem becomes identical to one where partial information concerning 
genetic (DNA/RNA) content (``a snippet'') is known and the question is one 
of finding the genetic sequence to which the snippet belongs.  Another way
to describe this is as a ``pattern-matching'' problem.  We are thus in the
position to create efficient algorithms capable of answering such questions 
about the embedding of component fields into superfields.

We wish to show how the adynkras of a theory in a lower spatial dimension
emerge from those of a higher dimension.  Our starting point will be the
adynkra for eleven dimensions.  Since we have previously given the answer
which emerges from the reduction to nine spatial dimensions, the new results in
this work will concentrate on the cases of 4 $\le$ D $\le$ 9.  For each of
these cases, we will derive the {\em unconstrained} scalar superfield adynkras that emerge
from the reduction from a higher dimension.

In Chapter two, we briefly review the construction of superspaces in various spacetime dimensions. Two approaches used to construct adynkrafields are reviewed as well. From Chapter three to Chapter eight, we present the following from 9D to 4D respectively, 
\vskip0.01in \indent
(a.) Minimal scalar superfield component decompositions into Lorentz representations,
\newline \indent
(b.) Adynkra and Adinkra diagrams,
\newline \indent
(c.) Young Tableaux descriptions of component fields, 
\newline \indent
(d.) Index structures and irreducible conditions of each component, and
\newline \indent
(e.) $(1,0)$ multiplet decompositions.
\vskip0.01in \noindent
Note that (e.) only applies to 8D and to 5D.

In each of the dimensions $\cal M$ of the associated Minkowski space under consideration, the adynkra $\cal {V_M}$ associated 
with the scalar superfield is explicitly presented.  This corresponds in every case to a library that explicitly
lists the Lorentz representations of all the component fields contained in the scalar superfield appropriate to that dimension.
Given the Dynkin Label description of any bosonic irrep (denoted by $\CMTB {\cal R}$) or any fermionic irrep (denoted by $\CMTR {\cal R}$),
the quantities ${\cal V}{}_{\CMTB {\cal R}} \,=\, {\CMTB {\cal R}} \otimes {\cal V_M}$ and ${\cal V}{}_{\CMTR {\cal R}} \,=\, 
{\CMTR {\cal R}} \otimes {\cal V_M}$ respective describe the Lorentz component field contents of {\it {any}} superfield
over the Minkowski space of dimension $\cal M$.

The content of the typical $\cal {V_M}$-library is a listing of Dynkin Labels and the associated heights at which the corresponding 
irreps appear.  For some heights, it is seen that a given Dynkin Label appears with a frequency greater than one.  This same 
phenomenon has been noted previously in the work of \cite{CGNN} where Dynkin Labels have been extensively used to describe the component field content of 11D superfields\footnote{In particular this is seen in the equation listed as (2.7) in this reference.}.

Chapter nine presents our conclusions.

There are two appendices included in this work. 
Appendix \ref{appen:irrep} contains a dictionary between Dynkin Labels and the corresponding representation 
dimensionalities of Lorentz groups in various spacetime dimensions. 
Appendix \ref{appen:proj} is devoted to present all necessary projection matrices for the branching rules used to develop minimal scalar superfields and $(1,0)$ superfields in various spacetime dimensions. 

The final portion of this paper includes our references.

\newpage
\section{Methodology}

A superspace consists of a set of spacetime coordinates and spinor coordinates. 
For each spacetime dimension D, there is a specific type of spinors and a minimal number of real components of the spinor coordinates that we call $d$. This information is listed in Table \ref{Tab:Clifford}.
We use the Minkowski signature $(-,+,+,\dots,+)$ in every dimension, and the corresponding Lorentz group is SO$(1,\rD-1)$.

\begin{table}[htp!]
\centering
\begin{tabular}{|c|c|c|c|}
\hline
Spacetime Dimension & Lorentz Group & Type of Spinors       & $d$ \\ \hline
11                  & SO(1,10)      & Majorana              & 32  \\ \hline
10                  & SO(1,9)       & Majorana-Weyl         & 16  \\ \hline
9                   & SO(1,8)       & Pseudo-Majorana       & 16  \\ \hline
8                   & SO(1,7)       & Pseudo-Majorana       & 16  \\ \hline
7                   & SO(1,6)       & SU(2)-Majorana        & 16  \\ \hline
6                   & SO(1,5)       & SU(2)-Majorana-Weyl   & 8   \\ \hline
5                   & SO(1,4)       & SU(2)-Majorana        & 8   \\ \hline
4                   & SO(1,3)       & Majorana/Weyl         & 4   \\ \hline
\end{tabular}
\caption{Summary of types of spinors in various dimensions \protect\cite{clifford} }
\label{Tab:Clifford}
\end{table}

In our previous research work \cite{CNT10d,CNT11d}, we decomposed each level of the 10D, ${\cal N}=1$, ${\cal N}=2$A, ${\cal N}=2$B, and 11D, ${\cal N}=1$ scalar superfields into Lorentz representations. 
A scalar superfield with $d$ spinor coordinates can be Taylor expanded into $(d+1)$ levels - from level-0 (zeroth power of spinor coordinates) to level-$d$ ($d$-th power of spinor coordinates).
In an arbitrary level-$n$, the decomposition of $\theta$-monomials translates to the antisymmetric product of $n$ spinor representation $\CMTred{\{d\}}$ in $\mathfrak{so}(1, \rD -1)$, which we denote by $[ \CMTred{\{d\}}^{\otimes \, n} ]_{A}$ or $\CMTred{\{d\}}^{\wedge \, n}$. The total degrees of freedom at level-$n$ is the binomial coefficient $\left(\begin{smallmatrix} d \\ n \end{smallmatrix}\right)$. Note that this is true only for {\em unconstrained} scalar superfields.

Two methods were used in obtaining the results of the antisymmetric products of each level in terms of direct sums of Lorentz irreducible representations. 
One utilizes the branching rules $\mathfrak{su}(d) \supset \mathfrak{so}(1, \rD -1)$ of the totally antisymmetric irreps in $\mathfrak{su}(d)$ constructed by the fundamental representation $\{d\}$. These calculations can be accomplished by either Susyno \cite{Susyno} or LieART \cite{LieART}. Both of them are availble online\footnote{The Susyno package is available at \url{https://renatofonseca.net/susyno}, while the LieART package is available at \url{https://lieart.hepforge.org/}.}. The relevant projection matrices used are listed in Appendix \ref{appen:proj}.
The other one involves the concept of plethysms, and the calculations are done by Susyno \cite{Susyno}. 
The details of these methods are explained explicitly in Chapter 4 of \cite{CNT11d}.

In \cite{nDx}, there is a complete methodological discussion on how to interpret both the bosonic and the spinorial irreps of $\mathfrak{so}(1,\rD-1)$, so that we can translate the scalar superfield components from Dynkin labels to Young Tableaux to field variables, and obtain the corresponding irreducible conditions. 
One can draw adynkra diagrams to encode all the information of a superfield. The links in the diagrams indicate sufficient but not necessary supersymmetry transformations between component fields, and the algorithm is described in Section 8.3 of \cite{nDx}.
These techniques are used in this paper.

\newpage
\section{9D Minimal Scalar Superfield Decomposition}

\subsection{Component Decompostion Results}

The 9D minimal superfield component decomposition results by Dynkin Labels are shown below.
\begin{itemize}
\item Level-0: $\CMTB{[0,0,0,0]}$
\item Level-1: $\CMTred{[0,0,0,1]}$
\item Level-2: $\CMTB{[0,1,0,0]}\oplus\CMTB{[0,0,1,0]}$
\item Level-3: $\CMTred{[1,0,0,1]}\oplus\CMTred{[0,1,0,1]}$
\item Level-4: $\CMTB{[2,0,0,0]}\oplus\CMTB{[0,0,0,2]}\oplus\CMTB{[1,1,0,0]}\oplus\CMTB{[0,2,0,0]}\oplus\CMTB{[1,0,0,2]}$
\item Level-5: $\CMTred{[1,0,0,1]}\oplus\CMTred{[0,1,0,1]}\oplus\CMTred{[2,0,0,1]}\oplus\CMTred{[0,0,0,3]}\oplus\CMTred{[1,1,0,1]}$
\item Level-6: $\CMTB{[0,1,0,0]}\oplus\CMTB{[0,0,1,0]}\oplus\CMTB{[1,1,0,0]}\oplus\CMTB{[1,0,1,0]}\oplus\CMTB{[2,1,0,0]}\oplus\CMTB{[1,0,0,2]}\oplus\CMTB{[2,0,1,0]}\oplus\CMTB{[0,1,0,2]}$
\item Level-7: $\CMTred{[0,0,0,1]}\oplus\CMTred{[1,0,0,1]}\oplus\CMTred{[0,1,0,1]}\oplus\CMTred{[2,0,0,1]}\oplus\CMTred{[0,0,1,1]}\oplus\CMTred{[3,0,0,1]}\oplus\CMTred{[1,1,0,1]}\oplus\CMTred{[1,0,1,1]}$
\item Level-8: $\CMTB{[0,0,0,0]}\oplus\CMTB{[1,0,0,0]}\oplus\CMTB{[2,0,0,0]}\oplus\CMTB{[0,0,1,0]}\oplus\CMTB{[0,0,0,2]}\oplus\CMTB{[3,0,0,0]}\oplus\CMTB{[4,0,0,0]}\oplus\CMTB{[0,2,0,0]}\oplus\CMTB{[1,0,1,0]}\oplus\CMTB{[1,0,0,2]}\oplus\CMTB{[0,1,1,0]}\oplus\CMTB{[0,0,2,0]}\oplus\CMTB{[2,0,1,0]}\oplus\CMTB{[2,0,0,2]}$
\item Level-9: $\CMTred{[0,0,0,1]}\oplus\CMTred{[1,0,0,1]}\oplus\CMTred{[0,1,0,1]}\oplus\CMTred{[2,0,0,1]}\oplus\CMTred{[0,0,1,1]}\oplus\CMTred{[3,0,0,1]}\oplus\CMTred{[1,1,0,1]}\oplus\CMTred{[1,0,1,1]}$
\item Level-10: $\CMTB{[0,1,0,0]}\oplus\CMTB{[0,0,1,0]}\oplus\CMTB{[1,1,0,0]}\oplus\CMTB{[1,0,1,0]}\oplus\CMTB{[2,1,0,0]}\oplus\CMTB{[1,0,0,2]}\oplus\CMTB{[2,0,1,0]}\oplus\CMTB{[0,1,0,2]}$
\item Level-11: $\CMTred{[1,0,0,1]}\oplus\CMTred{[0,1,0,1]}\oplus\CMTred{[2,0,0,1]}\oplus\CMTred{[0,0,0,3]}\oplus\CMTred{[1,1,0,1]}$
\item Level-12: $\CMTB{[2,0,0,0]}\oplus\CMTB{[0,0,0,2]}\oplus\CMTB{[1,1,0,0]}\oplus\CMTB{[0,2,0,0]}\oplus\CMTB{[1,0,0,2]}$
\item Level-13: $\CMTred{[1,0,0,1]}\oplus\CMTred{[0,1,0,1]}$
\item Level-14: $\CMTB{[0,1,0,0]}\oplus\CMTB{[0,0,1,0]}$
\item Level-15: $\CMTred{[0,0,0,1]}$
\item Level-16: $\CMTB{[0,0,0,0]}$
\end{itemize}

The component decompostion results by dimensions are shown below.
\begin{itemize}
    \item Level-0: $ \CMTB{\{1\}}$
    \item Level-1: $ \CMTred{\{16\}}$
    \item Level-2: $ \CMTB{\{36\}} \oplus \CMTB{\{84\}}$
    \item Level-3: $ \CMTred{\{128\}} \oplus \CMTred{\{432\}}$
    \item Level-4: $\CMTB{  \{44\}} \oplus \CMTB{\{126\}} \oplus \CMTB{\{231\}} \oplus \CMTB{\{495\}} \oplus \CMTB{\{924\}}$
    \item Level-5: $ \CMTred{\{128\}} \oplus \CMTred{\{432\}} \oplus \CMTred{\{576\}} \oplus \CMTred{\{672\}} \oplus \CMTred{\{2560\}}$
    \item Level-6: $ \CMTB{\{36\}} \oplus \CMTB{\{84\}} \oplus \CMTB{\{231\}} \oplus \CMTB{\{594\}} \oplus \CMTB{\{910\}} \
\oplus \CMTB{\{924\}} \oplus \CMTB{\{2457\}} \oplus \CMTB{\{2772\}}$
    \item Level-7: $ \CMTred{\{16\}} \oplus \CMTred{\{128\}} \oplus \CMTred{\{432\}} \oplus \CMTred{\{576\}} \oplus \CMTred{\{768\}} \
\oplus \CMTred{\{1920\}} \oplus \CMTred{\{2560\}} \oplus \CMTred{\{5040\}}$
    \item Level-8: $ \CMTB{\{1\}} \oplus \CMTB{\{9\}} \oplus \CMTB{  \{44\}} \oplus \CMTB{\{84\}} \oplus \CMTB{\{126\}} \oplus \
\CMTB{\{156\}} \oplus \CMTB{\{450\}} \oplus \CMTB{\{495\}} \oplus \CMTB{\{594\}} \oplus \CMTB{\{924\}} \
\oplus \CMTB{\{1650\}} \oplus \CMTB{\{1980\}} \oplus \CMTB{\{2457\}} \oplus \CMTB{\{3900\}}$
\item Level-9: $ \CMTred{\{16\}} \oplus \CMTred{\{128\}} \oplus \CMTred{\{432\}} \oplus \CMTred{\{576\}} \oplus \CMTred{\{768\}} \
\oplus \CMTred{\{1920\}} \oplus \CMTred{\{2560\}} \oplus \CMTred{\{5040\}}$
\item Level-10: $ \CMTB{\{36\}} \oplus \CMTB{\{84\}} \oplus \CMTB{\{231\}} \oplus \CMTB{\{594\}} \oplus \CMTB{\{910\}} \
\oplus \CMTB{\{924\}} \oplus \CMTB{\{2457\}} \oplus \CMTB{\{2772\}}$
 \item Level-11: $ \CMTred{\{128\}} \oplus \CMTred{\{432\}} \oplus \CMTred{\{576\}} \oplus \CMTred{\{672\}} \oplus \CMTred{\{2560\}}$
 \item Level-12: $\CMTB{  \{44\}} \oplus \CMTB{\{126\}} \oplus \CMTB{\{231\}} \oplus \CMTB{\{495\}} \oplus \CMTB{\{924\}}$
  \item Level-13: $ \CMTred{\{128\}} \oplus \CMTred{\{432\}}$
  \item Level-14: $ \CMTB{\{36\}} \oplus \CMTB{\{84\}}$
   \item Level-15: $ \CMTred{\{16\}}$
    \item Level-16: $\CMTB{\{1\}}$
\end{itemize}

\subsection{9D Minimal Adinkra Diagram}
\label{sec:9DAdinkra}

\begin{figure}[htp!]
\centering
\includegraphics[width=0.5\textwidth]{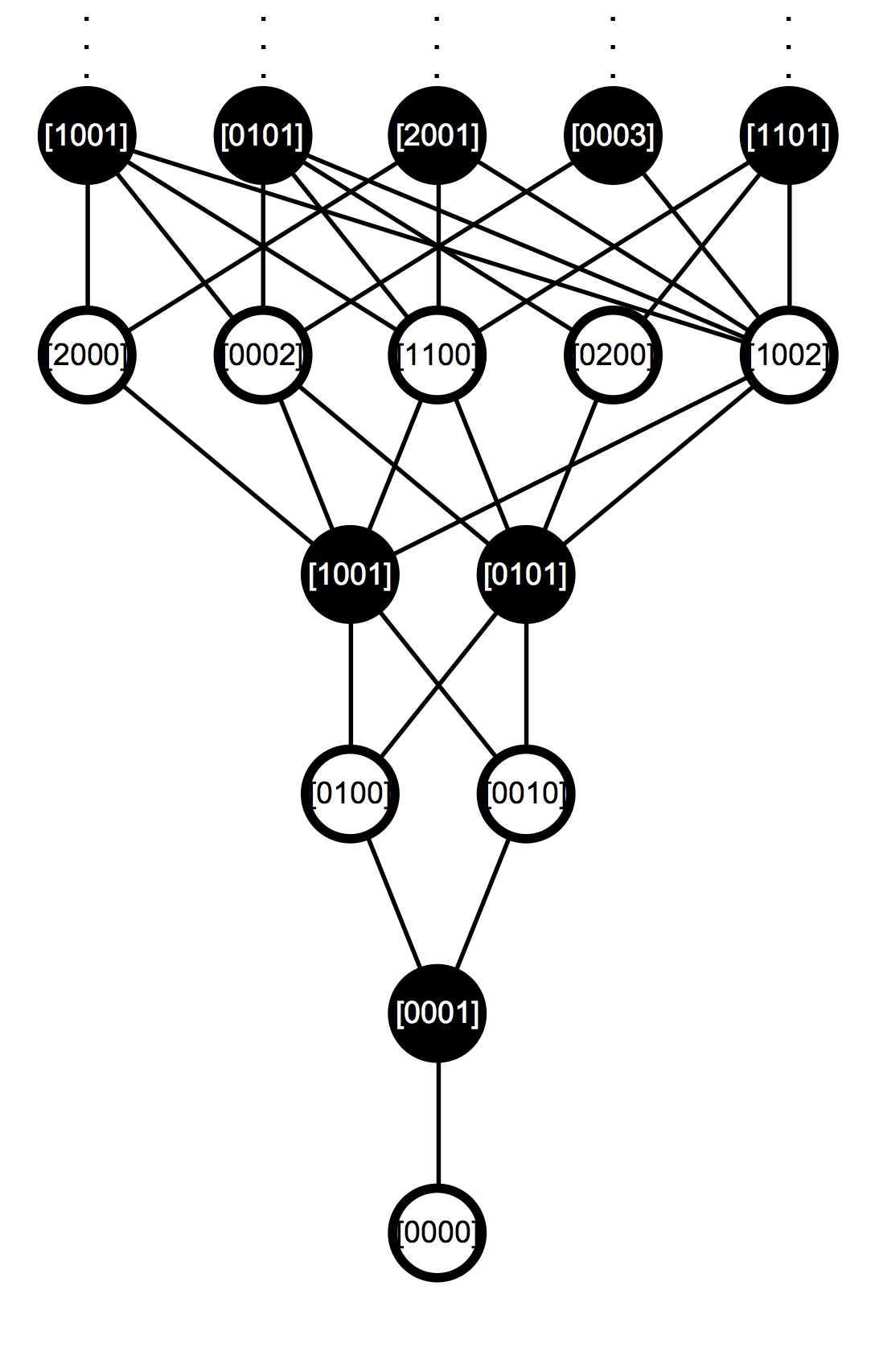}
\caption{Adynkra Diagram for 9D minimal scalar superfield}
\label{Fig:9D_Dynkin}
\end{figure}

\begin{figure}[htp!]
\centering
\includegraphics[width=0.5\textwidth]{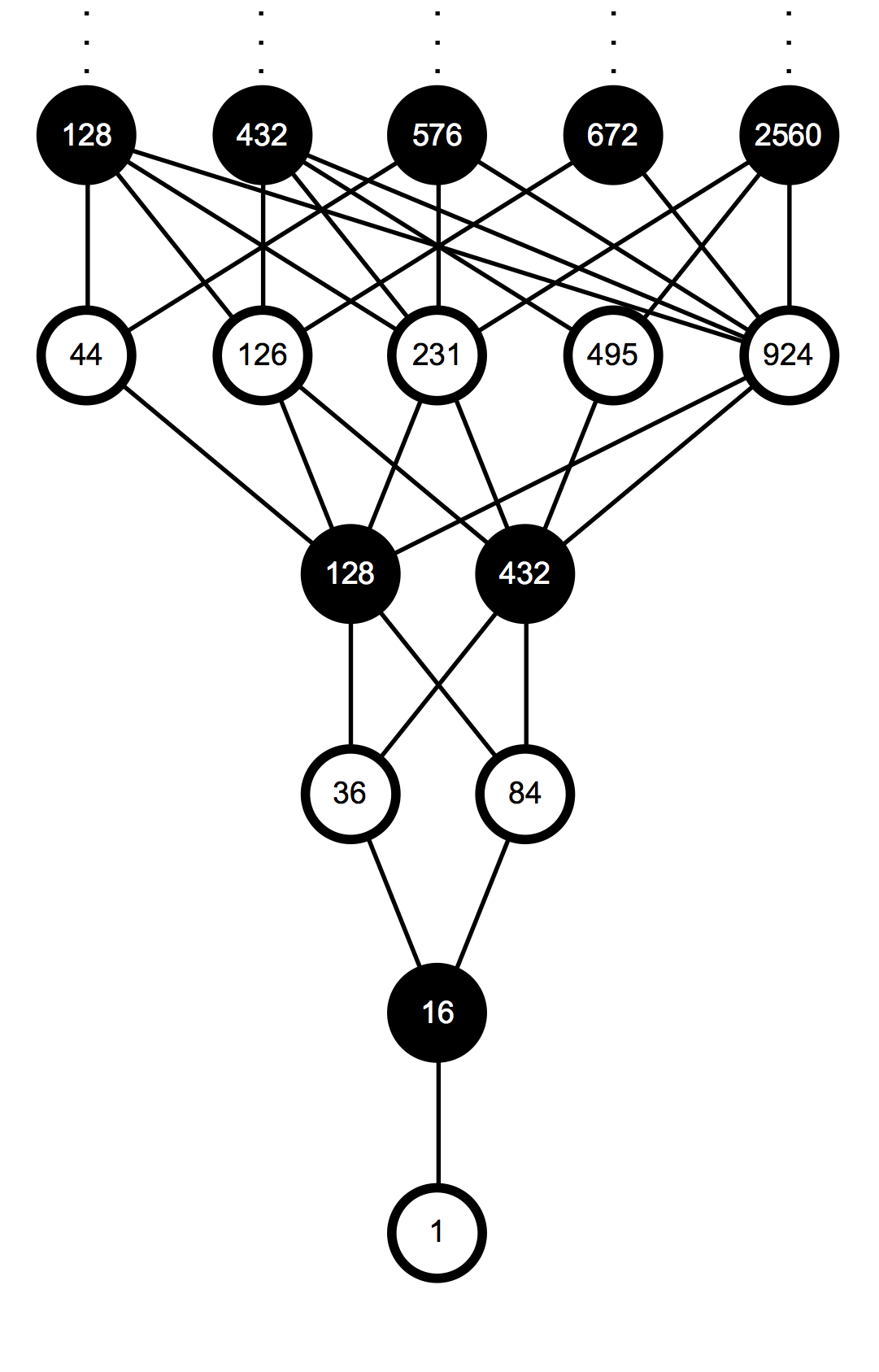}
\caption{Adinkra Diagram for 9D minimal scalar superfield}
\label{Fig:9D}
\end{figure}

The component content of 9D minimal scalar superfield (up to Level-5) is summarized in the Adynkra diagram in Figure~\ref{Fig:9D_Dynkin} and Adinkra diagram in Figure \ref{Fig:9D}. The definitions of Adinkra diagrams in spacetime dimension larger than one and Adynkra diagrams were presented in \cite{CNT10d,nDx}.

\newpage

\subsection{Young Tableaux Descriptions of Component Fields in 9D Minimal Scalar Superfield}

In order to establish a graphical language to describe bosonic irreducible representations
in $\mathfrak{so}(9)$, we define {\em irreducible} bosonic Young Tableaux as what we did in 10D~\cite{nDx}.

Consider the projection matrix for $\mathfrak{su}(9)\supset \mathfrak{so}(9)$~\cite{yamatsu2015},
\begin{equation}
P_{\mathfrak{su}(9)\supset \mathfrak{so}(9)} ~=~
\begin{pmatrix}
1 & 0 & 0 & 0 & 0 & 0 & 0 & 1\\
0 & 1 & 0 & 0  & 0 & 0 & 1 & 0\\
0 & 0 & 1 & 0  & 0 & 1 & 0 & 0\\
0 & 0 & 0 & 2  & 2 & 0 & 0 & 0\\
\end{pmatrix}~~~.
\end{equation}
The highest weight of a specified irrep of $\mathfrak{su}(9)$ is a row vector $[p_1,p_2,p_3,p_4,p_5,p_6,p_7,p_8]$,
where $p_1$ to $p_8$ are non-negative integers. Since the $\mathfrak{su}(9)$ YT with $n$ vertical 
boxes is the conjugate of the one with $9-n$ vertical boxes, we need only consider the $p_5=p_6=p_7=p_8=0$ case.

Starting from the weight vector $[p_1,p_2,p_3,p_4,0,0,0,0]$ in $\mathfrak{su}(9)$, we define its projected
weight vector $[p_1,p_2,p_3,2p_4]$ in $\mathfrak{so}(9)$ as the Dynkin Label of the corresponding
irreducible bosonic Young Tableau.
\begin{equation}
    [p_1,p_2,p_3,2p_4] ~=~ [p_1,p_2,p_3,p_4,0,0,0,0] \, P^T_{\mathfrak{su}(9)\supset\mathfrak{so}(9)} ~~~.
\end{equation}
Thus, given an irreducible bosonic Young Tableau with $p_1$ columns of one box,
$p_2$ columns of two vertical boxes, $p_3$ columns of three vertical boxes, and $p_4$ columns of
four vertical boxes, the Dynkin Label of its
corresponding bosonic irrep is $[p_1,p_2,p_3,2p_4]$. Then look at the congruence classes of a representation with 
Dynkin Label $[a,b,c,d]$ in $\mathfrak{so}(9)$,
\begin{equation}
    \begin{split}
        C_{c}(R) ~:=&~  d ~~ ({\rm mod} ~ 2)  ~~~.
    \end{split}
\end{equation}
$C_{c}(R)$ classifies the bosonic irreps and spinorial
irreps:  $C_{c}(R) = 0$ is bosonic and  $C_{c}(R)=1$ is spinorial.  Consequently, a
bosonic irrep satisfies  $d
= 0 ~~ ({\rm mod} ~ 2)$.

Therefore, we can reverse this process and show the
one-to-one correspondence between bosonic Dynkin Label irreps and irreducible bosonic Young Tableaux.
Namely, given a bosonic irrep with Dynkin Label $[a,b,c,d]$, its corresponding irreducible 
bosonic Young Tableau is composed of $a$ columns of one box, $b$ columns of two 
vertical boxes, $c$ columns of three vertical boxes, and $d/2$ columns of four vertical boxes.

The simplest examples,
also the fundamental building blocks of a BYT, are given below.
\begin{equation}
\begin{gathered}
    {\CMTB{\ydiagram{1}}}_{{\rm IR}} ~\equiv~ \CMTB{[1,0,0,0]}  ~~~,~~~
    {\CMTB{\ydiagram{1,1}}}_{{\rm IR}} ~\equiv~ \CMTB{[0,1,0,0]} ~~~, \\
    {\CMTB{\ydiagram{1,1,1}}}_{{\rm IR}} ~\equiv~  \CMTB{[0,0,1,0]} ~~~,~~~
    {\CMTB{\ydiagram{1,1,1,1}}}_{{\rm IR}} ~\equiv~ \CMTB{[0,0,0,2]} ~~~.
\end{gathered} \label{equ:BYTbasic_9D}
\end{equation}
The YT's with ``IR" subscript subject to certain conditions refer 
to irreducible representations of $\mathfrak{so}(9)$.
YT's without this subscript are reducible with respect to $\mathfrak{so}(9)$ while irreducible with respect to $\mathfrak{su}(9)$.
The irreducibility conditions are effectuated by the branching rules for $\mathfrak{su}(9) 
\supset \mathfrak{so}(9)$, which can also be derived by ``tying rules" invented in \cite{nDx}.

For spinorial irreps, the basic SYT is given by
\begin{equation}
    \CMTred{\ytableaushort{\tinysixteen}} ~\equiv~ \CMTred{[0,0,0,1]} ~~~.
\label{equ:SYTbasic_9D}
\end{equation}
We could translate the Dynkin Label of any spinorial irrep to a mixed YT (which contains a BYT part and a basic SYT above) with irreducible conditions by applying the same idea discussed in Chapter five of \cite{nDx}. 

Putting together the columns in (\ref{equ:BYTbasic_9D}) and (\ref{equ:SYTbasic_9D}) corresponds to adding their Dynkin Labels.

In summary, the irreducible Young Tableau descriptions of the 9D minimal scalar superfield decomposition is presented below.
\begin{equation}
\ytableausetup{boxsize=0.8em}
{\cal V} ~=~ \begin{cases}
~~{\rm {Level}}-0 \,~~~~~~~~~~ \CMTB{\cdot} ~~~,  \\
~~{\rm {Level}}-1 \,~~~~~~~~~~ \CMTred{\ytableaushort{\tinysixteen}} ~~~,  \\
~~{\rm {Level}}-2 \,~~~~~~~~~~ {\CMTB{\ydiagram{1,1}}}_{\rm IR}~\oplus~{\CMTB{\ydiagram{1,1,1}}}_{\rm IR} ~~~,  \\[15pt]
~~{\rm {Level}}-3 \,~~~~~~~~~~
{\CMTB{\ydiagram{1}}\CMTred{\ytableaushort{\tinysixteen}}}_{\rm IR}~\oplus~{\CMTB{\ydiagram{1,1}}\CMTred{\ytableaushort{\tinysixteen,\none}}}_{\rm IR}  ~~~,  \\[10pt]
~~{\rm {Level}}-4 \,~~~~~~~~~~ {\CMTB{\ydiagram{2}}}_{\rm IR} ~\oplus~{\CMTB{\ydiagram{2,1}}}_{\rm IR} ~\oplus~
{\CMTB{\ydiagram{2,2}}}_{\rm IR} ~\oplus~
{\CMTB{\ydiagram{1,1,1,1}}}_{\rm IR} ~\oplus~
{\CMTB{\ydiagram{2,1,1,1}}}_{\rm IR}
 ~~~,  \\
~~{\rm {Level}}-5 \,~~~~~~~~~~ {\CMTB{\ydiagram{1,1,1,1}}\CMTred{\ytableaushort{\tinysixteen,\none,\none,\none}}}_{{\rm IR}} ~\oplus~{\CMTB{\ydiagram{1}}\CMTred{\ytableaushort{\tinysixteen}}}_{\rm IR}~\oplus~ {\CMTB{\ydiagram{1,1}}\CMTred{\ytableaushort{\tinysixteen,\none}}}_{\rm IR}~\oplus~ {\CMTB{\ydiagram{2}}\CMTred{\ytableaushort{\tinysixteen}}}_{\rm IR}
~\oplus~ {\CMTB{\ydiagram{2,1}}\CMTred{\ytableaushort{\tinysixteen,\none}}}_{\rm IR}
~~~,  \\[20pt]
~~{\rm {Level}}-6 \,~~~~~~~~~~ {\CMTB{\ydiagram{2,1,1,1}}}_{{\rm IR}}  ~\oplus~ {\CMTB{\ydiagram{2,2,1,1}}}_{\rm IR}
~\oplus~ {\CMTB{\ydiagram{1,1}}}_{\rm IR}
~\oplus~ {\CMTB{\ydiagram{1,1,1}}}_{\rm IR}
~\oplus~ {\CMTB{\ydiagram{2,1}}}_{\rm IR}
~\oplus~ {\CMTB{\ydiagram{2,1,1}}}_{\rm IR}\\[20pt]
~~~~~~~~~~~~~~~~~~~~~~
~\oplus~ {\CMTB{\ydiagram{3,1}}}_{\rm IR}
~\oplus~ {\CMTB{\ydiagram{3,1,1}}}_{\rm IR}
~~~,  \\[20pt]
~~{\rm {Level}}-7 \,~~~~~~~~~~
\CMTred{\ytableaushort{\tinysixteen}}~\oplus~
{\CMTB{\ydiagram{1}}\CMTred{\ytableaushort{\tinysixteen}}}_{\rm IR}~\oplus~{\CMTB{\ydiagram{1,1}}\CMTred{\ytableaushort{\tinysixteen,\none}}}_{\rm IR}~\oplus~
{\CMTB{\ydiagram{2}}\CMTred{\ytableaushort{\tinysixteen}}}_{\rm IR} ~\oplus~
{\CMTB{\ydiagram{1,1,1}}\CMTred{\ytableaushort{\tinysixteen,\none,\none}}}_{\rm IR} ~\oplus~
{\CMTB{\ydiagram{3}}\CMTred{\ytableaushort{\tinysixteen}}}_{\rm IR} \\
~~~~~~~~~~~~~~~~~~~~~~
~\oplus~
{\CMTB{\ydiagram{2,1}}\CMTred{\ytableaushort{\tinysixteen,\none}}}_{\rm IR} ~\oplus~
{\CMTB{\ydiagram{2,1,1}}\CMTred{\ytableaushort{\tinysixteen,\none,\none}}}_{\rm IR}
~~~,  \\
~~{\rm {Level}}-8 \,~~~~~~~~~~ \CMTB{\cdot}~\oplus~
{\CMTB{\ydiagram{1}}}_{\rm IR} ~\oplus~
{\CMTB{\ydiagram{2}}}_{\rm IR} ~\oplus~
{\CMTB{\ydiagram{3}}}_{\rm IR} ~\oplus~
{\CMTB{\ydiagram{4}}}_{\rm IR} ~\oplus~
{\CMTB{\ydiagram{2,2}}}_{\rm IR} ~\oplus~
{\CMTB{\ydiagram{2,2,1}}}_{\rm IR} \\[15pt]
~~~~~~~~~~~~~~~~~~~~~~
~\oplus~{\CMTB{\ydiagram{2,2,2}}}_{\rm IR} ~\oplus~ 
{\CMTB{\ydiagram{1,1,1}}}_{\rm IR} ~\oplus~
{\CMTB{\ydiagram{1,1,1,1}}}_{\rm IR} ~\oplus~
{\CMTB{\ydiagram{2,1,1}}}_{\rm IR} ~\oplus~
{\CMTB{\ydiagram{2,1,1,1}}}_{\rm IR}~\oplus~
{\CMTB{\ydiagram{3,1,1}}}_{\rm IR} ~\oplus~
{\CMTB{\ydiagram{3,1,1,1}}}_{\rm IR} 
~~~,  \\
{~~~~~~}  {~~~~} \vdots  {~~~~~~~~~\,~~~~~~} \vdots
\end{cases}
\label{equ:V_9D}
\ytableausetup{boxsize=1.2em}
\end{equation}
where Level-9 to Level-16 have exactly the same expressions as Level-7 to Level-0.

\subsection{Index Structures and Irreducible Conditions of Component Fields in 9D Minimal Scalar Superfield}\label{sec:9D_Index}

In this section, we will translate the irreducible bosonic and mixed Young Tableaux into field variables. 
In order to express the index
structure clearly and efficiently, we introduce the following notational conventions for irreducible bosonic YT.
We use ``$|$'' to separate indices in YT with different heights and ``,'' to separate
indices in YT with the same heights. 

The vector index $\un{a}$ runs from 0 to 8. 
In fact, the $\{\}$-indices, irreducible bosonic Young
Tableaux, and Dynkin Labels are equivalent and have the one-to-one correspondence.

The general expression is as below in Equations (\ref{equ:index-notation1_9D}) and (\ref{equ:index-notation2_9D}),
\begin{equation}
    \begin{split}
       & \{ \un{a}_1,\dots,\un{a}_p ~|~ \un{b}_1\un{c}_1,\dots,\un{b}_q \un{c}_q ~|~  \un{d}_1\un{e}_1\un{f}_1,\dots,\un{d}_r\un{e}_r\un{f}_r  ~|~ \un{g}_1\un{h}_1\un{i}_1\un{j}_1,\dots, \un{g}_s\un{h}_s\un{i}_s\un{j}_s\}  \\[10pt]
      & ~{ \CMTB{{\ytableaushort{\aone}} } }~~~~~{ \CMTB{{\ytableaushort{\ap}} } }
      ~~~~{ \CMTB{{\ytableaushort{\bone,\cone}} } }~~~~~~~{ \CMTB{{\ytableaushort{\bq,\cunq}} } }
      ~~~~~~{ \CMTB{{\ytableaushort{\done,\eone,\fone}} } }~~~~~~~~~~~{ \CMTB{{\ytableaushort{\dr,\er,\fr}} } }
      ~~~~~~~~~{ \CMTB{{\ytableaushort{\gone,\hone,\ione,\jone}} } }~~~~~~~~~~~~~~{\CMTB{{\ytableaushort{\gs,\hs,\is,\js}} } }
      \\[10pt]
       & ~~ \CMTB{[p,0,0,0]} ~~~~~~ \CMTB{[0,q,0,0]} ~~~~~~~~~~~ \CMTB{[0,0,r,0]}  ~~~~~~~~~~~~~~~~~ \CMTB{[0,0,0,2s]}
    \end{split}
    \label{equ:index-notation1_9D}
\end{equation}
where above we have ``disassembled'' the YT to show how each column is affiliated with each type of
subscript structure. Below we have assembled all the columns into a proper YT.
\begin{equation}
\begin{gathered}
    { \CMTB{{\ytableaushort{\gone\dots\gs\done\dots\dr\bone\dots\bq\aone\dots\ap,\hone\dots\hs\eone\dots\er\cone\dots\cunq,\ione\dots\is\fone\dots\fr,\jone\dots\js}} } }_{\rm IR}\\[10pt]
    \CMTB{[p,q,r,2s]}
\end{gathered}
\label{equ:index-notation2_9D}
\end{equation}

As one moves from the YT's shown in Equation~(\ref{equ:index-notation1_9D}) to
Equation~(\ref{equ:index-notation2_9D}), it is clear the number of vertical boxes is tabulating the number of
1-forms, 2-forms, 3-forms, and 4-forms in the YT's.  These are the entries between
the vertical $|$ bars. These precisely correspond to the integers $p$, $q$, $r$, and $s$ in the Dynkin Labels.  
An example of the correspondence between the subscript conventions,
the affiliated YT, and Dynkin Label is shown in (\ref{equ:index-notation_ex1_9D}). 
\begin{equation}
    \{{\un a}_2 , {\un a}_3| {\un a}_1  {\un b}_1   {\un c}_1  {\un d}_1\} ~~\equiv~~
    { \CMTB{{\ytableaushort{\aone \atwo \athree,\bone,\cone,\done}} } }_{{\rm IR}}
    ~~\equiv~~ \CMTB{[2,0,0,2]} ~~~.
    \label{equ:index-notation_ex1_9D}
\end{equation}

The index structures as well as irreducible conditions of all bosonic and fermionic fields are identified below along 
with the level at which the fields occur in the adinkra of the superfield. The spinor index $\alpha$ runs from 1 to 16. 

\begin{itemize}
    \item Level-0: $\Phi(x)$~~~,
    %%%%%%%%%%%%%%%%%%%%%%%%%%%%%%%%%%%%%%%%
    \item Level-1: $\Psi_{\alpha}(x)$~~~,
    %%%%%%%%%%%%%%%%%%%%%%%%%%%%%%%%%%%%%%%%
    \item Level-2: $\Phi_{ \{\aone\bone\} }(x)$ ~~~,~~~ $\Phi_{ \{\aone\bone\cone\} }(x)$~~~,
    %%%%%%%%%%%%%%%%%%%%%%%%%%%%%%%%%%%%%%%%
    \item Level-3: 
    $\Psi_{\{\aone\}\alpha}(x):~ (\g^{\aone})_{\beta}{}^{\alpha}\Psi_{\{\aone\}\alpha}(x)~=~0 $~~~,~~~
    $\Psi_{\{\aone\bone\}\alpha}(x):~ (\g^{\aone})_{\beta}{}^{\alpha}\Psi_{\{\aone\bone\}\alpha}(x)~=~0 $~~~,
    %%%%%%%%%%%%%%%%%%%%%%%%%%%%%%%%%%%%%%%%
    \item Level-4: $\Phi_{ \{\aone,\un{a}_2 \} }(x):~ \eta^{\aone\un{a}_2}\Phi_{ \{\aone,\un{a}_2 \} }(x)~=~0$~~~,~~~
       $\Phi_{ \{\un{a}_2|\aone\bone \} }(x):~ \eta^{\aone\un{a}_2}\Phi_{ \{\un{a}_2|\aone\bone \} }(x)~=~0$~~~,\\[10pt]
       $ \Phi{}_{\{{\un a}_1  {\un b}_1,   {\un a}_2
{\un b}_2 \}}(x)$~~:~~ $\begin{cases}
\eta^{{\un a}_1  {\un a}_2}\Phi{}_{\{{\un a}_1  {\un b}_1,   {\un a}_2
{\un b}_2 \}}(x)~=~0~~~,\\
\eta^{{\un a}_1  {\un a}_2}\eta^{{\un b}_1  {\un b}_2}\Phi{}_{\{{\un a}_1  {\un b}_1,  
{\un a}_2 {\un b}_2 \}}(x)~=~0~~~,
\end{cases}$\\[10pt]
$\Phi_{ \{\aone\bone\cone\done\} }(x)$~~~,~~~
$\Phi_{ \{\un{a}_2|\aone\bone\cone\done\} }(x):~ \eta^{\aone\un{a}_2}\Phi_{ \{\un{a}_2|\aone\bone\cone\done\} }(x)~=~0$~~~,
%%%%%%%%%%%%%%%%%%%%%%%%%%%%%%%%%%%%%%%%
    \item Level-5:
    $\Psi_{\{\aone\bone\cone\done\}\alpha}(x):~ (\g^{\aone})_{\beta}{}^{\alpha}\Psi_{\{\aone\bone\cone\done\}\alpha}(x)~=~0 $~~~,\\[10pt]
    $\Psi_{\{\aone\}\alpha}(x):~ (\g^{\aone})_{\beta}{}^{\alpha}\Psi_{\{\aone\}\alpha}(x)~=~0 $~~~,~~~
    $\Psi_{\{\aone\bone\}\alpha}(x):~ (\g^{\aone})_{\beta}{}^{\alpha}\Psi_{\{\aone\bone\}\alpha}(x)~=~0 $~~~,\\[10pt]
    $\Psi_{\{\aone,\un{a}_2\}\alpha}(x):~\begin{cases}
    (\g^{\aone})_{\beta}{}^{\alpha}\Psi_{\{\aone,\un{a}_2\}\alpha}(x)~\equiv~ \psi_{\{\un{a}_2\}\beta} ~=~0~~~, \\
    (\g^{\un{a}_2})_{\g}{}^{\beta}\psi_{\{\un{a}_2\}\beta} ~=~0~~~,
    \end{cases}$\\[10pt]
    $\Psi_{\{\un{a}_2|\aone\bone\}\alpha}(x):~\begin{cases}
    (\g^{\bone})_{\beta}{}^{\alpha}\Psi_{\{\un{a}_2|\aone\bone\}\alpha}(x)~\equiv~ \psi_{\{\aone,\un{a}_2\}\beta} ~=~0 ~~~,\\
    (\g^{\un{a}_2})_{\beta}{}^{\alpha}\Psi_{\{\un{a}_2|\aone\bone\}\alpha}(x)~\equiv~ \psi_{\{\aone\bone\}\beta} ~=~0~~~, \\
    (\g^{\bone})_{\g}{}^{\beta}\psi_{\{\aone\bone\}\beta} ~=~0 ~~~,
    \end{cases}$
    %%%%%%%%%%%%%%%%%%%%%%%%%%%%%%%%%%%%%%%%
    \item Level-6: 
    $\Phi_{ \{\un{a}_2|\aone\bone\cone\done\} }(x):~ \eta^{\aone\un{a}_2}\Phi_{ \{\un{a}_2|\aone\bone\cone\done\} }(x)~=~0$~~~,\\[10pt]
    $\Phi_{ \{\un{a}_2\un{b}_2|\aone\bone\cone\done\} }(x):~ \begin{cases}
    \eta^{\aone\un{a}_2}\Phi_{ \{\un{a}_2\un{b}_2|\aone\bone\cone\done\} }(x)~=~0~~~,\\
    \eta^{\aone\un{a}_2}\eta^{{\un b}_1  {\un b}_2}\Phi_{ \{\un{a}_2\un{b}_2|\aone\bone\cone\done\} }(x)~=~0~~~,
    \end{cases}$\\[10pt]
    $\Phi_{ \{\aone\bone\} }(x)$~~~,~~~ 
    $\Phi_{ \{\aone\bone\cone\} }(x)$~~~, ~~~
    $\Phi_{ \{\un{a}_2|\aone\bone \} }(x):~ \eta^{\aone\un{a}_2}\Phi_{ \{\un{a}_2|\aone\bone \} }(x)~=~0$~~~,\\[10pt]
    $\Phi_{ \{\un{a}_2|\aone\bone\cone \} }(x):~ \eta^{\aone\un{a}_2}\Phi_{ \{\un{a}_2|\aone\bone\cone \} }(x)~=~0$~~~,\\[10pt]
    $\Phi_{ \{\un{a}_2,\un{a}_3|\aone\bone\} }(x):~ \begin{cases}
    \eta^{\un{a}_2\un{a}_3}\Phi_{ \{\un{a}_2,\un{a}_3|\aone\bone\} }(x)~=~0~~~,\\
    \eta^{\un{b}_1\un{a}_3}\Phi_{ \{\un{a}_2,\un{a}_3|\aone\bone\} }(x)~=~0~~~,
    \end{cases}$\\[10pt]
    $\Phi_{ \{\un{a}_2,\un{a}_3|\aone\bone\cone\} }(x):~ \begin{cases}
    \eta^{\un{a}_2\un{a}_3}\Phi_{ \{\un{a}_2,\un{a}_3|\aone\bone\cone\} }(x)~=~0~~~,\\
    \eta^{\un{c}_1\un{a}_3}\Phi_{ \{\un{a}_2,\un{a}_3|\aone\bone\cone\} }(x)~=~0~~~,
    \end{cases}$\\[10pt]
    %%%%%%%%%%%%%%%%%%%%%%%%%%%%%%%%%%%%%%%%
    \item Level-7: $\Psi_{\alpha}(x)$~~~,~~~
    $\Psi_{\{\aone\}\alpha}(x):~ (\g^{\aone})_{\beta}{}^{\alpha}\Psi_{\{\aone\}\alpha}(x)~=~0 $~~~,\\[10pt]
    $\Psi_{\{\aone\bone\}\alpha}(x):~ (\g^{\aone})_{\beta}{}^{\alpha}\Psi_{\{\aone\bone\}\alpha}(x)~=~0 $~~~,\\[10pt]
    $\Psi_{\{\aone,\un{a}_2\}\alpha}(x):~\begin{cases}
    (\g^{\aone})_{\beta}{}^{\alpha}\Psi_{\{\aone,\un{a}_2\}\alpha}(x)~\equiv~ \psi_{\{\un{a}_2\}\beta} ~=~0~~~, \\
    (\g^{\un{a}_2})_{\g}{}^{\beta}\psi_{\{\un{a}_2\}\beta} ~=~0~~~,
    \end{cases}$\\[10pt]
    $\Psi_{\{\aone\bone\cone\}\alpha}(x):~ (\g^{\aone})_{\beta}{}^{\alpha}\Psi_{\{\aone\bone\cone\}\alpha}(x)~=~0 $~~~,\\[10pt]
    $\Psi_{\{\aone,\un{a}_2,\un{a}_3\}\alpha}(x):~\begin{cases}
    (\g^{\aone})_{\beta}{}^{\alpha}\Psi_{\{\aone,\un{a}_2,\un{a}_3\}\alpha}(x)~\equiv~ \psi_{\{\un{a}_2,\un{a}_3\}\b} ~=~0~~~, \\
    (\g^{\un{a}_2})_{\g}{}^{\b}\psi_{\{\un{a}_2,\un{a}_3\}\beta} ~\equiv~ \psi_{\{\un{a}_3\}\g}~=~0~~~,\\
    (\g^{\un{a}_3})_{\e}{}^{\g}\psi_{\{\un{a}_3\}\g} ~=~0~~~,
    \end{cases}$\\[10pt]
    $\Psi_{\{\un{a}_2|\aone\bone\}\alpha}(x):~\begin{cases}
    (\g^{\bone})_{\beta}{}^{\alpha}\Psi_{\{\un{a}_2|\aone\bone\}\alpha}(x)~\equiv~ \psi_{\{\aone,\un{a}_2\}\beta} ~=~0~~~, \\
    (\g^{\un{a}_2})_{\beta}{}^{\alpha}\Psi_{\{\un{a}_2|\aone\bone\}\alpha}(x)~\equiv~ \psi_{\{\aone\bone\}\beta} ~=~0~~~, \\
    (\g^{\bone})_{\g}{}^{\beta}\psi_{\{\aone\bone\}\beta} ~=~0 ~~~,
    \end{cases}$\\[10pt]
    $\Psi_{\{\un{a}_2|\aone\bone\cone\}\alpha}(x):~\begin{cases}
    (\g^{\cone})_{\beta}{}^{\alpha}\Psi_{\{\un{a}_2|\aone\bone\cone\}\alpha}(x)~\equiv~ \psi_{\{\un{a}_2|\aone\bone\}\beta} ~=~0~~~, \\
    (\g^{\un{a}_2})_{\beta}{}^{\alpha}\Psi_{\{\un{a}_2|\aone\bone\cone\}\alpha}(x)~\equiv~ \psi_{\{\aone\bone\cone\}\beta} ~=~0~~~, \\
    (\g^{\cone})_{\g}{}^{\beta}\psi_{\{\aone\bone\cone\}\beta} ~=~0~~~, 
    \end{cases}$
    %%%%%%%%%%%%%%%%%%%%%%%%%%%%%%%%%%%%%%%%
    \item Level-8:
    $\Phi(x)$~~~,~~~
    $\Phi_{ \{\aone\} }(x)$~~~,~~~
    $\Phi_{ \{\aone,\un{a}_2 \} }(x):~ \eta^{\aone\un{a}_2}\Phi_{ \{\aone,\un{a}_2 \} }(x)~=~0$~~~,\\[10pt]
    $\Phi_{ \{\aone,\un{a}_2,\un{a}_3 \} }(x):~ \eta^{\aone\un{a}_2}\Phi_{ \{\aone,\un{a}_2,\un{a}_3 \} }(x)~=~0$~~~,\\[10pt]
    $\Phi_{ \{\aone,\un{a}_2,\un{a}_3,\un{a}_4\} }(x):~ \begin{cases}
    \eta^{\un{a}_1\un{a}_2}\Phi_{ \{\aone,\un{a}_2,\un{a}_3,\un{a}_4\} }(x)~=~0~~~,\\
     \eta^{\un{a}_1\un{a}_2}\eta^{\un{a}_3\un{a}_4}\Phi_{ \{\aone,\un{a}_2,\un{a}_3,\un{a}_4\} }(x)~=~0~~~,
    \end{cases}$\\[10pt]
    $ \Phi{}_{\{{\un a}_1  {\un b}_1,   {\un a}_2
{\un b}_2 \}}(x)$~~:~~ $\begin{cases}
\eta^{{\un a}_1  {\un a}_2}\Phi{}_{\{{\un a}_1  {\un b}_1,   {\un a}_2
{\un b}_2 \}}(x)~=~0~~~,\\
\eta^{{\un a}_1  {\un a}_2}\eta^{{\un b}_1  {\un b}_2}\Phi{}_{\{{\un a}_1  {\un b}_1,  
{\un a}_2 {\un b}_2 \}}(x)~=~0~~~,
\end{cases}$\\[10pt]
$\Phi_{ \{\un{a}_2\un{b}_2|\aone\bone\cone\} }(x):~ \begin{cases}
    \eta^{\aone\un{a}_2}\Phi_{ \{\un{a}_2\un{b}_2|\aone\bone\cone\} }(x)~=~0~~~,\\
    \eta^{\aone\un{a}_2}\eta^{{\un b}_1  {\un b}_2}\Phi_{ \{\un{a}_2\un{b}_2|\aone\bone\cone\} }(x)~=~0~~~,
    \end{cases}$\\[10pt]
    $\Phi_{ \{\aone\bone\cone,\un{a}_2\un{b}_2\un{c}_2\} }(x):~ \begin{cases}
    \eta^{\aone\un{a}_2}\Phi_{ \{\aone\bone\cone,\un{a}_2\un{b}_2\un{c}_2\} }(x)~=~0~~~,\\
    \eta^{\aone\un{a}_2}\eta^{{\un b}_1  {\un b}_2}\Phi_{ \{\aone\bone\cone,\un{a}_2\un{b}_2\un{c}_2\} }(x)~=~0~~~,\\
    \eta^{\aone\un{a}_2}\eta^{{\un b}_1  {\un b}_2}\eta^{{\un c}_1  {\un c}_2}\Phi_{ \{\aone\bone\cone,\un{a}_2\un{b}_2\un{c}_2\} }(x)~=~0~~~,
    \end{cases}$\\[10pt]
    $\Phi_{ \{\aone\bone\cone\} }(x)$~~~, ~~~
    $\Phi_{ \{\aone\bone\cone\done\} }(x)$~~~,~~~
    $\Phi_{ \{\un{a}_2|\aone\bone\cone \} }(x):~ \eta^{\aone\un{a}_2}\Phi_{ \{\un{a}_2|\aone\bone\cone \} }(x)~=~0$~~~,\\[10pt]
    $\Phi_{ \{\un{a}_2|\aone\bone\cone\done \} }(x):~ \eta^{\aone\un{a}_2}\Phi_{ \{\un{a}_2|\aone\bone\cone\done \} }(x)~=~0$~~~,\\[10pt]
    $\Phi_{ \{\un{a}_2,\un{a}_3|\aone\bone\cone\} }(x):~ \begin{cases}
    \eta^{\un{a}_2\un{a}_3}\Phi_{ \{\un{a}_2,\un{a}_3|\aone\bone\cone\} }(x)~=~0~~~,\\
    \eta^{\un{c}_1\un{a}_3}\Phi_{ \{\un{a}_2,\un{a}_3|\aone\bone\cone\} }(x)~=~0~~~,
    \end{cases}$\\[10pt]
    $\Phi_{ \{\un{a}_2,\un{a}_3|\aone\bone\cone\done\} }(x):~ \begin{cases}
    \eta^{\un{a}_2\un{a}_3}\Phi_{ \{\un{a}_2,\un{a}_3|\aone\bone\cone\done\} }(x)~=~0~~~,\\
    \eta^{\un{d}_1\un{a}_3}\Phi_{ \{\un{a}_2,\un{a}_3|\aone\bone\cone\done\} }(x)~=~0~~~.
    \end{cases}$
\end{itemize}
Level-9 to Level-16 have exactly the same expressions as Level-7 to Level-0.

\newpage
\section{8D Minimal Scalar Superfield Decomposition}

\subsection{Component Decompostion Results}

The 8D minimal superfield component decomposition results in terms of Dynkin Labels are shown below.
\begin{itemize}
\item Level-0: $\CMTB{[0,0,0,0]}$
\item Level-1: $ \CMTred{[0,0,0,1]} \oplus \CMTred{[0,0,1,0]}$
\item Level-2: $\CMTB{[1,0,0,0]} \oplus (2)\CMTB{[0,1,0,0]} \oplus \CMTB{[0,0,1,1]} $
\item Level-3: $\CMTred{[0,0,0,1]} \oplus \CMTred{[0,0,1,0]} \oplus (2)\CMTred{[1,0,1,0]} \oplus (2)\CMTred{[1,0,0,1]} \oplus \CMTred{[0,1,0,1]} \oplus \CMTred{[0,1,1,0]} $
\item Level-4: $ \CMTB{[0,0,0,0]} \oplus (2)\CMTB{[1,0,0,0]} \oplus \CMTB{[0,1,0,0]} \oplus (3)\CMTB{[2,0,0,0]} \oplus (2)\CMTB{[0,0,2,0]} \oplus (2)\CMTB{[0,0,0,2]} \oplus (2)\CMTB{[0,0,1,1]} \oplus (2)\CMTB{[1,1,0,0]} \oplus \CMTB{[1,0,2,0]} \oplus \CMTB{[1,0,0,2]} \oplus \CMTB{[0,2,0,0]} \oplus \CMTB{[1,0,1,1]}$
\item Level-5: $(2)\CMTred{[0,0,0,1]} \oplus (2)\CMTred{[0,0,1,0]} \oplus (4)\CMTred{[1,0,1,0]} \oplus (4)\CMTred{[1,0,0,1]} \oplus \CMTred{[0,0,0,3]} \oplus \CMTred{[0,0,3,0]} \oplus (2)\CMTred{[0,1,0,1]} \oplus (2)\CMTred{[0,1,1,0]} \oplus (2)\CMTred{[2,0,0,1]} \oplus \CMTred{[0,0,2,1]} \oplus (2)\CMTred{[2,0,1,0]} \oplus \CMTred{[0,0,1,2]} \oplus \CMTred{[1,1,1,0]} \oplus \CMTred{[1,1,0,1]} $
\item Level-6: $(3)\CMTB{[1,0,0,0]} \oplus (6)\CMTB{[0,1,0,0]} \oplus (2)\CMTB{[2,0,0,0]} \oplus \CMTB{[0,0,2,0]} \oplus \CMTB{[0,0,0,2]} \oplus (4)\CMTB{[0,0,1,1]} \oplus \CMTB{[3,0,0,0]} \oplus (4)\CMTB{[1,1,0,0]} \oplus (2)\CMTB{[1,0,2,0]} \oplus (2)\CMTB{[1,0,0,2]} \oplus (4)\CMTB{[1,0,1,1]} \oplus (2)\CMTB{[2,1,0,0]} \oplus \CMTB{[0,1,2,0]} \oplus \CMTB{[0,1,0,2]} \oplus \CMTB{[0,1,1,1]} \oplus \CMTB{[2,0,1,1]} $
\item Level-7: $(4)\CMTred{[0,0,0,1]} \oplus (4)\CMTred{[0,0,1,0]} \oplus (5)\CMTred{[1,0,1,0]} \oplus (5)\CMTred{[1,0,0,1]} \oplus (4)\CMTred{[0,1,0,1]} \oplus (4)\CMTred{[0,1,1,0]} \oplus (3)\CMTred{[2,0,0,1]} \oplus (2)\CMTred{[0,0,2,1]} \oplus (3)\CMTred{[2,0,1,0]} \oplus (2)\CMTred{[0,0,1,2]} \oplus \CMTred{[3,0,1,0]} \oplus \CMTred{[3,0,0,1]} \oplus (2)\CMTred{[1,1,1,0]} \oplus (2)\CMTred{[1,1,0,1]} \oplus \CMTred{[1,0,1,2]} \oplus \CMTred{[1,0,2,1]} $
\item Level-8: $(5)\CMTB{[0,0,0,0]} \oplus (4)\CMTB{[1,0,0,0]} \oplus (3)\CMTB{[0,1,0,0]} \oplus (4)\CMTB{[2,0,0,0]} \oplus (3)\CMTB{[0,0,2,0]} \oplus (3)\CMTB{[0,0,0,2]} \oplus (6)\CMTB{[0,0,1,1]} \oplus (2)\CMTB{[3,0,0,0]} \oplus (4)\CMTB{[1,1,0,0]} \oplus (2)\CMTB{[1,0,2,0]} \oplus (2)\CMTB{[1,0,0,2]} \oplus \CMTB{[4,0,0,0]} \oplus (3)\CMTB{[0,2,0,0]} \oplus (5)\CMTB{[1,0,1,1]} \oplus \CMTB{[2,1,0,0]} \oplus (2)\CMTB{[0,1,1,1]} \oplus \CMTB{[2,0,2,0]} \oplus \CMTB{[2,0,0,2]} \oplus \CMTB{[0,0,2,2]} \oplus (2)\CMTB{[2,0,1,1]} $
\item Level-9: $(4)\CMTred{[0,0,0,1]} \oplus (4)\CMTred{[0,0,1,0]} \oplus (5)\CMTred{[1,0,1,0]} \oplus (5)\CMTred{[1,0,0,1]} \oplus (4)\CMTred{[0,1,0,1]} \oplus (4)\CMTred{[0,1,1,0]} \oplus (3)\CMTred{[2,0,0,1]} \oplus (2)\CMTred{[0,0,2,1]} \oplus (3)\CMTred{[2,0,1,0]} \oplus (2)\CMTred{[0,0,1,2]} \oplus \CMTred{[3,0,1,0]} \oplus \CMTred{[3,0,0,1]} \oplus (2)\CMTred{[1,1,1,0]} \oplus (2)\CMTred{[1,1,0,1]} \oplus \CMTred{[1,0,1,2]} \oplus \CMTred{[1,0,2,1]} $
\item Level-10: $(3)\CMTB{[1,0,0,0]} \oplus (6)\CMTB{[0,1,0,0]} \oplus (2)\CMTB{[2,0,0,0]} \oplus \CMTB{[0,0,2,0]} \oplus \CMTB{[0,0,0,2]} \oplus (4)\CMTB{[0,0,1,1]} \oplus \CMTB{[3,0,0,0]} \oplus (4)\CMTB{[1,1,0,0]} \oplus (2)\CMTB{[1,0,2,0]} \oplus (2)\CMTB{[1,0,0,2]} \oplus (4)\CMTB{[1,0,1,1]} \oplus (2)\CMTB{[2,1,0,0]} \oplus \CMTB{[0,1,2,0]} \oplus \CMTB{[0,1,0,2]} \oplus \CMTB{[0,1,1,1]} \oplus \CMTB{[2,0,1,1]} $
\item Level-11: $(2)\CMTred{[0,0,0,1]} \oplus (2)\CMTred{[0,0,1,0]} \oplus (4)\CMTred{[1,0,1,0]} \oplus (4)\CMTred{[1,0,0,1]} \oplus \CMTred{[0,0,0,3]} \oplus \CMTred{[0,0,3,0]} \oplus (2)\CMTred{[0,1,0,1]} \oplus (2)\CMTred{[0,1,1,0]} \oplus (2)\CMTred{[2,0,0,1]} \oplus \CMTred{[0,0,2,1]} \oplus (2)\CMTred{[2,0,1,0]} \oplus \CMTred{[0,0,1,2]} \oplus \CMTred{[1,1,1,0]} \oplus \CMTred{[1,1,0,1]} $
\item Level-12: $ \CMTB{[0,0,0,0]} \oplus (2)\CMTB{[1,0,0,0]} \oplus \CMTB{[0,1,0,0]} \oplus (3)\CMTB{[2,0,0,0]} \oplus (2)\CMTB{[0,0,2,0]} \oplus (2)\CMTB{[0,0,0,2]} \oplus (2)\CMTB{[0,0,1,1]} \oplus (2)\CMTB{[1,1,0,0]} \oplus \CMTB{[1,0,2,0]} \oplus \CMTB{[1,0,0,2]} \oplus \CMTB{[0,2,0,0]} \oplus \CMTB{[1,0,1,1]}$
\item Level-13: $\CMTred{[0,0,0,1]} \oplus \CMTred{[0,0,1,0]} \oplus (2)\CMTred{[1,0,1,0]} \oplus (2)\CMTred{[1,0,0,1]} \oplus \CMTred{[0,1,0,1]} \oplus \CMTred{[0,1,1,0]} $
\item Level-14: $\CMTB{[1,0,0,0]} \oplus (2)\CMTB{[0,1,0,0]} \oplus \CMTB{[0,0,1,1]} $
\item Level-15: $ \CMTred{[0,0,0,1]} \oplus \CMTred{[0,0,1,0]}$
\item Level-16: $\CMTB{[0,0,0,0]}$
\end{itemize}

The component decompostion results by dimensions are shown below.
\begin{itemize}
\item Level-0: $\CMTB{\{1\}}$
\item Level-1: $ \CMTred{\{8_{s}\}} \oplus \CMTred{\{8_{c}\}}$
\item Level-2: $\CMTB{\{8_{v}\}} \oplus (2) \CMTB{\{28\}} \oplus \CMTB{\{56_{v}\}} $
\item Level-3: $\CMTred{\{8_{s}\}} \oplus \CMTred{\{8_{c}\}} \oplus (2) \CMTred{\{56_{s}\}} \oplus (2) \CMTred{\{56_{c}\}} \oplus \CMTred{\{160_{s}\}} \oplus \CMTred{\{160_{c}\}} $
\item Level-4: $ \CMTB{\{1\}} \oplus (2) \CMTB{\{8_{v}\}} \oplus \CMTB{\{28\}} \oplus (3) \CMTB{\{35_{v}\}} \oplus (2) \CMTB{ \{35_{c}\}} \oplus (2) \CMTB{ \{35_{s}\}} \oplus (2) \CMTB{\{56_{v}\}} \oplus (2) \CMTB{\{160_{v}\}} \oplus \CMTB{\{224_{cv}\}} \oplus \CMTB{\{224_{sv}\}} \oplus \CMTB{\{300\}} \oplus \CMTB{\{350\}}$
\item Level-5: $(2) \CMTred{\{8_{s}\}} \oplus (2) \CMTred{\{8_{c}\}} \oplus (4) \CMTred{\{56_{s}\}} \oplus (4) \CMTred{\{56_{c}\}} \oplus \CMTred{\{112_{s}\}} \oplus \CMTred{\{112_{c}\}} \oplus (2) \CMTred{\{160_{s}\}} \oplus (2) \CMTred{\{160_{c}\}} \oplus (2) \CMTred{\{224_{vs}\}} \oplus \CMTred{\{224_{cs}\}} \oplus (2) \CMTred{\{224_{vc}\}} \oplus \CMTred{\{224_{sc}\}} \oplus \CMTred{\{840_{s}\}} \oplus \CMTred{\{840_{c}\}} $
\item Level-6: $(3) \CMTB{\{8_{v}\}} \oplus (6) \CMTB{\{28\}} \oplus (2) \CMTB{ \{35_{v}\}} \oplus \CMTB{ \{35_{c}\}} \oplus \CMTB{ \{35_{s}\}} \oplus (4) \CMTB{\{56_{v}\}} \oplus \CMTB{\{112_{v}\}} \oplus (4) \CMTB{\{160_{v}\}} \oplus (2) \CMTB{\{224_{cv}\}} \oplus (2) \CMTB{\{224_{sv}\}} \oplus (4) \CMTB{\{350\}} \oplus (2) \CMTB{\{567_{v}\}} \oplus \CMTB{\{567_{c}\}} \oplus \CMTB{\{567_{s}\}} \oplus \CMTB{\{840_{v}\}} \oplus \CMTB{\{1296_{v}\}} $
\item Level-7: $(4) \CMTred{\{8_{s}\}} \oplus (4) \CMTred{\{8_{c}\}} \oplus (5) \CMTred{\{56_{s}\}} \oplus (5) \CMTred{\{56_{c}\}} \oplus (4) \CMTred{\{160_{s}\}} \oplus (4) \CMTred{\{160_{c}\}} \oplus (3) \CMTred{\{224_{vs}\}} \oplus (2) \CMTred{\{224_{cs}\}} \oplus (3) \CMTred{\{224_{vc}\}} \oplus (2) \CMTred{\{224_{sc}\}} \oplus \CMTred{\{672_{vc}\}} \oplus \CMTred{\{672_{vs}\}} \oplus (2) \CMTred{\{840_{s}\}} \oplus (2) \CMTred{\{840_{c}\}} \oplus \CMTred{\{1296_{s}\}} \oplus \CMTred{\{1296_{c}\}} $
\item Level-8: $(5) \CMTB{\{1\}} \oplus(4)\CMTB{\{8_{v}\}} \oplus (3) \CMTB{\{28\}} \oplus (4) \CMTB{ \{35_{v}\}} \oplus (3) \CMTB{ \{35_{c}\}} \oplus (3) \CMTB{ \{35_{s}\}} \oplus (6) \CMTB{\{56_{v}\}} \oplus (2) \CMTB{\{112_{v}\}} \oplus (4) \CMTB{\{160_{v}\}} \oplus (2) \CMTB{\{224_{cv}\}} \oplus (2) \CMTB{\{224_{sv}\}} \oplus \CMTB{\{294_{v}\}} \oplus (3) \CMTB{\{300\}} \oplus (5) \CMTB{\{350\}} \oplus \CMTB{\{567_{v}\}} \oplus (2) \CMTB{\{840_{v}\}} \oplus \CMTB{\{840'_{s}\}} \oplus \CMTB{\{840'_{c}\}} \oplus \CMTB{\{840'_{v}\}} \oplus (2) \CMTB{\{1296_{v}\}} $
\item Level-9: $(4) \CMTred{\{8_{s}\}} \oplus (4) \CMTred{\{8_{c}\}} \oplus (5) \CMTred{\{56_{s}\}} \oplus (5) \CMTred{\{56_{c}\}} \oplus (4) \CMTred{\{160_{s}\}} \oplus (4) \CMTred{\{160_{c}\}} \oplus (3) \CMTred{\{224_{vs}\}} \oplus (2) \CMTred{\{224_{cs}\}} \oplus (3) \CMTred{\{224_{vc}\}} \oplus (2) \CMTred{\{224_{sc}\}} \oplus \CMTred{\{672_{vc}\}} \oplus \CMTred{\{672_{vs}\}} \oplus (2) \CMTred{\{840_{s}\}} \oplus (2) \CMTred{\{840_{c}\}} \oplus \CMTred{\{1296_{s}\}} \oplus \CMTred{\{1296_{c}\}} $
\item Level-10: $(3) \CMTB{\{8_{v}\}} \oplus (6) \CMTB{\{28\}} \oplus (2) \CMTB{ \{35_{v}\}} \oplus \CMTB{ \{35_{c}\}} \oplus \CMTB{ \{35_{s}\}} \oplus (4) \CMTB{\{56_{v}\}} \oplus \CMTB{\{112_{v}\}} \oplus (4) \CMTB{\{160_{v}\}} \oplus (2) \CMTB{\{224_{cv}\}} \oplus (2) \CMTB{\{224_{sv}\}} \oplus (4) \CMTB{\{350\}} \oplus (2) \CMTB{\{567_{v}\}} \oplus \CMTB{\{567_{c}\}} \oplus \CMTB{\{567_{s}\}} \oplus \CMTB{\{840_{v}\}} \oplus \CMTB{\{1296_{v}\}} $
\item Level-11: $(2) \CMTred{\{8_{s}\}} \oplus (2) \CMTred{\{8_{c}\}} \oplus (4) \CMTred{\{56_{s}\}} \oplus (4) \CMTred{\{56_{c}\}} \oplus \CMTred{\{112_{s}\}} \oplus \CMTred{\{112_{c}\}} \oplus (2) \CMTred{\{160_{s}\}} \oplus (2) \CMTred{\{160_{c}\}} \oplus (2) \CMTred{\{224_{vs}\}} \oplus \CMTred{\{224_{cs}\}} \oplus (2) \CMTred{\{224_{vc}\}} \oplus \CMTred{\{224_{sc}\}} \oplus \CMTred{\{840_{s}\}} \oplus \CMTred{\{840_{c}\}} $
\item Level-12: $ \CMTB{\{1\}} \oplus (2) \CMTB{\{8_{v}\}} \oplus \CMTB{\{28\}} \oplus (3) \CMTB{ \{35_{v}\}} \oplus (2) \CMTB{ \{35_{c}\}} \oplus (2) \CMTB{ \{35_{s}\}} \oplus (2) \CMTB{\{56_{v}\}} \oplus (2) \CMTB{\{160_{v}\}} \oplus \CMTB{\{224_{cv}\}} \oplus \CMTB{\{224_{sv}\}} \oplus \CMTB{\{300\}} \oplus \CMTB{\{350\}}$
\item Level-13: $\CMTred{\{8_{s}\}} \oplus \CMTred{\{8_{c}\}} \oplus (2) \CMTred{\{56_{s}\}} \oplus (2) \CMTred{\{56_{c}\}} \oplus \CMTred{\{160_{s}\}} \oplus \CMTred{\{160_{c}\}} $
\item Level-14: $\CMTB{\{8_{v}\}} \oplus (2) \CMTB{\{28\}} \oplus \CMTB{\{56_{v}\}} $
\item Level-15: $ \CMTred{\{8_{s}\}} \oplus \CMTred{\{8_{c}\}}$
\item Level-16: $\CMTB{\{1\}}$
\end{itemize}

\subsection{8D Minimal Adinkra Diagram}
\begin{figure}[htp!]
\centering
\includegraphics[width=0.8\textwidth]{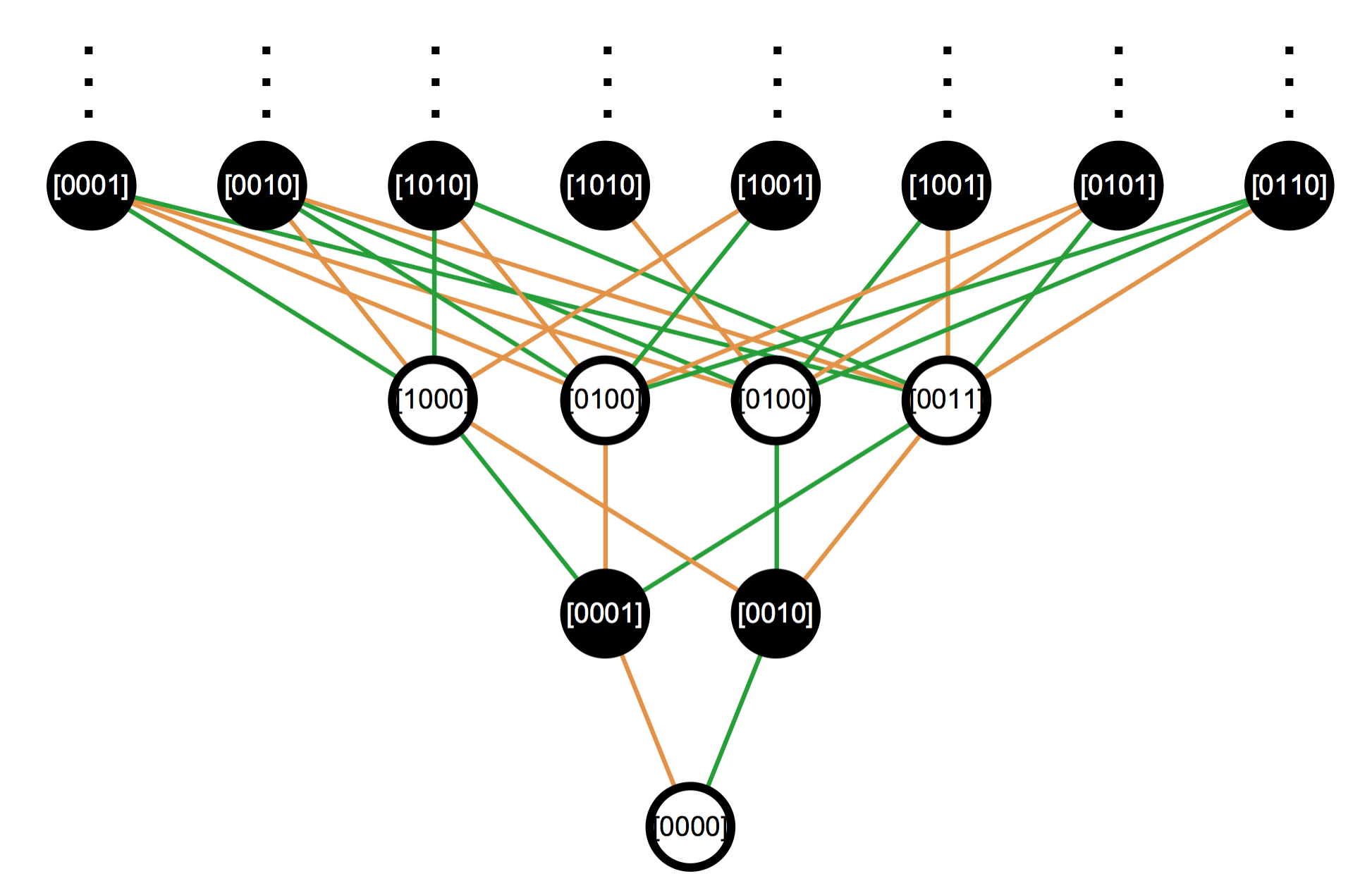}
\caption{Adynkra Diagram for 8D minimal scalar superfield}
\label{Fig:8D_Dynkin}
\end{figure}

\begin{figure}[htp!]
\centering
\includegraphics[width=0.8\textwidth]{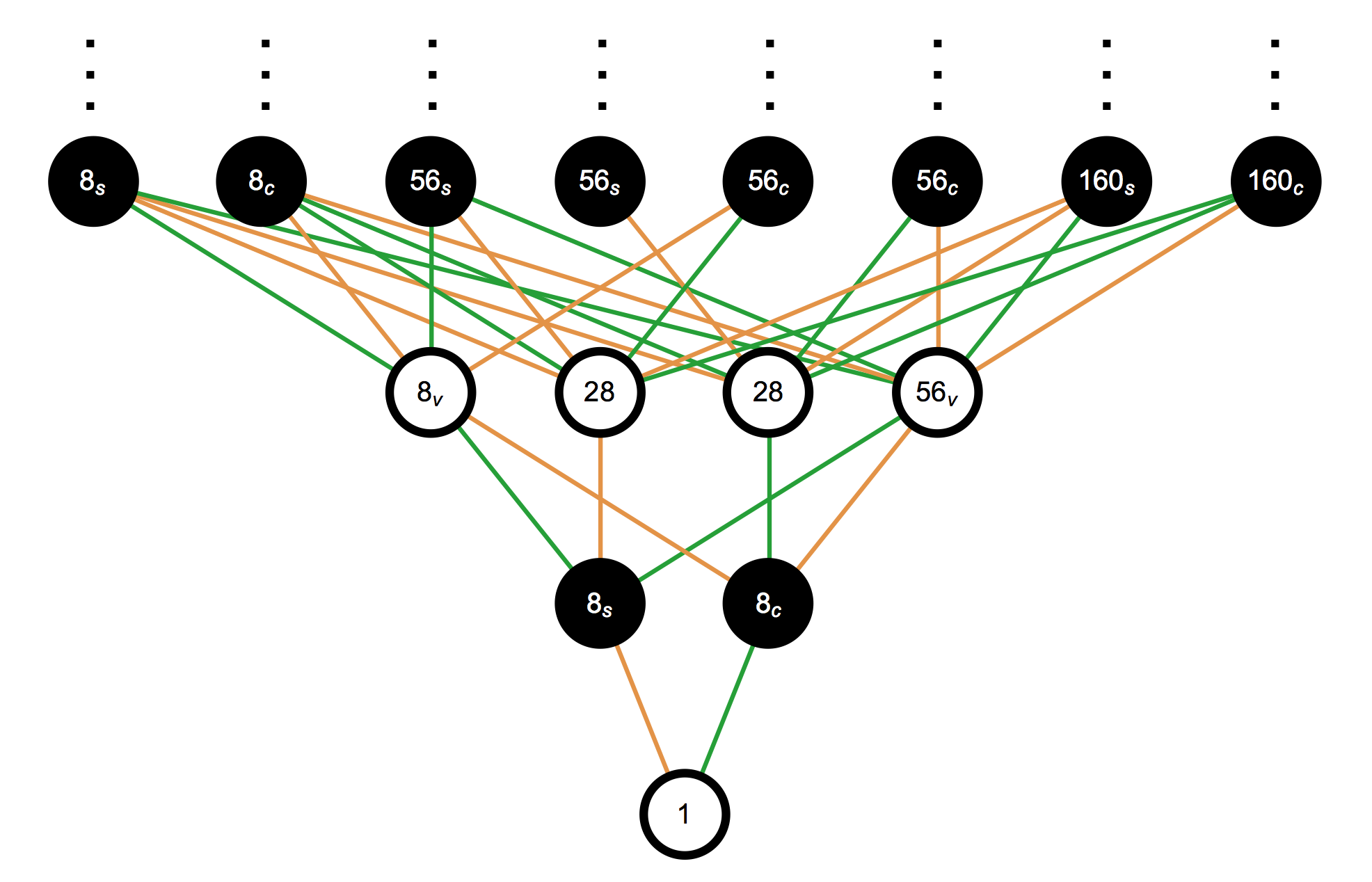}
\caption{Adinkra Diagram for 8D minimal scalar superfield}
\label{Fig:8D}
\end{figure}

The Adynkra and Adinkra diagrams for 8D minimal scalar superfield (up to Level-3) are Figure~\ref{Fig:8D_Dynkin} and \ref{Fig:8D}. 
In eight dimensions, we have two types of spinors having opposite chirality. Thus we have two types of spinorial derivatives $\CMTorg{{\rm D}_{\Dot\alpha}}$ and $\CMTG{{\rm D}_{\alpha}}$. In Figures \ref{Fig:8D_Dynkin} and \ref{Fig:8D} we use orange links to denote the supersymmetry transformations carried by $\CMTorg{{\rm D}_{\Dot\alpha}}$ and green links for $\CMTG{{\rm D}_{\alpha}}$. 

\subsection{Young Tableaux Descriptions of Component Fields in 8D Minimal Scalar Superfield}\label{sec:8D_component}

Consider the projection matrix for $\mathfrak{su}(8)\supset \mathfrak{so}(8)$~\cite{yamatsu2015},
\begin{equation}
P_{\mathfrak{su}(8)\supset \mathfrak{so}(8)} ~=~
\begin{pmatrix}
1 & 0 & 0 & 0 & 0 & 0 & 1\\
0 & 1 & 0 & 0 & 0 & 1 & 0\\
0 & 0 & 1 & 0 & 1 & 0 & 0\\
0 & 0 & 1 & 2 & 1 & 0 & 0\\
\end{pmatrix}~~~.
\label{eqn:Psu8so8vector}
\end{equation}
The highest weight of a specified irrep of $\mathfrak{su}(8)$ is a row vector $[p_1,p_2,p_3,p_4,p_5,p_6,p_7]$,
where $p_1$ to $p_7$ are non-negative integers. Since the $\mathfrak{su}(8)$ YT with $n$ vertical 
boxes is the conjugate of the one with $8-n$ vertical boxes, we need only consider the $p_5=p_6=p_7=0$ case.

Starting from the weight vector $[p_1,p_2,p_3,p_4,0,0,0]$ in $\mathfrak{su}(8)$, we define its projected
weight vector $[p_1,p_2,p_3,p_3+2p_4]$ in $\mathfrak{so}(8)$ as the Dynkin Label of the corresponding
irreducible bosonic Young Tableau.
\begin{equation}
    [p_1,p_2,p_3,p_3+2p_4] ~=~ [p_1,p_2,p_3,p_4,0,0,0] \, P^T_{\mathfrak{su}(8)\supset\mathfrak{so}(8)} ~~~.
\end{equation}
Note that in 8D, duality condition needs to be considered for the four-form. Similar as the situation in 10D discussed in \cite{nDx}, the Dynkin Label $[p_1,p_2,p_3+2p_4,p_3]$ carries the same dimensionality and corresponds to the same YT shape as $[p_1,p_2,p_3,p_3+2p_4]$, although there's no complex conjugation in $\mathfrak{so}(8)$ and all irreps are real. 
Then look at the congruence classes of a representation with 
Dynkin Label $[a,b,c,d]$ in $\mathfrak{so}(8)$,
\begin{equation}
    \begin{split}
        C_{c1}(R) ~:=&~ c + d ~~ ({\rm mod} ~ 2) ~~~,\\
        C_{c2}(R) ~:=&~ 2a + 2c + 4d ~~ ({\rm mod} ~ 4) ~~~.
    \end{split}
\end{equation}
Based on the above equations, there are totally four congruence classes in $\mathfrak{so}(8)$,
\begin{equation}
    \Big[ C_{c1}, C_{c2} \Big](R) ~=~
    \begin{cases}
    ~[0,0]\\
    ~[0,2]\\
    ~[1,0]\\
    ~[1,2]
    \end{cases} ~~~.
\end{equation}
$C_{c1}(R)$ actually classifies the bosonic irreps and spinorial
irreps:  $C_{c1}(R) = 0$ is bosonic and  $C_{c1}(R)=1$ is spinorial.  Consequently, a
bosonic irrep satisfies  $c-d
= 0 ~~ ({\rm mod} ~ 2)$.

Thus, given an irreducible bosonic Young Tableau with $p_1$ columns of one box,
$p_2$ columns of two vertical boxes, $p_3$ columns of three vertical boxes, and $p_4$ columns of
four vertical boxes, the Dynkin Label of its
corresponding bosonic irrep is $[p_1,p_2,p_3,p_3+2p_4]$ (or $[p_1,p_2,p_3+2p_4,p_3]$). Reversely, given a bosonic irrep with Dynkin Label $[a,b,c,d]$, its corresponding irreducible 
bosonic Young Tableau is composed of $a$ columns of one box, $b$ columns of two 
vertical boxes, $\min\{c,d\}$ columns of three vertical boxes, and $|d-c|/2$ columns of four vertical boxes. 

The simplest examples,
also the fundamental building blocks of an irreducible BYT, are given below.
\begin{equation}
\begin{gathered}
    {\CMTB{\ydiagram{1}}}_{{\rm IR}} ~\equiv~ \CMTB{[1,0,0,0]}  ~~~,~~~
    {\CMTB{\ydiagram{1,1}}}_{{\rm IR}} ~\equiv~ \CMTB{[0,1,0,0]} ~~~,~~~
    {\CMTB{\ydiagram{1,1,1}}}_{{\rm IR}} ~\equiv~  \CMTB{[0,0,1,1]} ~~~, \\ 
    {\CMTB{\ydiagram{1,1,1,1}}}_{{\rm IR, S}} ~\equiv~ \CMTB{[0,0,0,2]} ~~~,~~~  
    {\CMTB{\ydiagram{1,1,1,1}}}_{{\rm IR, C}} ~\equiv~ \CMTB{[0,0,2,0]} ~~~,
\end{gathered} \label{equ:BYTbasic_8D}
\end{equation}

For spinorial irreps, the basic SYTs are given by
\begin{equation}
\begin{split}
    \CMTred{\ytableaushort{\tinyeights}} ~\equiv~~ \CMTred{[0,0,0,1]} ~~~, \\
    \CMTred{\ytableaushort{\tinyeightc}} ~\equiv~~ \CMTred{[0,0,1,0]} ~~~.
\end{split}
\label{equ:SYTbasic_8D}
\end{equation}
We could translate the Dynkin Label of any spinorial irrep to a mixed YT (which contains a BYT part and one of the basic SYT above) with irreducible conditions by applying the same idea discussed in Chapter five in \cite{nDx}. 

Putting together the columns in (\ref{equ:BYTbasic_8D}) and (\ref{equ:SYTbasic_8D}) corresponds to adding their Dynkin Labels. 
We follow the well-known (VCS) convention for labelling $\mathfrak{so}(8)$ irreps to name our irreducible YTs. 

In summary, the irreducible Young Tableau descriptions of the 8D minimal scalar superfield decomposition is presented below.
\begin{itemize} \ytableausetup{boxsize=0.8em}
\item Level-0: $\CMTB{\cdot} ~~~,$ 
%%%%%%%%%%%%%%%%%%%%%%%%%%%%%%%%%%%%%%%%%%
\item Level-1: $\CMTred{\ytableaushort{\tinyeights}}
~\oplus~\CMTred{\ytableaushort{\tinyeightc}}~~~,$
%%%%%%%%%%%%%%%%%%%%%%%%%%%%%%%%%%%%%%%%%%
\item Level-2: ${\CMTB{\ydiagram{1}}}_{\rm IR}~\oplus~
(2)\,{\CMTB{\ydiagram{1,1}}}_{\rm IR}~\oplus~{\CMTB{\ydiagram{1,1,1}}}_{\rm IR} ~~~,$
%%%%%%%%%%%%%%%%%%%%%%%%%%%%%%%%%%%%%%%%%%
\item Level-3: $\CMTred{\ytableaushort{\tinyeights}}
~\oplus~\CMTred{\ytableaushort{\tinyeightc}} ~\oplus~
(2)\,{\CMTB{\ydiagram{1}}\CMTred{\ytableaushort{\tinyeightc}}}_{\rm IR}~\oplus~
(2)\,{\CMTB{\ydiagram{1}}\CMTred{\ytableaushort{\tinyeights}}}_{\rm IR}~\oplus~
{\CMTB{\ydiagram{1,1}}\CMTred{\ytableaushort{\tinyeights,\none}}}_{\rm IR}
~\oplus~{\CMTB{\ydiagram{1,1}}\CMTred{\ytableaushort{\tinyeightc,\none}}}_{\rm IR}~~~,$
%%%%%%%%%%%%%%%%%%%%%%%%%%%%%%%%%%%%%%%%%%
\item Level-4: $\CMTB{\cdot}~\oplus~
(2)\,{\CMTB{\ydiagram{1}}}_{\rm IR}~\oplus~
{\CMTB{\ydiagram{1,1}}}_{\rm IR}~\oplus~
(3)\,{\CMTB{\ydiagram{2}}}_{\rm IR} ~\oplus~
(2)\,{\CMTB{\ydiagram{1,1,1,1}}}_{\rm IR,C}~\oplus~
(2)\,{\CMTB{\ydiagram{1,1,1,1}}}_{\rm IR,S}\\[20pt]
~~~~~~~~~~~~~
~\oplus~
(2)\,{\CMTB{\ydiagram{1,1,1}}}_{\rm IR}~\oplus~
(2)\,{\CMTB{\ydiagram{2,1}}}_{\rm IR} ~\oplus~
{\CMTB{\ydiagram{2,1,1,1}}}_{\rm IR,C} ~\oplus~
{\CMTB{\ydiagram{2,1,1,1}}}_{\rm IR,S} ~\oplus~
{\CMTB{\ydiagram{2,2}}}_{\rm IR} ~\oplus~
{\CMTB{\ydiagram{2,1,1}}}_{\rm IR}
 ~~~, $
 %%%%%%%%%%%%%%%%%%%%%%%%%%%%%%%%%%%%%%%%%%
 \item Level-5: $(2)\,\CMTred{\ytableaushort{\tinyeights}}~\oplus~
(2)\,\CMTred{\ytableaushort{\tinyeightc}} ~\oplus~
(4)\,{\CMTB{\ydiagram{1}}\CMTred{\ytableaushort{\tinyeightc}}}_{\rm IR}~\oplus~
(4)\,{\CMTB{\ydiagram{1}}\CMTred{\ytableaushort{\tinyeights}}}_{\rm IR}~\oplus~
{\CMTB{\ydiagram{1,1,1,1}}\CMTred{\ytableaushort{\tinyeights,\none,\none,\none}}}_{\rm IR,S}~\oplus~
{\CMTB{\ydiagram{1,1,1,1}}\CMTred{\ytableaushort{\tinyeightc,\none,\none,\none}}}_{\rm IR,C} \\[20pt]
~~~~~~~~~~~~~
~\oplus~
(2)\,{\CMTB{\ydiagram{1,1}}\CMTred{\ytableaushort{\tinyeights,\none}}}_{\rm IR}~\oplus~
(2)\,{\CMTB{\ydiagram{1,1}}\CMTred{\ytableaushort{\tinyeightc,\none}}}_{\rm IR}~\oplus~
(2)\,{\CMTB{\ydiagram{2}}\CMTred{\ytableaushort{\tinyeights}}}_{\rm IR}~\oplus~
{\CMTB{\ydiagram{1,1,1,1}}\CMTred{\ytableaushort{\tinyeights,\none,\none,\none}}}_{\rm IR,C}~\oplus~
(2)\,{\CMTB{\ydiagram{2}}\CMTred{\ytableaushort{\tinyeightc}}}_{\rm IR}
\\
~~~~~~~~~~~~~
~\oplus~
{\CMTB{\ydiagram{1,1,1,1}}\CMTred{\ytableaushort{\tinyeightc,\none,\none,\none}}}_{\rm IR,S}
~\oplus~
{\CMTB{\ydiagram{2,1}}\CMTred{\ytableaushort{\tinyeightc,\none}}}_{\rm IR}
~\oplus~
{\CMTB{\ydiagram{2,1}}\CMTred{\ytableaushort{\tinyeights,\none}}}_{\rm IR}
~~~,$
%%%%%%%%%%%%%%%%%%%%%%%%%%%%%%%%%%%%%%%%%%
 \item Level-6: $(3)\,{\CMTB{\ydiagram{1}}}_{\rm IR}~\oplus~
(6)\,{\CMTB{\ydiagram{1,1}}}_{\rm IR}~\oplus~
(2)\,{\CMTB{\ydiagram{2}}}_{\rm IR} ~\oplus~
{\CMTB{\ydiagram{1,1,1,1}}}_{\rm IR,C}~\oplus~
{\CMTB{\ydiagram{1,1,1,1}}}_{\rm IR,S}~\oplus~
(4)\,{\CMTB{\ydiagram{1,1,1}}}_{\rm IR}\\[20pt]
~~~~~~~~~~~~~
~\oplus~{\CMTB{\ydiagram{3}}}_{{\rm IR}} 
~\oplus~ (4)\,{\CMTB{\ydiagram{2,1}}}_{\rm IR}
~\oplus~ (2)\,{\CMTB{\ydiagram{2,1,1,1}}}_{\rm IR,C}
~\oplus~ (2)\,{\CMTB{\ydiagram{2,1,1,1}}}_{\rm IR,S}
~\oplus~ (4)\,{\CMTB{\ydiagram{2,1,1}}}_{\rm IR}
\\[20pt]
~~~~~~~~~~~~~
~\oplus~ (2)\,{\CMTB{\ydiagram{3,1}}}_{\rm IR}
~\oplus~ {\CMTB{\ydiagram{2,2,1,1}}}_{\rm IR,C}
~\oplus~ {\CMTB{\ydiagram{2,2,1,1}}}_{\rm IR,S}
~\oplus~ {\CMTB{\ydiagram{2,2,1}}}_{\rm IR}
~\oplus~ {\CMTB{\ydiagram{3,1,1}}}_{\rm IR}
~~~, $
%%%%%%%%%%%%%%%%%%%%%%%%%%%%%%%%%%%%%%%%%%
 \item Level-7: $(4)\,\CMTred{\ytableaushort{\tinyeights}}~\oplus~
(4)\,\CMTred{\ytableaushort{\tinyeightc}} ~\oplus~
(5)\,{\CMTB{\ydiagram{1}}\CMTred{\ytableaushort{\tinyeightc}}}_{\rm IR}~\oplus~
(5)\,{\CMTB{\ydiagram{1}}\CMTred{\ytableaushort{\tinyeights}}}_{\rm IR}~\oplus~
(4)\,{\CMTB{\ydiagram{1,1}}\CMTred{\ytableaushort{\tinyeights,\none}}}_{\rm IR}~\oplus~
(4)\,{\CMTB{\ydiagram{1,1}}\CMTred{\ytableaushort{\tinyeightc,\none}}}_{\rm IR}\\[20pt]
~~~~~~~~~~~~~
~\oplus~(3)\,{\CMTB{\ydiagram{2}}\CMTred{\ytableaushort{\tinyeights}}}_{\rm IR}~\oplus~
(2)\,{\CMTB{\ydiagram{1,1,1,1}}\CMTred{\ytableaushort{\tinyeights,\none,\none,\none}}}_{\rm IR,C}~\oplus~
(3)\,{\CMTB{\ydiagram{2}}\CMTred{\ytableaushort{\tinyeightc}}}_{\rm IR}~\oplus~
(2)\,{\CMTB{\ydiagram{1,1,1,1}}\CMTred{\ytableaushort{\tinyeightc,\none,\none,\none}}}_{\rm IR,S}~\oplus~
{\CMTB{\ydiagram{3}}\CMTred{\ytableaushort{\tinyeightc}}}_{\rm IR}
\\[20pt]
~~~~~~~~~~~~~
~\oplus~
{\CMTB{\ydiagram{3}}\CMTred{\ytableaushort{\tinyeights}}}_{\rm IR}
~\oplus~
(2)\,{\CMTB{\ydiagram{2,1}}\CMTred{\ytableaushort{\tinyeightc,\none}}}_{\rm IR}
~\oplus~
(2)\,{\CMTB{\ydiagram{2,1}}\CMTred{\ytableaushort{\tinyeights,\none}}}_{\rm IR}
~\oplus~
{\CMTB{\ydiagram{2,1,1,1}}\CMTred{\ytableaushort{\tinyeightc,\none,\none,\none}}}_{\rm IR,S}
~\oplus~
{\CMTB{\ydiagram{2,1,1,1}}\CMTred{\ytableaushort{\tinyeights,\none,\none,\none}}}_{\rm IR,C}
~~~,$
%%%%%%%%%%%%%%%%%%%%%%%%%%%%%%%%%%%%%%%%%%
 \item Level-8: $(5)\,\CMTB{\cdot}~\oplus~
(4)\,{\CMTB{\ydiagram{1}}}_{\rm IR} ~\oplus~
(3)\,{\CMTB{\ydiagram{1,1}}}_{\rm IR} ~\oplus~
(4)\,{\CMTB{\ydiagram{2}}}_{\rm IR} ~\oplus~
(3)\,{\CMTB{\ydiagram{1,1,1,1}}}_{\rm IR,C} ~\oplus~
(3)\,{\CMTB{\ydiagram{1,1,1,1}}}_{\rm IR,S}  \\[15pt]
~~~~~~~~~~~~~
~\oplus~
(6)\,{\CMTB{\ydiagram{1,1,1}}}_{\rm IR}~\oplus~
(2)\,{\CMTB{\ydiagram{3}}}_{\rm IR} ~\oplus~
(4)\,{\CMTB{\ydiagram{2,1}}}_{\rm IR} ~\oplus~
(2)\,{\CMTB{\ydiagram{2,1,1,1}}}_{\rm IR,C} ~\oplus~
(2)\,{\CMTB{\ydiagram{2,1,1,1}}}_{\rm IR,S}  \\[15pt]
~~~~~~~~~~~~~
~\oplus~
{\CMTB{\ydiagram{4}}}_{\rm IR}~\oplus~
(3)\,{\CMTB{\ydiagram{2,2}}}_{\rm IR} ~\oplus~ 
(5)\,{\CMTB{\ydiagram{2,1,1}}}_{\rm IR} ~\oplus~
{\CMTB{\ydiagram{3,1}}}_{\rm IR} ~\oplus~
(2)\,{\CMTB{\ydiagram{2,2,1}}}_{\rm IR} \\[15pt]
~~~~~~~~~~~~~
~\oplus~
(3)\,{\CMTB{\ydiagram{3,1,1,1}}}_{\rm IR,C}~\oplus~
(3)\,{\CMTB{\ydiagram{3,1,1,1}}}_{\rm IR,S}~\oplus~
{\CMTB{\ydiagram{2,2,2}}}_{\rm IR} ~\oplus~
(2)\,{\CMTB{\ydiagram{3,1,1}}}_{\rm IR} 
~~~, $
\ytableausetup{boxsize=1.2em}
\end{itemize}
where Level-9 to Level-16 have exactly the same expressions as Level-7 to Level-0.

\subsection{Index Structures and Irreducible Conditions of Component Fields in 8D Minimal Scalar Superfield}\label{sec:8D_Index}

In this section, we will translate the irreducible bosonic and mixed Young Tableaux into field variables. 
We follow the same $\{\}$-indices notation as well as ``$|$'' to separate indices in YT with different heights and ``,'' to separate
indices in YT with the same heights. 

The vector index $\un{a}$ runs from 0 to 7. 
The $\{\}$-indices, irreducible bosonic Young
Tableaux, and Dynkin Labels are equivalent and have the one-to-one correspondence.

The general expression is as below in Equations (\ref{equ:index-notation1_8D}) and (\ref{equ:index-notation2_8D}),
\begin{equation}
    \begin{split}
       & \{ \un{a}_1,\dots,\un{a}_p ~|~ \un{b}_1\un{c}_1,\dots,\un{b}_q \un{c}_q ~|~  \un{d}_1\un{e}_1\un{f}_1,\dots,\un{d}_r\un{e}_r\un{f}_r  ~|~ \un{g}_1\un{h}_1\un{i}_1\un{j}_1,\dots, \un{g}_s\un{h}_s\un{i}_s\un{j}_s\}^{S/C}  \\[10pt]
      & ~{ \CMTB{{\ytableaushort{\aone}} } }~~~~~{ \CMTB{{\ytableaushort{\ap}} } }
      ~~~~{ \CMTB{{\ytableaushort{\bone,\cone}} } }~~~~~~~{ \CMTB{{\ytableaushort{\bq,\cunq}} } }
      ~~~~~~{ \CMTB{{\ytableaushort{\done,\eone,\fone}} } }~~~~~~~~~~~{ \CMTB{{\ytableaushort{\dr,\er,\fr}} } }
      ~~~~~~~~~{ \CMTB{{\ytableaushort{\gone,\hone,\ione,\jone}} } }~~~~~~~~~~~~~~{\CMTB{{\ytableaushort{\gs,\hs,\is,\js}} } }
      \\[10pt]
      & ~~ \CMTB{[p,0,0,0]} ~~~~~ \CMTB{[0,q,0,0]} ~~~~~~~~~~~ \CMTB{[0,0,r,r]}  ~~~~~~~ \begin{matrix} \CMTB{[0,0,0,2s]}~~\text{with superscript } S \\
       \CMTB{[0,0,2s,0]}~~\text{with superscript } C \end{matrix}
    \end{split}
    \label{equ:index-notation1_8D}
\end{equation}
where above we have ``disassembled'' the YT to show how each column is affiliated with each type of
subscript structure. Below we have assembled all the columns into a proper YT.
\begin{equation}
\begin{split}
    & { \CMTB{{\ytableaushort{\gone\dots\gs\done\dots\dr\bone\dots\bq\aone\dots\ap,\hone\dots\hs\eone\dots\er\cone\dots\cunq,\ione\dots\is\fone\dots\fr,\jone\dots\js}} } }_{\rm IR,S/C}\\[10pt]
    &~~~\begin{matrix} \CMTB{[p,q,r,r+2s]}~~\text{with subscript } S \\
    \CMTB{[p,q,r+2s,r]}~~\text{with subscript } C \end{matrix}
\end{split}
\label{equ:index-notation2_8D}
\end{equation}

As one moves from the YT's shown in Equation~(\ref{equ:index-notation1_8D}) to
Equation~(\ref{equ:index-notation2_8D}), it is clear the number of vertical boxes is tabulating the number of
1-forms, 2-forms, 3-forms, and 4-forms in the YT's.  These are the entries between
the vertical $|$ bars. These precisely correspond to the integers $p$, $q$, $r$, and $s$ that appeared in Dynkin Labels.  
An example of the correspondence between the subscript conventions, the affiliated YT, and Dynkin Label is shown in (\ref{equ:index-notation_ex1_8D}). 
\begin{equation}
    \{{\un a}_2 , {\un a}_3| {\un a}_1  {\un b}_1   {\un c}_1 \} ~~\equiv~~
    { \CMTB{{\ytableaushort{\aone \atwo \athree,\bone,\cone}} } }_{{\rm IR}}
    ~~\equiv~~ \CMTB{[2,0,1,1]} ~~~.
    \label{equ:index-notation_ex1_8D}
\end{equation}

In 8D, we have two types of spinor indices corresponding to $\CMTred{\ytableaushort{\tinyeights}}$ and $\CMTred{\ytableaushort{\tinyeightc}}$ respectively.
We define the field $\Psi^{\alpha}$ or $\Psi_{\Dot{\alpha}}$ correpsonds to $\CMTred{\ytableaushort{\tinyeights}}$ and the field $\Psi_{\alpha}$ or $\Psi^{\Dot{\alpha}}$ correpsonds to $\CMTred{\ytableaushort{\tinyeightc}}$
The spinor indices $\alpha$ and $\Dot{\alpha}$ run from 1 to 8. 

The index structures as well as irreducible conditions of all bosonic and fermionic fields are identified below along 
with the level at which the fields occur in the adinkra of the scalar superfield. 

\begin{itemize}
    \item Level-0: $\Phi(x)$
    %%%%%%%%%%%%%%%%%%%%%%%%%%%%%%%%%%%%%
    \item Level-1: $\Psi^{\a}(x)$~~~,~~~
    $\Psi_{\a}(x)$~~~,
    %%%%%%%%%%%%%%%%%%%%%%%%%%%%%%%%%%%%%
    \item Level-2: $\Phi_{\{{\un a}_1\}}(x)$~~~,~~~
    $(2)\,\Phi_{\{{\un a}_1\bone\}}(x)$~~~,~~~
     $\Phi_{\{{\un a}_1\bone\cone\}}(x)$~~~,
    %%%%%%%%%%%%%%%%%%%%%%%%%%%%%%%%%%%%%
    \item Level-3: $\Psi^{\a}(x)$~~~,~~~
    $\Psi_{\a}(x)$~~~,~~~\\[10pt]
    $(2)\, \Psi{}_{\{{\un a}_1 \}} {}_{\a}(x)$~:~ $(\sigma^{{\un a}_1})^{\a\b}\Psi{}_{
\{{\un a}_1 \}} {}_{\a}(x) = 0$~~~,~~~
$(2)\, \Psi{}_{\{{\un a}_1 \}} {}^{\a}(x)$~:~ $(\sigma^{{\un a}_1})_{\a\b}\Psi{}_{
\{{\un a}_1 \}} {}^{\a}(x) = 0$~~~,\\[10pt]
    $ \Psi{}_{\{{\un a}_1  {\un b}_1 \}} {}^{\a}(x)$~:~ $(\sigma^{{\un a}_1})_{\a\b}\Psi{}_{
\{{\un a}_1  {\un b}_1 \}} {}^{\a}(x) = 0$~~~,\\[10pt]
$ \Psi{}_{\{{\un a}_1  {\un b}_1 \}} {}_{\a}(x)$~~:~~ $(\sigma^{{\un a}_1})^{\a\b}\Psi{}_{
\{{\un a}_1  {\un b}_1 \}} {}_{\a}(x) = 0$~~~,
    %%%%%%%%%%%%%%%%%%%%%%%%%%%%%%%%%%%%%
    \item Level-4: $\Phi(x)$~~~,~~~
    $(2)\,\Phi_{\{{\un a}_1\}}(x)$~~~,~~~
    $\Phi_{\{{\un a}_1\bone\}}(x)$~~~,~~~
    (3)\,$\Phi_{ \{\aone,\un{a}_2 \} }(x):~ \eta^{\aone\un{a}_2}\Phi_{ \{\aone,\un{a}_2 \} }(x)~=~0$~~~,~~~\\[10pt]
    (2)\,$\Phi_{ \{\aone\bone\cone\done\}^{\rm C} }(x):~\Phi{}_{\{{\un a}_1   {\un b}_1 {\un c}_1    {\un d}_1   \}{}^{\rm C}}(x)~=~
-\frac{1}{4!}\epsilon_{{\un a}_1 {\un b}_1 {\un c}_1    {\un d}_1  }{}^{{\un e}_1  
{\un f}_1 {\un g}_1    {\un h}_1}\Phi{}_{\{ {\un e}_1   {\un f}_1
{\un g}_1    {\un h}_1 \}{}^{\rm C}}(x)$~~~,~~~\\[10pt]
    (2)\,$\Phi_{ \{\aone\bone\cone\done\}^{\rm S} }(x):~\Phi{}_{\{{\un a}_1   {\un b}_1 {\un c}_1    {\un d}_1   \}{}^{\rm S}}(x)~=~
\frac{1}{4!}\epsilon_{{\un a}_1 {\un b}_1 {\un c}_1    {\un d}_1  }{}^{{\un e}_1  
{\un f}_1 {\un g}_1    {\un h}_1}\Phi{}_{\{ {\un e}_1   {\un f}_1
{\un g}_1    {\un h}_1 \}{}^{\rm S}}(x)$~~~,~~~\\[10pt]
    (2)\,$\Phi_{ \{\aone\bone\cone\} }(x)$~~~,~~~
    (2)\,$\Phi_{ \{\un{a}_2|\aone\bone \} }(x):~ \eta^{\aone\un{a}_2}\Phi_{ \{\un{a}_2|\aone\bone \} }(x)~=~0$~~~,\\[10pt]
    $ \Phi{}_{\{{\un a}_2| {\un a}_1   {\un b}_1 {\un c}_1    {\un d}_1   \}{}^{\rm C}}(x)
$~~:~~$\begin{cases}
\eta^{{\un a}_1  {\un a}_2}\Phi{}_{\{{\un a}_2| {\un a}_1   {\un b}_1 {\un c}_1  {\un d}_1     \}{}^{\rm C}}(x)~=~0~~~,\\
\Phi{}_{\{{\un a}_2| {\un a}_1   {\un b}_1 {\un c}_1    {\un d}_1   \}{}^{\rm C}}(x)~=~
-\frac{1}{4!}\epsilon_{{\un a}_1 {\un b}_1 {\un c}_1    {\un d}_1  }{}^{{\un e}_1  
{\un f}_1 {\un g}_1    {\un h}_1}\Phi{}_{\{{\un a}_2| {\un e}_1   {\un f}_1
{\un g}_1    {\un h}_1 \}{}^{\rm C}}(x)~~~.
\end{cases}$\\[10pt]
$ \Phi{}_{\{{\un a}_2| {\un a}_1   {\un b}_1 {\un c}_1    {\un d}_1   \}{}^{\rm S}}(x)
$~~:~~$\begin{cases}
\eta^{{\un a}_1  {\un a}_2}\Phi{}_{\{{\un a}_2| {\un a}_1   {\un b}_1 {\un c}_1  {\un d}_1     \}{}^{\rm S}}(x)~=~0~~~,\\
\Phi{}_{\{{\un a}_2| {\un a}_1   {\un b}_1 {\un c}_1    {\un d}_1   \}{}^{\rm S}}(x)~=~
\frac{1}{4!}\epsilon_{{\un a}_1 {\un b}_1 {\un c}_1    {\un d}_1  }{}^{{\un e}_1  
{\un f}_1 {\un g}_1    {\un h}_1}\Phi{}_{\{{\un a}_2| {\un e}_1   {\un f}_1
{\un g}_1    {\un h}_1 \}{}^{\rm S}}(x)~~~.
\end{cases}$\\[10pt]
    $ \Phi{}_{\{{\un a}_1  {\un b}_1,   {\un a}_2
{\un b}_2 \}}(x)$~~:~~ $\begin{cases}
\eta^{{\un a}_1  {\un a}_2}\Phi{}_{\{{\un a}_1  {\un b}_1,   {\un a}_2
{\un b}_2 \}}(x)~=~0~~~,\\
\eta^{{\un a}_1  {\un a}_2}\eta^{{\un b}_1  {\un b}_2}\Phi{}_{\{{\un a}_1  {\un b}_1,  
{\un a}_2 {\un b}_2 \}}(x)~=~0~~~,
\end{cases}$\\[10pt]
$\Phi_{ \{\un{a}_2|\aone\bone\cone\} }(x):~ \eta^{\aone\un{a}_2}\Phi_{ \{\un{a}_2|\aone\bone\cone\} }(x)~=~0$~~~,
    %%%%%%%%%%%%%%%%%%%%%%%%%%%%%%%%%%%%%
    \item Level-5: $(2)\,\Psi^{\a}(x)$~~~,~~~
    $(2)\,\Psi_{\a}(x)$~~~,~~~\\[10pt]
    $(4)\, \Psi{}_{\{{\un a}_1 \}} {}_{\a}(x)$~:~ $(\sigma^{{\un a}_1})^{\a\b}\Psi{}_{
\{{\un a}_1 \}} {}_{\a}(x) = 0$~~~,~~~
$(4)\, \Psi{}_{\{{\un a}_1 \}} {}^{\a}(x)$~:~ $(\sigma^{{\un a}_1})_{\a\b}\Psi{}_{
\{{\un a}_1 \}} {}^{\a}(x) = 0$~~~,\\[10pt]
$ \Psi{}_{\{{\un a}_1   {\un b}_1 {\un c}_1   {\un d}_1      \}{}^{\rm S}}{}^{\a}(x)$~~:~~  
$\begin{cases}
(\sigma^{{\un a}_1   {\un b}_1})_{\a}{}^{\d}\Psi{}_{\{{\un a}_1   {\un b}_1 {\un c}_1  {\un d}_1    
  \}{}^{\rm S}}{}^{\a}(x) ~\equiv~  \psi{}_{\{ {\un c}_1   {\un d}_1    \}}{}^{\d}
 (x)~=~ 0~~~,\\
(\sigma^{{\un c}_1})_{\e\d}\psi{}_{\{ {\un c}_1   {\un d}_1     \}}{}^{\d}(x) ~\equiv
~\psi{}_{\{   {\un d}_1    \}\e}(x) ~=~ 0~~~,\\
(\sigma^{{\un d}_1 })^{\tau\e}\psi{}_{\{  {\un d}_1     \}\e}(x) ~=~ 0~~~,
\end{cases}$\\[10pt]
$ \Psi{}_{\{{\un a}_1 {\un b}_1 {\un c}_1 {\un d}_1  \}{}^{\rm C}}{}_{\a}(x)$~~:~~  
$\begin{cases}
(\sigma^{{\un a}_1   {\un b}_1})_{\d}{}^{\a}\Psi{}_{\{{\un a}_1   {\un b}_1 {\un c}_1  {\un
d}_1 \}{}^{\rm C}\a}(x) ~\equiv~  \psi{}_{\{ {\un c}_1   {\un d}_1      \}\d}(x)
~=~ 0~~~,\\
(\sigma^{{\un c}_1})^{\e\d}\psi{}_{\{ {\un c}_1   {\un d}_1      \}\d}(x) ~\equiv~\psi{
}_{\{  {\un d}_1      \}}{}^{\e}(x) ~=~ 0~~~,\\
(\sigma^{{\un d}_1 })_{\tau\e}\psi{}_{\{  {\un d}_1      \}}{}^{\e}(x) ~=~ 0~~~,
\end{cases}$\\[10pt]
    $(2)\, \Psi{}_{\{{\un a}_1  {\un b}_1 \}} {}^{\a}(x)$~:~ $(\sigma^{{\un a}_1})_{\a\b}\Psi{}_{
\{{\un a}_1  {\un b}_1 \}} {}^{\a}(x) = 0$~~~,\\[10pt]
$(2)\, \Psi{}_{\{{\un a}_1  {\un b}_1 \}} {}_{\a}(x)$~~:~~ $(\sigma^{{\un a}_1})^{\a\b}\Psi{}_{
\{{\un a}_1  {\un b}_1 \}} {}_{\a}(x) = 0$~~~,\\[10pt]
$(2)\, \Psi{}_{\{{\un a}_1,  {\un a}_2  \}}{}^{\a}(x)$~~:~~   $\begin{cases}
(\s^{{\un a}_2})_{\a\d}\Psi{}_{\{{\un a}_1,  {\un a}_2  \}}{}^{\a}(x) ~\equiv~\psi{
}_{\{{\un a}_1\}}{}_{\d}(x)~=~0~~~,\\
(\s^{{\un a}_1})^{\e\d}\psi{}_{\{{\un a}_1\}}{}_{\d}(x) ~=~0~~~,
\end{cases}$\\[10pt]
$ \Psi{}_{\{{\un a}_1   {\un b}_1 {\un c}_1   {\un d}_1      \}{}^{\rm C}}{}^{\a}(x)$~~:~~  
$\begin{cases}
(\sigma^{{\un a}_1   {\un b}_1 {\un c}_1})_{\a\d}\Psi{}_{\{{\un a}_1   {\un b}_1 {\un c}_1  {\un d}_1    
  \}{}^{\rm C}}{}^{\a}(x) ~\equiv~  \psi{}_{\{ {\un d}_1    \}}{}_{\d}
 (x)~=~ 0~~~,\\
(\sigma^{{\un d}_1})^{\e\d}\psi{}_{\{  {\un d}_1     \}}{}_{\d}(x)  ~=~ 0~~~,
\end{cases}$\\[10pt]
$(2)\, \Psi{}_{\{{\un a}_1,  {\un a}_2  \}}{}_{\a}(x)$~~:~~   $\begin{cases}
(\s^{{\un a}_2})^{\a\d}\Psi{}_{\{{\un a}_1,  {\un a}_2  \}}{}_{\a}(x) ~\equiv~\psi{
}_{\{{\un a}_1\}}{}^{\d}(x)~=~0~~~,\\
(\s^{{\un a}_1})_{\e\d}\psi{}_{\{{\un a}_1\}}{}^{\d}(x) ~=~0~~~,
\end{cases}$\\[10pt]
$ \Psi{}_{\{{\un a}_1   {\un b}_1 {\un c}_1   {\un d}_1      \}{}^{\rm S}}{}_{\a}(x)$~~:~~  
$\begin{cases}
(\sigma^{{\un a}_1   {\un b}_1 {\un c}_1})^{\a\d}\Psi{}_{\{{\un a}_1   {\un b}_1 {\un c}_1  {\un d}_1    
  \}{}^{\rm S}}{}_{\a}(x) ~\equiv~  \psi{}_{\{ {\un d}_1    \}}{}^{\d}
 (x)~=~ 0~~~,\\
(\sigma^{{\un d}_1})_{\e\d}\psi{}_{\{  {\un d}_1     \}}{}^{\d}(x)  ~=~ 0~~~,
\end{cases}$\\[10pt]
$ \Psi{}_{\{{\un a}_2|   {\un a}_1 {\un b}_1  \}}{}_{\a}(x)$~~:~~ $\begin{cases}
(\s^{{\un b}_1})^{\a\d}\Psi{}_{\{{\un a}_2|   {\un a}_1 {\un b}_1  \}\a}(x) ~\equiv~\psi{}_{\{{
\un a}_1,   {\un a}_2
\}}{}^{\d}(x)~=~0~~~,\\
(\s^{{\un a}_2})^{\a\d}\Psi{}_{\{{\un a}_2|   {\un a}_1 {\un b}_1  \}\a}(x) ~\equiv~\psi{}_{\{{\un
a}_1  {\un b}_1  \}}{}^{\d}(x)~=~0~~~,\\
(\s^{{\un b}_1})_{\d\e}\psi{}_{\{{\un a}_1  {\un b}_1 \}}{}^{\d}(x)~=~0~~~,
\end{cases}$\\[10pt]
$ \Psi{}_{\{{\un a}_2|   {\un a}_1 {\un b}_1  \}}{}^{\a}(x)$~~:~~ $\begin{cases}
(\s^{{\un b}_1})_{\a\d}\Psi{}_{\{{\un a}_2|   {\un a}_1 {\un b}_1  \}}{}^{\a}(x) ~\equiv~
\psi{}_{\{{\un a}_1,   {\un a}_2  \}\d}(x)~=~0~~~,\\
(\s^{{\un a}_2})_{\a\d}\Psi{}_{\{{\un a}_2|   {\un a}_1 {\un b}_1  \}}{}^{\a}(x) ~\equiv~
\psi{}_{\{{\un a}_1  {\un b}_1  \}\d}(x)~=~0~~~,\\
(\s^{{\un b}_1})^{\d\e}\psi{}_{\{{\un a}_1  {\un b}_1
\}\d}(x)~=~0~~~,
\end{cases}$
    %%%%%%%%%%%%%%%%%%%%%%%%%%%%%%%%%%%%%
    \item Level-6:
     $(3)\,\Phi_{\{{\un a}_1\}}(x)$~~~,~~~
    $(6)\,\Phi_{\{{\un a}_1\bone\}}(x)$~~~,~~~
    (2)\,$\Phi_{ \{\aone,\un{a}_2 \} }(x):~ \eta^{\aone\un{a}_2}\Phi_{ \{\aone,\un{a}_2 \} }(x)~=~0$~~~,~~~\\[10pt]
    $\Phi_{ \{\aone\bone\cone\done\}^{\rm C} }(x):~\Phi{}_{\{{\un a}_1   {\un b}_1 {\un c}_1    {\un d}_1   \}{}^{\rm C}}(x)~=~
-\frac{1}{4!}\epsilon_{{\un a}_1 {\un b}_1 {\un c}_1    {\un d}_1  }{}^{{\un e}_1  
{\un f}_1 {\un g}_1    {\un h}_1}\Phi{}_{\{ {\un e}_1   {\un f}_1
{\un g}_1    {\un h}_1 \}{}^{\rm C}}(x)$~~~,~~~\\[10pt]
    $\Phi_{ \{\aone\bone\cone\done\}^{\rm S} }(x):~\Phi{}_{\{{\un a}_1   {\un b}_1 {\un c}_1    {\un d}_1   \}{}^{\rm S}}(x)~=~
\frac{1}{4!}\epsilon_{{\un a}_1 {\un b}_1 {\un c}_1    {\un d}_1  }{}^{{\un e}_1  
{\un f}_1 {\un g}_1    {\un h}_1}\Phi{}_{\{ {\un e}_1   {\un f}_1
{\un g}_1    {\un h}_1 \}{}^{\rm S}}(x)$~~~,~~~\\[10pt]
    (4)\,$\Phi_{ \{\aone\bone\cone\} }(x)$~~~,~~~
    $\Phi_{ \{\aone,\un{a}_2,\un{a}_3 \} }(x):~ \eta^{\aone\un{a}_2}\Phi_{ \{\aone,\un{a}_2,\un{a}_3 \} }(x)~=~0$~~~,\\[10pt]
    (4)\,$\Phi_{ \{\un{a}_2|\aone\bone \} }(x):~ \eta^{\aone\un{a}_2}\Phi_{ \{\un{a}_2|\aone\bone \} }(x)~=~0$~~~,\\[10pt]
    $(2)\, \Phi{}_{\{{\un a}_2| {\un a}_1   {\un b}_1 {\un c}_1    {\un d}_1   \}{}^{\rm C}}(x)
$~~:~~$\begin{cases}
\eta^{{\un a}_1  {\un a}_2}\Phi{}_{\{{\un a}_2| {\un a}_1   {\un b}_1 {\un c}_1  {\un d}_1     \}{}^{\rm C}}(x)~=~0~~~,\\
\Phi{}_{\{{\un a}_2| {\un a}_1   {\un b}_1 {\un c}_1    {\un d}_1   \}{}^{\rm C}}(x)~=~
-\frac{1}{4!}\epsilon_{{\un a}_1 {\un b}_1 {\un c}_1    {\un d}_1  }{}^{{\un e}_1  
{\un f}_1 {\un g}_1    {\un h}_1}\Phi{}_{\{{\un a}_2| {\un e}_1   {\un f}_1
{\un g}_1    {\un h}_1 \}{}^{\rm C}}(x)~~~.
\end{cases}$\\[10pt]
$(2)\, \Phi{}_{\{{\un a}_2| {\un a}_1   {\un b}_1 {\un c}_1    {\un d}_1   \}{}^{\rm S}}(x)
$~~:~~$\begin{cases}
\eta^{{\un a}_1  {\un a}_2}\Phi{}_{\{{\un a}_2| {\un a}_1   {\un b}_1 {\un c}_1  {\un d}_1     \}{}^{\rm S}}(x)~=~0~~~,\\
\Phi{}_{\{{\un a}_2| {\un a}_1   {\un b}_1 {\un c}_1    {\un d}_1   \}{}^{\rm S}}(x)~=~
\frac{1}{4!}\epsilon_{{\un a}_1 {\un b}_1 {\un c}_1    {\un d}_1  }{}^{{\un e}_1  
{\un f}_1 {\un g}_1    {\un h}_1}\Phi{}_{\{{\un a}_2| {\un e}_1   {\un f}_1
{\un g}_1    {\un h}_1 \}{}^{\rm S}}(x)~~~.
\end{cases}$\\[10pt]
$(4)\,\Phi_{ \{\un{a}_2|\aone\bone\cone\} }(x):~ \eta^{\aone\un{a}_2}\Phi_{ \{\un{a}_2|\aone\bone\cone\} }(x)~=~0$~~~,\\[10pt]
(2)\,$\Phi_{ \{\un{a}_2,\un{a}_3|\aone\bone\} }(x):~ \begin{cases}
    \eta^{\un{a}_2\un{a}_3}\Phi_{ \{\un{a}_2,\un{a}_3|\aone\bone\} }(x)~=~0~~~,\\
    \eta^{\un{b}_1\un{a}_3}\Phi_{ \{\un{a}_2,\un{a}_3|\aone\bone\} }(x)~=~0~~~,
    \end{cases}$\\[10pt]
$ \Phi{}_{\{{\un a}_2  {\un b}_2|  {\un a}_1 {\un b}_1 {\un c}_1  {\un d}_1 \}{}^{\rm C
}}(x)$~~:~~ $\begin{cases}
\eta^{{\un a}_1{\un a}_2} \Phi{}_{\{{\un a}_2  {\un b}_2|  {\un a}_1 {\un b}_1 {\un c}_1    
{\un d}_1     \}{}^{\rm C}}(x) = 0~~~,\\
\eta^{{\un a}_1{\un a}_2}\eta^{{\un b}_1{\un b}_2} \Phi{}_{\{{\un a}_2  {\un b}_2|  {\un a}_1
{\un b}_1 {\un c}_1    {\un d}_1     \}{}^{\rm C}}(x) = 0~~~,\\
\Phi{}_{\{{\un a}_2  {\un b}_2|  {\un a}_1 {\un b}_1 {\un c}_1 {\un d}_1\}{}^{\rm C}}(x)
= -\frac{1}{4!}\epsilon_{{\un a}_1 {\un b}_1 {\un c}_1  {\un d}_1}{}^{{\un f}_1 {\un
g}_1 {\un h}_1    {\un i}_1   }\Phi{}_{\{{\un a}_2  {\un b}_2|  {\un f}_1 {\un g}_1
{\un h}_1    {\un i}_1    \}{}^{\rm C}}(x)~~~.
\end{cases}$\\[10pt]
$ \Phi{}_{\{{\un a}_2  {\un b}_2|  {\un a}_1 {\un b}_1 {\un c}_1  {\un d}_1 \}{}^{\rm S
}}(x)$~~:~~ $\begin{cases}
\eta^{{\un a}_1{\un a}_2} \Phi{}_{\{{\un a}_2  {\un b}_2|  {\un a}_1 {\un b}_1 {\un c}_1    
{\un d}_1     \}{}^{\rm S}}(x) = 0~~~,\\
\eta^{{\un a}_1{\un a}_2}\eta^{{\un b}_1{\un b}_2} \Phi{}_{\{{\un a}_2  {\un b}_2|  {\un a}_1
{\un b}_1 {\un c}_1    {\un d}_1     \}{}^{\rm S}}(x) = 0~~~,\\
\Phi{}_{\{{\un a}_2  {\un b}_2|  {\un a}_1 {\un b}_1 {\un c}_1 {\un d}_1\}{}^{\rm S}}(x)
= \frac{1}{4!}\epsilon_{{\un a}_1 {\un b}_1 {\un c}_1  {\un d}_1}{}^{{\un f}_1 {\un
g}_1 {\un h}_1    {\un i}_1   }\Phi{}_{\{{\un a}_2  {\un b}_2|  {\un f}_1 {\un g}_1
{\un h}_1    {\un i}_1     \}{}^{\rm S}}(x)~~~.
\end{cases}$\\[10pt]
$\Phi_{ \{\un{a}_2\un{b}_2|\aone\bone\cone\} }(x):~ \begin{cases}
    \eta^{\aone\un{a}_2}\Phi_{ \{\un{a}_2\un{b}_2|\aone\bone\cone\} }(x)~=~0~~~,\\
    \eta^{\aone\un{a}_2}\eta^{{\un b}_1  {\un b}_2}\Phi_{ \{\un{a}_2\un{b}_2|\aone\bone\cone\} }(x)~=~0~~~,
    \end{cases}$\\[10pt]
    $\Phi_{ \{\un{a}_2,\un{a}_3|\aone\bone\cone\} }(x):~ \begin{cases}
    \eta^{\un{a}_2\un{a}_3}\Phi_{ \{\un{a}_2,\un{a}_3|\aone\bone\cone\} }(x)~=~0~~~,\\
    \eta^{\un{c}_1\un{a}_3}\Phi_{ \{\un{a}_2,\un{a}_3|\aone\bone\cone\} }(x)~=~0~~~,
    \end{cases}$
    %%%%%%%%%%%%%%%%%%%%%%%%%%%%%%%%%%%%%
    \item Level-7: $(4)\,\Psi^{\a}(x)$~~~,~~~
    $(4)\,\Psi_{\a}(x)$~~~,~~~\\[10pt]
    $(5)\, \Psi{}_{\{{\un a}_1 \}} {}_{\a}(x)$~:~ $(\sigma^{{\un a}_1})^{\a\b}\Psi{}_{
\{{\un a}_1 \}} {}_{\a}(x) = 0$~~~,~~~
$(5)\, \Psi{}_{\{{\un a}_1 \}} {}^{\a}(x)$~:~ $(\sigma^{{\un a}_1})_{\a\b}\Psi{}_{
\{{\un a}_1 \}} {}^{\a}(x) = 0$~~~,\\[10pt]
    $(4)\, \Psi{}_{\{{\un a}_1  {\un b}_1 \}} {}^{\a}(x)$~:~ $(\sigma^{{\un a}_1})_{\a\b}\Psi{}_{
\{{\un a}_1  {\un b}_1 \}} {}^{\a}(x) = 0$~~~,\\[10pt]
$(4)\, \Psi{}_{\{{\un a}_1  {\un b}_1 \}} {}_{\a}(x)$~~:~~ $(\sigma^{{\un a}_1})^{\a\b}\Psi{}_{
\{{\un a}_1  {\un b}_1 \}} {}_{\a}(x) = 0$~~~,\\[10pt]
$(3)\, \Psi{}_{\{{\un a}_1,  {\un a}_2  \}}{}^{\a}(x)$~~:~~   $\begin{cases}
(\s^{{\un a}_2})_{\a\d}\Psi{}_{\{{\un a}_1,  {\un a}_2  \}}{}^{\a}(x) ~\equiv~\psi{
}_{\{{\un a}_1\}}{}_{\d}(x)~=~0~~~,\\
(\s^{{\un a}_1})^{\e\d}\psi{}_{\{{\un a}_1\}}{}_{\d}(x) ~=~0~~~,
\end{cases}$\\[10pt]
$(2)\, \Psi{}_{\{{\un a}_1   {\un b}_1 {\un c}_1   {\un d}_1      \}{}^{\rm C}}{}^{\a}(x)$~~:~~  
$\begin{cases}
(\sigma^{{\un a}_1   {\un b}_1 {\un c}_1})_{\a\d}\Psi{}_{\{{\un a}_1   {\un b}_1 {\un c}_1  {\un d}_1    
  \}{}^{\rm C}}{}^{\a}(x) ~\equiv~  \psi{}_{\{ {\un d}_1    \}}{}_{\d}
 (x)~=~ 0~~~,\\
(\sigma^{{\un d}_1})^{\e\d}\psi{}_{\{  {\un d}_1     \}}{}_{\d}(x)  ~=~ 0~~~,
\end{cases}$\\[10pt]
$(3)\, \Psi{}_{\{{\un a}_1,  {\un a}_2  \}}{}_{\a}(x)$~~:~~   $\begin{cases}
(\s^{{\un a}_2})^{\a\d}\Psi{}_{\{{\un a}_1,  {\un a}_2  \}}{}_{\a}(x) ~\equiv~\psi{
}_{\{{\un a}_1\}}{}^{\d}(x)~=~0~~~,\\
(\s^{{\un a}_1})_{\e\d}\psi{}_{\{{\un a}_1\}}{}^{\d}(x) ~=~0~~~,
\end{cases}$\\[10pt]
$(2)\, \Psi{}_{\{{\un a}_1   {\un b}_1 {\un c}_1   {\un d}_1      \}{}^{\rm S}}{}_{\a}(x)$~~:~~  
$\begin{cases}
(\sigma^{{\un a}_1   {\un b}_1 {\un c}_1})^{\a\d}\Psi{}_{\{{\un a}_1   {\un b}_1 {\un c}_1  {\un d}_1    
  \}{}^{\rm S}}{}_{\a}(x) ~\equiv~  \psi{}_{\{ {\un d}_1    \}}{}^{\d}
 (x)~=~ 0~~~,\\
(\sigma^{{\un d}_1})_{\e\d}\psi{}_{\{  {\un d}_1     \}}{}^{\d}(x)  ~=~ 0~~~,
\end{cases}$\\[10pt]
$ \Psi{}_{\{{\un a}_1,  {\un a}_2, {\un a}_3  \}}{}_{\a}(x)$~~:~~   $\begin{cases}
(\s^{{\un a}_3})^{\a\d}\Psi{}_{\{{\un a}_1,  {\un a}_2, {\un a}_3  \}}{}_{\a}(x) ~\equiv~\psi{
}_{\{{\un a}_1,   {\un a}_2  \}}{}^{\d}(x)~=~0~~~,\\
(\s^{{\un a}_2})_{\e\d}\psi{}_{\{{\un a}_1,   {\un a}_2 \}}{}^{\d}(x) ~\equiv~\psi{}_{\{{\un a}_1  
 \}\e}(x)~=~0~~~,\\
 (\s^{{\un a}_1})^{\tau\e}\psi{}_{\{{\un a}_1  \}\e}(x)~=~0~~~.
\end{cases}$\\[10pt]
$ \Psi{}_{\{{\un a}_1,  {\un a}_2, {\un a}_3  \}}{}^{\a}(x)$~~:~~  $\begin{cases}
(\s^{{\un a}_3})_{\a\d}\Psi{}_{\{{\un a}_1,  {\un a}_2 {\un a}_3  \}}{}^{\a}(x) ~\equiv~\psi{
}_{\{{\un a}_1,   {\un a}_2  \}}{}_{\d}(x)~=~0~~~,\\
(\s^{{\un a}_2})^{\e\d}\psi{}_{\{{\un a}_1,   {\un a}_2  \}}{}_{\d}(x) ~\equiv~\psi{}_{\{{\un a}_1  
\}}{}^{\e}(x)~=~0~~~,\\
(\s^{{\un a}_1})_{\tau\e}\psi{}_{\{{\un a}_1  \}}{}^{\e}(x)~=~0~~~.
\end{cases}$\\[10pt]
$(2)\, \Psi{}_{\{{\un a}_2|   {\un a}_1 {\un b}_1  \}}{}_{\a}(x)$~~:~~ $\begin{cases}
(\s^{{\un b}_1})^{\a\d}\Psi{}_{\{{\un a}_2|   {\un a}_1 {\un b}_1  \}\a}(x) ~\equiv~\psi{}_{\{{
\un a}_1,   {\un a}_2
\}}{}^{\d}(x)~=~0~~~,\\
(\s^{{\un a}_2})^{\a\d}\Psi{}_{\{{\un a}_2|   {\un a}_1 {\un b}_1  \}\a}(x) ~\equiv~\psi{}_{\{{\un
a}_1  {\un b}_1  \}}{}^{\d}(x)~=~0~~~,\\
(\s^{{\un b}_1})_{\d\e}\psi{}_{\{{\un a}_1  {\un b}_1 \}}{}^{\d}(x)~=~0~~~,
\end{cases}$\\[10pt]
$(2)\, \Psi{}_{\{{\un a}_2|   {\un a}_1 {\un b}_1  \}}{}^{\a}(x)$~~:~~ $\begin{cases}
(\s^{{\un b}_1})_{\a\d}\Psi{}_{\{{\un a}_2|   {\un a}_1 {\un b}_1  \}}{}^{\a}(x) ~\equiv~
\psi{}_{\{{\un a}_1,   {\un a}_2  \}\d}(x)~=~0~~~,\\
(\s^{{\un a}_2})_{\a\d}\Psi{}_{\{{\un a}_2|   {\un a}_1 {\un b}_1  \}}{}^{\a}(x) ~\equiv~
\psi{}_{\{{\un a}_1  {\un b}_1  \}\d}(x)~=~0~~~,\\
(\s^{{\un b}_1})^{\d\e}\psi{}_{\{{\un a}_1  {\un b}_1
\}\d}(x)~=~0~~~,
\end{cases}$\\[10pt]
$ \Psi{}_{\{{{\un a}_2| \un a}_1   {\un b}_1 {\un c}_1   {\un d}_1      \}{}^{\rm S}}{}_{\a}(x)$~~:~~  
$\begin{cases}
(\sigma^{{\un a}_2})^{\a\d}\Psi{}_{\{{\un a}_2|{\un a}_1   {\un b}_1 {\un c}_1  {\un d}_1    
  \}{}^{\rm S}}{}_{\a}(x) ~\equiv~  \psi{}_{\{{\un a}_1   {\un b}_1  {\un c}_1   {\un d}_1    \}{}^{\rm S}}{}^{\d}(x)~=~ 0~~~,\\
(\sigma^{{\un b}_1{\un c}_1{\un d}_1})^{\a\d}\Psi{}_{\{{\un a}_2|{\un a}_1   {\un b}_1 {\un c}_1  {\un d}_1    
  \}{}^{\rm S}}{}_{\a}(x) ~\equiv
~\psi{}_{\{   {\un a}_1,{\un a}_2    \}}{}^{\d}(x) ~=~ 0~~~,\\
(\sigma^{{\un a}_2 })_{\d\e}\psi{}_{\{   {\un a}_1,{\un a}_2    \}}{}^{\d}(x) ~=~ 0~~~,
\end{cases}$\\[10pt]
$ \Psi{}_{\{{{\un a}_2| \un a}_1   {\un b}_1 {\un c}_1   {\un d}_1      \}{}^{\rm C}}{}^{\a}(x)$~~:~~  
$\begin{cases}
(\sigma^{{\un a}_2})_{\a\d}\Psi{}_{\{{\un a}_2|{\un a}_1   {\un b}_1 {\un c}_1  {\un d}_1    
  \}{}^{\rm C}}{}^{\a}(x) ~\equiv~  \psi{}_{\{{\un a}_1   {\un b}_1  {\un c}_1   {\un d}_1    \}{}^{\rm C}}{}_{\d}(x)~=~ 0~~~,\\
(\sigma^{{\un b}_1{\un c}_1{\un d}_1})_{\a\d}\Psi{}_{\{{\un a}_2|{\un a}_1   {\un b}_1 {\un c}_1  {\un d}_1    
  \}{}^{\rm C}}{}^{\a}(x) ~\equiv
~\psi{}_{\{   {\un a}_1,{\un a}_2    \}}{}_{\d}(x) ~=~ 0~~~,\\
(\sigma^{{\un a}_2 })^{\d\e}\psi{}_{\{   {\un a}_1,{\un a}_2    \}}{}_{\d}(x) ~=~ 0~~~,
\end{cases}$
    %%%%%%%%%%%%%%%%%%%%%%%%%%%%%%%%%%%%%
    \item Level-8: $(5)\,\Phi(x)$~~~,~~~
    $(4)\,\Phi_{\{{\un a}_1\}}(x)$~~~,~~~
    $(3)\,\Phi_{\{{\un a}_1\bone\}}(x)$~~~,~~~
    (4)\,$\Phi_{ \{\aone,\un{a}_2 \} }(x):~ \eta^{\aone\un{a}_2}\Phi_{ \{\aone,\un{a}_2 \} }(x)~=~0$~~~,~~~\\[10pt]
    $(3)\,\Phi_{ \{\aone\bone\cone\done\}^{\rm C} }(x):~\Phi{}_{\{{\un a}_1   {\un b}_1 {\un c}_1    {\un d}_1   \}{}^{\rm C}}(x)~=~
-\frac{1}{4!}\epsilon_{{\un a}_1 {\un b}_1 {\un c}_1    {\un d}_1  }{}^{{\un e}_1  
{\un f}_1 {\un g}_1    {\un h}_1}\Phi{}_{\{ {\un e}_1   {\un f}_1
{\un g}_1    {\un h}_1 \}{}^{\rm C}}(x)$~~~,~~~\\[10pt]
    $(3)\,\Phi_{ \{\aone\bone\cone\done\}^{\rm S} }(x):~\Phi{}_{\{{\un a}_1   {\un b}_1 {\un c}_1    {\un d}_1   \}{}^{\rm S}}(x)~=~
\frac{1}{4!}\epsilon_{{\un a}_1 {\un b}_1 {\un c}_1    {\un d}_1  }{}^{{\un e}_1  
{\un f}_1 {\un g}_1    {\un h}_1}\Phi{}_{\{ {\un e}_1   {\un f}_1
{\un g}_1    {\un h}_1 \}{}^{\rm S}}(x)$~~~,~~~\\[10pt]
    (6)\,$\Phi_{ \{\aone\bone\cone\} }(x)$~~~,~~~
    $(2)\,\Phi_{ \{\aone,\un{a}_2,\un{a}_3 \} }(x):~ \eta^{\aone\un{a}_2}\Phi_{ \{\aone,\un{a}_2,\un{a}_3 \} }(x)~=~0$~~~,\\[10pt]
    (4)\,$\Phi_{ \{\un{a}_2|\aone\bone \} }(x):~ \eta^{\aone\un{a}_2}\Phi_{ \{\un{a}_2|\aone\bone \} }(x)~=~0$~~~,\\[10pt]
    $(2)\, \Phi{}_{\{{\un a}_2| {\un a}_1   {\un b}_1 {\un c}_1    {\un d}_1   \}{}^{\rm C}}(x)
$~~:~~$\begin{cases}
\eta^{{\un a}_1  {\un a}_2}\Phi{}_{\{{\un a}_2| {\un a}_1   {\un b}_1 {\un c}_1  {\un d}_1     \}{}^{\rm C}}(x)~=~0~~~,\\
\Phi{}_{\{{\un a}_2| {\un a}_1   {\un b}_1 {\un c}_1    {\un d}_1   \}{}^{\rm C}}(x)~=~
-\frac{1}{4!}\epsilon_{{\un a}_1 {\un b}_1 {\un c}_1    {\un d}_1  }{}^{{\un e}_1  
{\un f}_1 {\un g}_1    {\un h}_1}\Phi{}_{\{{\un a}_2| {\un e}_1   {\un f}_1
{\un g}_1    {\un h}_1 \}{}^{\rm C}}(x)~~~.
\end{cases}$\\[10pt]
$(2)\, \Phi{}_{\{{\un a}_2| {\un a}_1   {\un b}_1 {\un c}_1    {\un d}_1   \}{}^{\rm S}}(x)
$~~:~~$\begin{cases}
\eta^{{\un a}_1  {\un a}_2}\Phi{}_{\{{\un a}_2| {\un a}_1   {\un b}_1 {\un c}_1  {\un d}_1     \}{}^{\rm S}}(x)~=~0~~~,\\
\Phi{}_{\{{\un a}_2| {\un a}_1   {\un b}_1 {\un c}_1    {\un d}_1   \}{}^{\rm S}}(x)~=~
\frac{1}{4!}\epsilon_{{\un a}_1 {\un b}_1 {\un c}_1    {\un d}_1  }{}^{{\un e}_1  
{\un f}_1 {\un g}_1    {\un h}_1}\Phi{}_{\{{\un a}_2| {\un e}_1   {\un f}_1
{\un g}_1    {\un h}_1 \}{}^{\rm S}}(x)~~~.
\end{cases}$\\[10pt]
$\Phi_{ \{\aone,\un{a}_2,\un{a}_3,\un{a}_4\} }(x):~ \begin{cases}
    \eta^{\un{a}_1\un{a}_2}\Phi_{ \{\aone,\un{a}_2,\un{a}_3,\un{a}_4\} }(x)~=~0~~~,\\
     \eta^{\un{a}_1\un{a}_2}\eta^{\un{a}_3\un{a}_4}\Phi_{ \{\aone,\un{a}_2,\un{a}_3,\un{a}_4\} }(x)~=~0~~~,
    \end{cases}$\\[10pt]
    (3)\,$ \Phi{}_{\{{\un a}_1  {\un b}_1,   {\un a}_2
{\un b}_2 \}}(x)$~~:~~ $\begin{cases}
\eta^{{\un a}_1  {\un a}_2}\Phi{}_{\{{\un a}_1  {\un b}_1,   {\un a}_2
{\un b}_2 \}}(x)~=~0~~~,\\
\eta^{{\un a}_1  {\un a}_2}\eta^{{\un b}_1  {\un b}_2}\Phi{}_{\{{\un a}_1  {\un b}_1,  
{\un a}_2 {\un b}_2 \}}(x)~=~0~~~,
\end{cases}$\\[10pt]
$(5)\,\Phi_{ \{\un{a}_2|\aone\bone\cone\} }(x):~ \eta^{\aone\un{a}_2}\Phi_{ \{\un{a}_2|\aone\bone\cone\} }(x)~=~0$~~~,\\[10pt]
$\Phi_{ \{\un{a}_2,\un{a}_3|\aone\bone\} }(x):~ \begin{cases}
    \eta^{\un{a}_2\un{a}_3}\Phi_{ \{\un{a}_2,\un{a}_3|\aone\bone\} }(x)~=~0~~~,\\
    \eta^{\un{b}_1\un{a}_3}\Phi_{ \{\un{a}_2,\un{a}_3|\aone\bone\} }(x)~=~0~~~,
    \end{cases}$\\[10pt]
$(2)\,\Phi_{ \{\un{a}_2\un{b}_2|\aone\bone\cone\} }(x):~ \begin{cases}
    \eta^{\aone\un{a}_2}\Phi_{ \{\un{a}_2\un{b}_2|\aone\bone\cone\} }(x)~=~0~~~,\\
    \eta^{\aone\un{a}_2}\eta^{{\un b}_1  {\un b}_2}\Phi_{ \{\un{a}_2\un{b}_2|\aone\bone\cone\} }(x)~=~0~~~,
    \end{cases}$\\[10pt]
    $\Phi_{ \{\un{a}_2,\un{a}_3|\aone\bone\cone\done\}^{\rm C} }(x):~ \begin{cases}
    \eta^{\un{a}_2\un{a}_3}\Phi_{ \{\un{a}_2,\un{a}_3|\aone\bone\cone\done\}^{\rm C} }(x)~=~0~~~,\\
    \eta^{\un{d}_1\un{a}_3}\Phi_{ \{\un{a}_2,\un{a}_3|\aone\bone\cone\done\}^{\rm C} }(x)~=~0~~~,\\
    \Phi_{ \{\un{a}_2,\un{a}_3|\aone\bone\cone\done\}^{\rm C} }(x)~=~-\frac{1}{4!}\epsilon_{\aone\bone\cone\done}{}^{\un{e}_1\un{f}_1\un{g}_1\un{h}_1}\Phi_{ \{\un{a}_2,\un{a}_3|\un{e}_1\un{f}_1\un{g}_1\un{h}_1\}^{\rm C} }(x)~~~,
    \end{cases}$\\[10pt]
    $\Phi_{ \{\un{a}_2,\un{a}_3|\aone\bone\cone\done\}^{\rm S} }(x):~ \begin{cases}
    \eta^{\un{a}_2\un{a}_3}\Phi_{ \{\un{a}_2,\un{a}_3|\aone\bone\cone\done\}^{\rm S} }(x)~=~0~~~,\\
    \eta^{\un{d}_1\un{a}_3}\Phi_{ \{\un{a}_2,\un{a}_3|\aone\bone\cone\done\}^{\rm S} }(x)~=~0~~~,\\
    \Phi_{ \{\un{a}_2,\un{a}_3|\aone\bone\cone\done\}^{\rm S} }(x)~=~\frac{1}{4!}\epsilon_{\aone\bone\cone\done}{}^{\un{e}_1\un{f}_1\un{g}_1\un{h}_1}\Phi_{ \{\un{a}_2,\un{a}_3|\un{e}_1\un{f}_1\un{g}_1\un{h}_1\}^{\rm S} }(x)~~~,
    \end{cases}$\\[10pt]
    $\Phi_{ \{\aone\bone\cone,\un{a}_2\un{b}_2\un{c}_2\} }(x):~ \begin{cases}
    \eta^{\aone\un{a}_2}\Phi_{ \{\aone\bone\cone,\un{a}_2\un{b}_2\un{c}_2\} }(x)~=~0~~~,\\
    \eta^{\aone\un{a}_2}\eta^{{\un b}_1  {\un b}_2}\Phi_{ \{\aone\bone\cone,\un{a}_2\un{b}_2\un{c}_2\} }(x)~=~0~~~,\\
    \eta^{\aone\un{a}_2}\eta^{{\un b}_1  {\un b}_2}\eta^{{\un c}_1  {\un c}_2}\Phi_{ \{\aone\bone\cone,\un{a}_2\un{b}_2\un{c}_2\} }(x)~=~0~~~,
    \end{cases}$\\[10pt]
    $(2)\,\Phi_{ \{\un{a}_2,\un{a}_3|\aone\bone\cone\} }(x):~ \begin{cases}
    \eta^{\un{a}_2\un{a}_3}\Phi_{ \{\un{a}_2,\un{a}_3|\aone\bone\cone\} }(x)~=~0~~~,\\
    \eta^{\un{c}_1\un{a}_3}\Phi_{ \{\un{a}_2,\un{a}_3|\aone\bone\cone\} }(x)~=~0~~~,
    \end{cases}$
\end{itemize}
Level-9 to Level-16 have exactly the same expressions as Level-7 to Level-0.

\subsection{$(1,0)$ Multiplet Decompositions}

Since in eight spacetime dimensions, we actually have chiral spinors corresponding to $\CMTred{[0,0,0,1]}$ representation. So in principle, we can study the component decompositions in the $(1,0)$ supermultiplet, namely only consider the $\CMTred{[0,0,0,1]}$ spinor in the $\theta$-expansion. The dimension of $\CMTred{[0,0,0,1]}$ equals eight implies the $(1,0)$ decomposition has 9 levels in total. Level-$n$ in this case is the antisymmetric product decomposition of $n$ $\CMTred{[0,0,0,1]}$'s, which can be direcly calculated by Plethysm function as discussed in \cite{CNT11d,Susyno}. Branching rules of $\mathfrak{su(8)}\supset\mathfrak{so(8)}$ also give the same results. 
The antisymmetric product decomposition of $(\CMTred{[0,0,0,1]})^{\wedge n}$ is equivalent to the branching rule of $[0,...,1,...,0]$ irrep in $\mathfrak{su(8)}$ to $\mathfrak{so(8)}$, where the $n$-th integer is 1 and all others are zero. 
The projection matrix of $\mathfrak{su(8)}\supset\mathfrak{so(8)}$ will be discussed in Appendix \ref{appen:Proj_Mat_SU8}. 

The $(1,0)$ decompostion results are as below. They are the subset of the complete decomposition results. 
\begin{itemize}
    \item Level-0: $\CMTB{[0,0,0,0]}~(\CMTB{\{1\}})$
    \item Level-1: $\CMTred{[0,0,0,1]}~(\CMTred{\{8_s\}})$
    \item Level-2: $\CMTB{[0,1,0,0]}~(\CMTB{\{28\}})$
    \item Level-3: $\CMTred{[1,0,1,0]}~(\CMTred{\{56_s\}})$
    \item Level-4: $\CMTB{[0,0,2,0]}~(\CMTB{\{35_c\}})~\oplus~\CMTB{[2,0,0,0]}~(\CMTB{\{35_v\}})$
    \item Level-5: $\CMTred{[1,0,1,0]}~(\CMTred{\{56_s\}})$
    \item Level-6: $\CMTB{[0,1,0,0]}~(\CMTB{\{28\}})$
    \item Level-7: $\CMTred{[0,0,0,1]}~(\CMTred{\{8_s\}})$
    \item Level-8: $\CMTB{[0,0,0,0]}~(\CMTB{\{1\}})$
\end{itemize}

Starting from the $(1,0)$ multiplet decomposition, we can reproduce the scalar superfield decomposition using similar ideas when we constructed 10D Type IIA scalar superfield from 10D Type I superfield in \cite{CNT10d}. Basically we can label spinors as $\CMTgrn{\theta^{\alpha}}$ and $\CMTorg{\theta^{\Dot{\alpha}}}$ corresponding to $\CMTred{[0,0,0,1]}$ and $\CMTred{[0,0,1,0]}$ respectively. Then expand the scalar superfield only with respect to $\CMTorg{\theta^{\Dot{\alpha}}}$ first where the accompanying component fields are actually $(1,0)$ superfields. Namely, 
\begin{equation}
\begin{split}
    \mathcal{V}(x,\CMTgrn{\theta^{\alpha}},\CMTorg{\theta^{\Dot{\alpha}}})~=&~ \mathcal{V}(x,\CMTgrn{\theta^{\alpha}}) ~+~ \CMTorg{\theta^{\Dot{\alpha}}}\mathcal{V}_{\Dot{\alpha}}(x,\CMTgrn{\theta^{\alpha}}) ~+~ \CMTorg{\theta^{\Dot{\alpha}}}\CMTorg{\theta^{\Dot{\b}}}\mathcal{V}_{\Dot{\alpha}\Dot{\b}}(x,\CMTgrn{\theta^{\alpha}})~+~\cdots \\
    ~=&~ \mathcal{V}^{(0)}(x)~+~ \CMTgrn{\theta^{\alpha}}\mathcal{V}^{(1)}_{\a}(x) ~+~ \CMTgrn{\theta^{\alpha}}\CMTgrn{\theta^{\b}}\mathcal{V}^{(2)}_{\a\b}(x) ~+~ \cdots \\
    &~+~ \CMTorg{\theta^{\Dot{\alpha}}}\Big[\mathcal{V}^{(0)}_{\Dot{\alpha}}(x)~+~ \CMTgrn{\theta^{\alpha}}\mathcal{V}^{(1)}_{\Dot{\alpha}\a}(x) ~+~ \CMTgrn{\theta^{\alpha}}\CMTgrn{\theta^{\b}}\mathcal{V}^{(2)}_{\Dot{\alpha}\a\b}(x) ~+~ \cdots\Big] \\
    &~+~ \CMTorg{\theta^{\Dot{\alpha}}}\CMTorg{\theta^{\Dot{\b}}}\Big[\mathcal{V}_{\Dot{\alpha}\Dot{\b}}^{(0)}(x)~+~ \CMTgrn{\theta^{\alpha}}\mathcal{V}^{(1)}_{{\Dot{\alpha}\Dot{\b}}\a}(x) ~+~ \CMTgrn{\theta^{\alpha}}\CMTgrn{\theta^{\b}}\mathcal{V}^{(2)}_{{\Dot{\alpha}\Dot{\b}}\a\b}(x) ~+~ \cdots\Big] \\
    &~+~\cdots
\end{split}  
\end{equation}
which implies that for example, Level-2 in the scalar superfield decomposition is nothing but $\CMTred{\{8_s\}}\wedge\CMTred{\{8_s\}}~\oplus~\CMTred{\{8_c\}}\wedge\CMTred{\{8_c\}}~\oplus~\CMTred{\{8_s\}}\otimes\CMTred{\{8_c\}}~=~\CMTB{\{8_{v}\}} \oplus (2) \CMTB{\{28\}} \oplus \CMTB{\{56_{v}\}}$.

\begin{figure}[htp!]
\centering
\begin{minipage}{0.46\textwidth}
    \centering
    \includegraphics[width=0.3\textwidth]{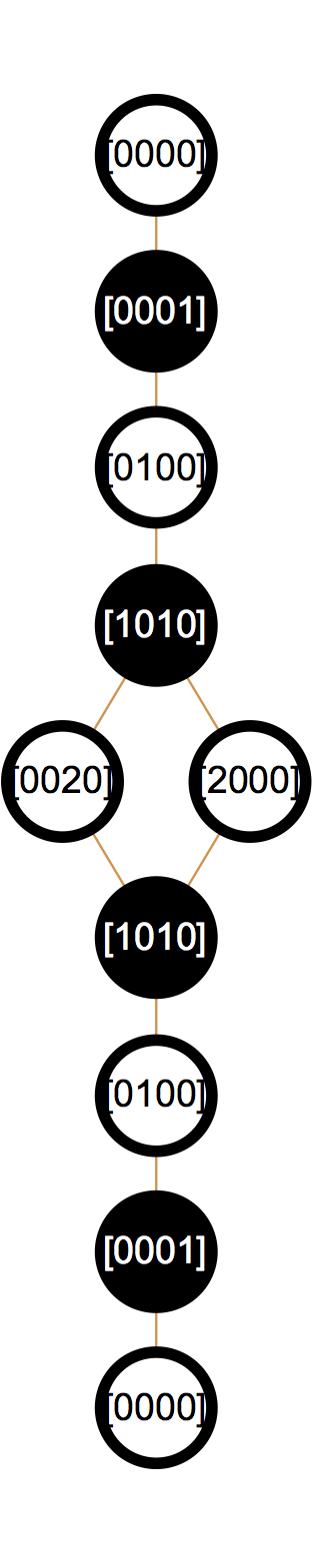}
    \caption{Adynkra Diagram for 8D $(1,0)$ Multiplet}
    \label{Fig:8Dchiral_Dynkin}
\end{minipage}
\begin{minipage}{0.46\textwidth}
    \centering
    \includegraphics[width=0.3\textwidth]{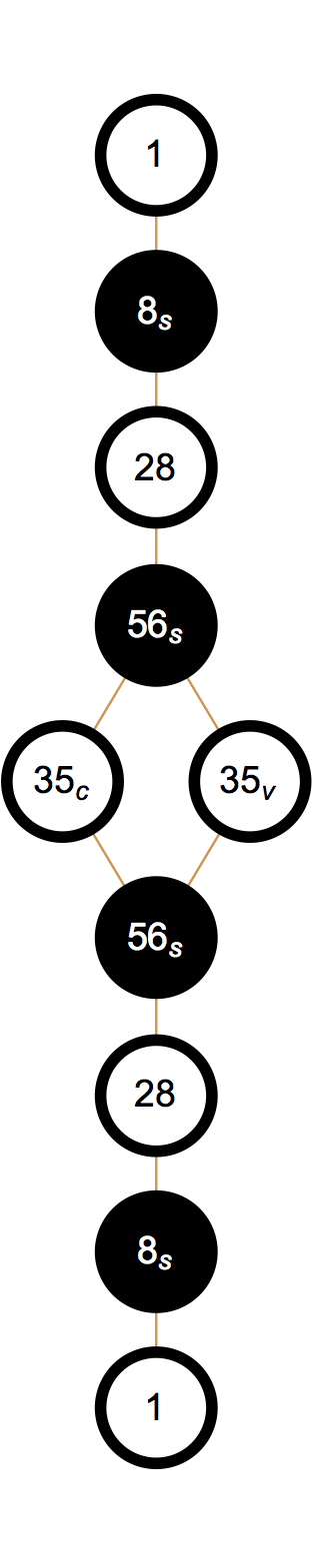}
    \caption{Adinkra Diagram for 8D $(1,0)$ Multiplet}
   \label{Fig:8Dchiral}
\end{minipage}
\end{figure}

Last but not least, we can draw the adynkra and adinkra diagrams corresponding to the $(1,0)$ multiplet, which are Figures \ref{Fig:8Dchiral_Dynkin} and \ref{Fig:8Dchiral}. 

\newpage
\section{7D Minimal Scalar Superfield Decomposition}

\subsection{Component Decompostion Results}

The 7D minimal superfield component decompostion results by Dynkin Labels are given below.
\begin{itemize}
\item Level-0: $\CMTB{[0,0,0]}$
\item Level-1: $(2) \CMTred{[0,0,1]}$
\item Level-2: $\CMTB{[0,0,0]} \oplus (3) \CMTB{[1,0,0]} \oplus (3) \CMTB{[0,1,0]} \oplus \CMTB{[0,0,2]} $
\item Level-3: $(6) \CMTred{[0,0,1]} \oplus (6) \CMTred{[1,0,1]} \oplus (2) \CMTred{[0,1,1]} $
\item Level-4: $(6) \CMTB{[0,0,0]} \oplus (8) \CMTB{[1,0,0]} \oplus (6) \CMTB{[0,1,0]} \oplus (6) \CMTB{[2,0,0]} \oplus (9) \CMTB{[0,0,2]} \oplus (4) \CMTB{[1,1,0]} \oplus \CMTB{[0,2,0]} \oplus (3) \CMTB{[1,0,2]} $
\item Level-5: $(16) \CMTred{[0,0,1]} \oplus (18) \CMTred{[1,0,1]} \oplus (8) \CMTred{[0,1,1]} \oplus (4) \CMTred{[0,0,3]} \oplus (6) \CMTred{[2,0,1]} \oplus (2) \CMTred{[1,1,1]} $
\item Level-6: $(6) \CMTB{[0,0,0]} \oplus (18) \CMTB{[1,0,0]} \oplus (21) \CMTB{[0,1,0]} \oplus (9) \CMTB{[2,0,0]} \oplus (15) \CMTB{[0,0,2]} \oplus (3) \CMTB{[3,0,0]} \oplus (12) \CMTB{[1,1,0]} \oplus \CMTB{[0,2,0]} \oplus (12) \CMTB{[1,0,2]} \oplus (3) \CMTB{[2,1,0]} \oplus (3) \CMTB{[0,1,2]} \oplus \CMTB{[2,0,2]} $
\item Level-7: $(26) \CMTred{[0,0,1]} \oplus (30) \CMTred{[1,0,1]} \oplus (18) \CMTred{[0,1,1]} \oplus (6) \CMTred{[0,0,3]} \oplus (12) \CMTred{[2,0,1]} \oplus (2) \CMTred{[3,0,1]} \oplus (6) \CMTred{[1,1,1]} \oplus (2) \CMTred{[1,0,3]} $
\item Level-8: $(16) \CMTB{[0,0,0]} \oplus (19) \CMTB{[1,0,0]} \oplus (21) \CMTB{[0,1,0]} \oplus (15) \CMTB{[2,0,0]} \oplus (25) \CMTB{[0,0,2]} \oplus (4) \CMTB{[3,0,0]} \oplus (17) \CMTB{[1,1,0]} \oplus (6) \CMTB{[0,2,0]} \oplus \CMTB{[4,0,0]} \oplus (15) \CMTB{[1,0,2]} \oplus \CMTB{[0,0,4]} \oplus (3) \CMTB{[2,1,0]} \oplus (3) \CMTB{[0,1,2]} \oplus (4) \CMTB{[2,0,2]}$
\item Level-9: $(26) \CMTred{[0,0,1]} \oplus (30) \CMTred{[1,0,1]} \oplus (18) \CMTred{[0,1,1]} \oplus (6) \CMTred{[0,0,3]} \oplus (12) \CMTred{[2,0,1]} \oplus (2) \CMTred{[3,0,1]} \oplus (6) \CMTred{[1,1,1]} \oplus (2) \CMTred{[1,0,3]} $
\item Level-10: $(6) \CMTB{[0,0,0]} \oplus (18) \CMTB{[1,0,0]} \oplus (21) \CMTB{[0,1,0]} \oplus (9) \CMTB{[2,0,0]} \oplus (15) \CMTB{[0,0,2]} \oplus (3) \CMTB{[3,0,0]} \oplus (12) \CMTB{[1,1,0]} \oplus \CMTB{[0,2,0]} \oplus (12) \CMTB{[1,0,2]} \oplus (3) \CMTB{[2,1,0]} \oplus (3) \CMTB{[0,1,2]} \oplus \CMTB{[2,0,2]} $
\item Level-11: $(16) \CMTred{[0,0,1]} \oplus (18) \CMTred{[1,0,1]} \oplus (8) \CMTred{[0,1,1]} \oplus (4) \CMTred{[0,0,3]} \oplus (6) \CMTred{[2,0,1]} \oplus (2) \CMTred{[1,1,1]} $
\item Level-12: $(6) \CMTB{[0,0,0]} \oplus (8) \CMTB{[1,0,0]} \oplus (6) \CMTB{[0,1,0]} \oplus (6) \CMTB{[2,0,0]} \oplus (9) \CMTB{[0,0,2]} \oplus (4) \CMTB{[1,1,0]} \oplus \CMTB{[0,2,0]} \oplus (3) \CMTB{[1,0,2]} $
\item Level-13: $(6) \CMTred{[0,0,1]} \oplus (6) \CMTred{[1,0,1]} \oplus (2) \CMTred{[0,1,1]} $
\item Level-14: $\CMTB{[0,0,0]} \oplus (3) \CMTB{[1,0,0]} \oplus (3) \CMTB{[0,1,0]} \oplus \CMTB{[0,0,2]} $
\item Level-15: $(2) \CMTred{[0,0,1]}$
\item Level-16: $\CMTB{[0,0,0]}$
\end{itemize}

Here are the component decompostion results by dimensions.
\begin{itemize}
\item Level-0: $\CMTB{\{1\}}$
\item Level-1: $  (2) \CMTred{\{8\}}$
\item Level-2: $ \CMTB{\{1\}} \oplus (3) \CMTB{\{7\}} \oplus (3) \CMTB{\{21\}} \oplus \CMTB{\{35\}} $
\item Level-3: $ (6) \CMTred{\{8\}} \oplus (6) \CMTred{\{48\}} \oplus (2) \CMTred{\{112\}} $
\item Level-4: $(6) \CMTB{\{1\}} \oplus (8) \CMTB{\{7\}} \oplus (6) \CMTB{\{21\}} \oplus (6) \CMTB{\{27\}} \oplus (9) \CMTB{\{35\}} \oplus (4) \CMTB{\{105\}} \oplus \CMTB{\{168'\}} \oplus (3) \CMTB{\{189\}} $
\item Level-5: $(16) \CMTred{\{8\}} \oplus (18) \CMTred{\{48\}} \oplus (8) \CMTred{\{112\}} \oplus (4) \CMTred{\{112'\}} \oplus (6) \CMTred{\{168\}} \oplus (2) \CMTred{\{512\}} $
\item Level-6: $ (6) \CMTB{\{1\}} \oplus (18) \CMTB{\{7\}} \oplus (21) \CMTB{\{21\}} \oplus (9) \CMTB{\{27\}} \oplus (15) \CMTB{\{35\}} \oplus (3) \CMTB{\{77\}} \oplus (12) \CMTB{\{105\}} \oplus \CMTB{\{168'\}} \oplus (12) \CMTB{\{189\}} \oplus (3) \CMTB{\{330\}} \oplus (3) \CMTB{\{378\}} \oplus \CMTB{\{616\}} $
\item Level-7: $ (26) \CMTred{\{8\}} \oplus (30) \CMTred{\{48\}} \oplus (18) \CMTred{\{112\}} \oplus (6) \CMTred{\{112'\}} \oplus (12) \CMTred{\{168\}} \oplus (2) \CMTred{\{448\}} \oplus (6) \CMTred{\{512\}} \oplus (2) \CMTred{\{560\}} $
\item Level-8: $  (16) \CMTB{\{1\}} \oplus (19) \CMTB{\{7\}} \oplus (21) \CMTB{\{21\}} \oplus (15) \CMTB{\{27\}} \oplus (25) \CMTB{\{35\}} \oplus (4) \CMTB{\{77\}} \oplus (17) \CMTB{\{105\}} \oplus (6) \CMTB{\{168'\}} \oplus \CMTB{\{182\}} \oplus (15) \CMTB{\{189\}} \oplus \CMTB{\{294\}} \oplus (3) \CMTB{\{330\}} \oplus (3) \CMTB{\{378\}} \oplus (4) \CMTB{\{616\}}$
\item Level-9: $ (26) \CMTred{\{8\}} \oplus (30) \CMTred{\{48\}} \oplus (18) \CMTred{\{112\}} \oplus (6) \CMTred{\{112'\}} \oplus (12) \CMTred{\{168\}} \oplus (2) \CMTred{\{448\}} \oplus (6) \CMTred{\{512\}} \oplus (2) \CMTred{\{560\}} $
\item Level-10: $ (6) \CMTB{\{1\}} \oplus (18) \CMTB{\{7\}} \oplus (21) \CMTB{\{21\}} \oplus (9) \CMTB{\{27\}} \oplus (15) \CMTB{\{35\}} \oplus (3) \CMTB{\{77\}} \oplus (12) \CMTB{\{105\}} \oplus \CMTB{\{168'\}} \oplus (12) \CMTB{\{189\}} \oplus (3) \CMTB{\{330\}} \oplus (3) \CMTB{\{378\}} \oplus \CMTB{\{616\}} $
\item Level-11: $(16) \CMTred{\{8\}} \oplus (18) \CMTred{\{48\}} \oplus (8) \CMTred{\{112\}} \oplus (4) \CMTred{\{112'\}} \oplus (6) \CMTred{\{168\}} \oplus (2) \CMTred{\{512\}} $
\item Level-12: $(6) \CMTB{\{1\}} \oplus (8) \CMTB{\{7\}} \oplus (6) \CMTB{\{21\}} \oplus (6) \CMTB{\{27\}} \oplus (9) \CMTB{\{35\}} \oplus (4) \CMTB{\{105\}} \oplus \CMTB{\{168'\}} \oplus (3) \CMTB{\{189\}} $
\item Level-13: $ (6) \CMTred{\{8\}} \oplus (6) \CMTred{\{48\}} \oplus (2) \CMTred{\{112\}} $
\item Level-14: $ \CMTB{\{1\}} \oplus (3) \CMTB{\{7\}} \oplus (3) \CMTB{\{21\}} \oplus \CMTB{\{35\}} $
\item Level-15: $  (2) \CMTred{\{8\}}$
\item Level-16: $\CMTB{\{1\}}$
\end{itemize}

\subsection{7D Minimal Adinkra Diagram}
\begin{figure}[htp!]
\centering
\includegraphics[width=0.8\textwidth]{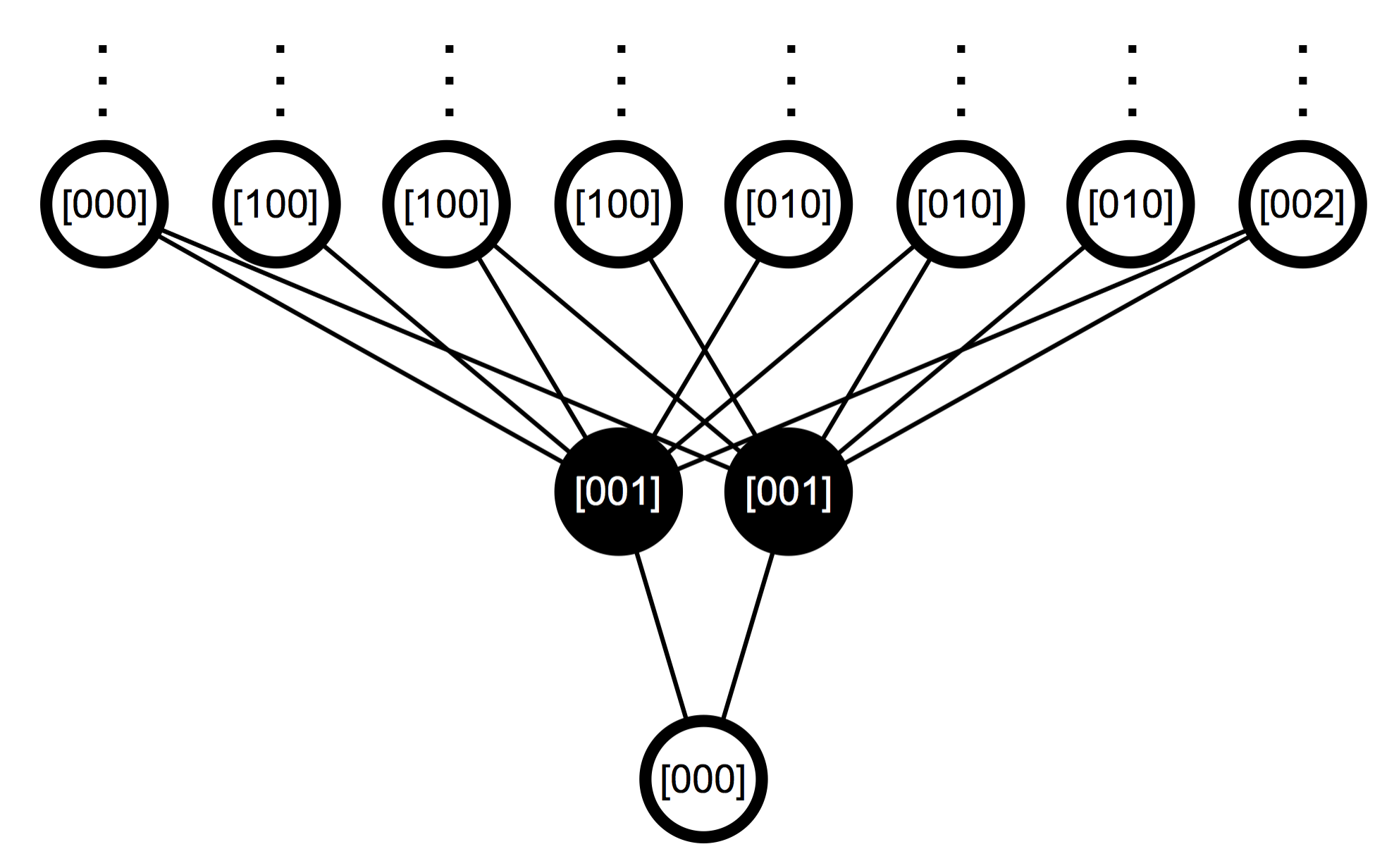}
\caption{Adynkra Diagram for 7D minimal scalar superfield}
\label{Fig:7D_Dynkin}
\end{figure}

\begin{figure}[htp!]
\centering
\includegraphics[width=0.8\textwidth]{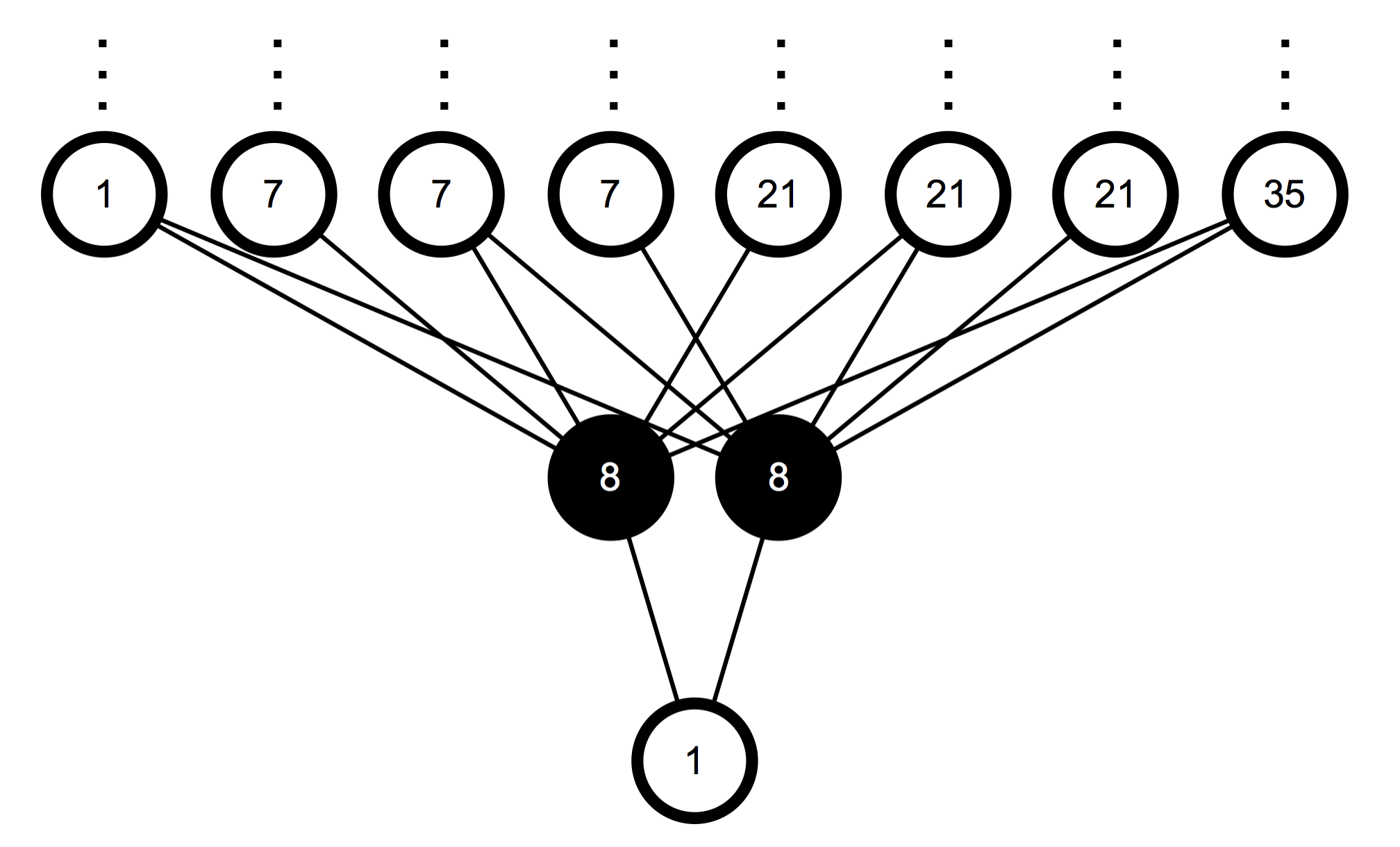}
\caption{Adinkra Diagram for 7D minimal scalar superfield}
\label{Fig:7D}
\end{figure}

The Adynkra and Adinkra diagrams for 7D minimal scalar superfield (up to Level-2) are shown in Figures \ref{Fig:7D_Dynkin} and \ref{Fig:7D}. 

\subsection{Young Tableaux Descriptions of Component Fields in 7D Minimal Scalar Superfield}

Consider the projection matrix for $\mathfrak{su}(7)\supset \mathfrak{so}(7)$~\cite{yamatsu2015},
\begin{equation}
P_{\mathfrak{su}(7)\supset \mathfrak{so}(7)} ~=~
\begin{pmatrix}
1 & 0 & 0 & 0 & 0 & 1\\
0 & 1 & 0 & 0 & 1 & 0\\
0 & 0 & 2 & 2 & 0 & 0\\
\end{pmatrix}~~~.
\end{equation}
The highest weight of a specified irrep of $\mathfrak{su}(7)$ is a row vector $[p_1,p_2,p_3,p_4,p_5,p_6]$,
where $p_1$ to $p_6$ are non-negative integers. Since the $\mathfrak{su}(7)$ YT with $n$ vertical 
boxes is the conjugate of the one with $7-n$ vertical boxes, we only need to consider the $p_4=p_5=p_6=0$ case.

Starting from the weight vector $[p_1,p_2,p_3,0,0,0]$ in $\mathfrak{su}(7)$, we define its projected
weight vector $[p_1,p_2,2p_3]$ in $\mathfrak{so}(7)$ as the Dynkin Label of the corresponding
irreducible bosonic Young Tableau.
\begin{equation}
    [p_1,p_2,2p_3] ~=~ [p_1,p_2,p_3,0,0,0] \, P^T_{\mathfrak{su}(7)\supset\mathfrak{so}(7)} ~~~.
\end{equation}
Then look at the congruence classes of a representation with 
Dynkin Label $[a,b,c]$ in $\mathfrak{so}(7)$,
\begin{equation}
    \begin{split}
        C_{c}(R) ~:=&~  c ~~ ({\rm mod} ~ 2)  ~~~.
    \end{split}
\end{equation}
$C_{c}(R)$ classifies the bosonic irreps and spinorial
irreps:  $C_{c}(R) = 0$ is bosonic and  $C_{c}(R)=1$ is spinorial.  Consequently, a
bosonic irrep satisfies  $c
= 0 ~~ ({\rm mod} ~ 2)$.

Therefore, given an irreducible bosonic Young Tableau with $p_1$ columns of one box,
$p_2$ columns of two vertical boxes, and $p_3$ columns of three vertical boxes, the Dynkin Label of its
corresponding bosonic irrep is $[p_1,p_2,2p_3]$. Reversely, given a bosonic irrep with Dynkin Label $[a,b,c]$, its corresponding irreducible 
bosonic Young Tableau is composed of $a$ columns of one box, $b$ columns of two 
vertical boxes, and $c/2$ columns of three vertical boxes.

The simplest examples,
also the fundamental building blocks of a BYT, are given below.
\begin{equation}
\begin{gathered}
    {\CMTB{\ydiagram{1}}}_{{\rm IR}} ~\equiv~ \CMTB{[1,0,0]}  ~~~,~~~
    {\CMTB{\ydiagram{1,1}}}_{{\rm IR}} ~\equiv~ \CMTB{[0,1,0]} ~~~,~~~
    {\CMTB{\ydiagram{1,1,1}}}_{{\rm IR}} ~\equiv~  \CMTB{[0,0,2]} ~~~.
\end{gathered} \label{equ:BYTbasic_7D}
\end{equation}

For spinorial irreps, the basic SYT is given by
\begin{equation}
    \CMTred{\ytableaushort{\tinyeight}} ~\equiv~~ \CMTred{[0,0,1]} ~~~.
\label{equ:SYTbasic_7D}
\end{equation}
We could translate the Dynkin Label of any spinorial irrep to a mixed YT (which contains a BYT part and a basic SYT above) with irreducible conditions by applying the same idea discussed in Chapter five in \cite{nDx}. 

Putting together the columns in (\ref{equ:BYTbasic_7D}) and (\ref{equ:SYTbasic_7D}) corresponds to adding their Dynkin Labels.

In summary, the irreducible Young Tableau descriptions of the 7D minimal scalar superfield decomposition is presented below.
\begin{equation}
\ytableausetup{boxsize=0.8em}
{\cal V} ~=~ \begin{cases}
~~{\rm {Level}}-0 \,~~~~~~~ \CMTB{\cdot} ~~~,  \\
%%%%%%%%%%%%%%%%%%%%%%%%%%%%%%%%%%%%%%%%%%%%%%%%
~~{\rm {Level}}-1 \,~~~~~~~ (2)\,\CMTred{\ytableaushort{\tinyeight}} ~~~,  \\
%%%%%%%%%%%%%%%%%%%%%%%%%%%%%%%%%%%%%%%%%%%%%%%%
~~{\rm {Level}}-2 \,~~~~~~~ \CMTB{\cdot}~\oplus~
(3)\,{\CMTB{\ydiagram{1}}}_{\rm IR}~\oplus~
(3)\,{\CMTB{\ydiagram{1,1}}}_{\rm IR}~\oplus~{\CMTB{\ydiagram{1,1,1}}}_{\rm IR} ~~~,  \\[15pt]
%%%%%%%%%%%%%%%%%%%%%%%%%%%%%%%%%%%%%%%%%%%%%%%%
~~{\rm {Level}}-3 \,~~~~~~~
(6)\,\CMTred{\ytableaushort{\tinyeight}}~\oplus~
(6)\,{\CMTB{\ydiagram{1}}\CMTred{\ytableaushort{\tinyeight}}}_{\rm IR}~\oplus~
(2)\,{\CMTB{\ydiagram{1,1}}\CMTred{\ytableaushort{\tinyeight,\none}}}_{\rm IR}  ~~~,  \\[10pt]
%%%%%%%%%%%%%%%%%%%%%%%%%%%%%%%%%%%%%%%%%%%%%%%%
~~{\rm {Level}}-4 \,~~~~~~~ (6)\,\CMTB{\cdot}~\oplus~
(8)\,{\CMTB{\ydiagram{1}}}_{\rm IR}~\oplus~
(6)\,{\CMTB{\ydiagram{1,1}}}_{\rm IR}~\oplus~
(6)\,{\CMTB{\ydiagram{2}}}_{\rm IR} ~\oplus~(9)\,{\CMTB{\ydiagram{1,1,1}}}_{\rm IR} \\
~~~~~~~~~~~~~~~~~~~
~\oplus~
(4)\,{\CMTB{\ydiagram{2,1}}}_{\rm IR} ~\oplus~
{\CMTB{\ydiagram{2,2}}}_{\rm IR} ~\oplus~
(3)\,{\CMTB{\ydiagram{2,1,1}}}_{\rm IR}
 ~~~,  \\[20pt]
 %%%%%%%%%%%%%%%%%%%%%%%%%%%%%%%%%%%%%%%%%%%%%%%%
~~{\rm {Level}}-5 \,~~~~~~
(16)\,\CMTred{\ytableaushort{\tinyeight}}~\oplus~
(18)\,{\CMTB{\ydiagram{1}}\CMTred{\ytableaushort{\tinyeight}}}_{\rm IR}~\oplus~
(8)\,{\CMTB{\ydiagram{1,1}}\CMTred{\ytableaushort{\tinyeight,\none}}}_{\rm IR}~\oplus~
(4)\,{\CMTB{\ydiagram{1,1,1}}\CMTred{\ytableaushort{\tinyeight,\none,\none}}}_{{\rm IR}} ~\oplus~(6)\,{\CMTB{\ydiagram{2}}\CMTred{\ytableaushort{\tinyeight}}}_{\rm IR}~\oplus~  (2)\,{\CMTB{\ydiagram{2,1}}\CMTred{\ytableaushort{\tinyeight,\none}}}_{\rm IR}
~~~,  \\[20pt]
%%%%%%%%%%%%%%%%%%%%%%%%%%%%%%%%%%%%%%%%%%%%%%%%
~~{\rm {Level}}-6 \,~~~~~~ (6)\,\CMTB{\cdot}~\oplus~
(18)\,{\CMTB{\ydiagram{1}}}_{\rm IR}~\oplus~
(21)\,{\CMTB{\ydiagram{1,1}}}_{\rm IR}~\oplus~
(9)\,{\CMTB{\ydiagram{2}}}_{{\rm IR}}  ~\oplus~ (15)\,{\CMTB{\ydiagram{1,1,1}}}_{\rm IR}
~\oplus~ (3)\,{\CMTB{\ydiagram{3}}}_{\rm IR}\\[20pt]
~~~~~~~~~~~~~~~~~~~
~\oplus~ (12)\,{\CMTB{\ydiagram{2,1}}}_{\rm IR}
~\oplus~ {\CMTB{\ydiagram{2,2}}}_{\rm IR}
~\oplus~  (12)\,{\CMTB{\ydiagram{2,1,1}}}_{\rm IR}
~\oplus~  (3)\,{\CMTB{\ydiagram{3,1}}}_{\rm IR}
~\oplus~  (3)\,{\CMTB{\ydiagram{2,2,1}}}_{\rm IR}
~\oplus~ {\CMTB{\ydiagram{3,1,1}}}_{\rm IR}
~~~,  \\[20pt]
%%%%%%%%%%%%%%%%%%%%%%%%%%%%%%%%%%%%%%%%%%%%%%%%
~~{\rm {Level}}-7 \,~~~~~~
(26)\,\CMTred{\ytableaushort{\tinyeight}}~\oplus~
(30)\,{\CMTB{\ydiagram{1}}\CMTred{\ytableaushort{\tinyeight}}}_{\rm IR}~\oplus~
(18)\,{\CMTB{\ydiagram{1,1}}\CMTred{\ytableaushort{\tinyeight,\none}}}_{\rm IR}~\oplus~
(6)\,{\CMTB{\ydiagram{1,1,1}}\CMTred{\ytableaushort{\tinyeight,\none,\none}}}_{{\rm IR}} ~\oplus~(12)\,{\CMTB{\ydiagram{2}}\CMTred{\ytableaushort{\tinyeight}}}_{\rm IR}\\[20pt]
~~~~~~~~~~~~~~~~~~~
~\oplus~ 
(2)\,{\CMTB{\ydiagram{3}}\CMTred{\ytableaushort{\tinyeight}}}_{\rm IR}~\oplus~
(6)\,{\CMTB{\ydiagram{2,1}}\CMTred{\ytableaushort{\tinyeight,\none}}}_{\rm IR}
~\oplus~
(2)\,{\CMTB{\ydiagram{2,1,1}}\CMTred{\ytableaushort{\tinyeight,\none,\none}}}_{{\rm IR}}
~~~,  \\[20pt]
%%%%%%%%%%%%%%%%%%%%%%%%%%%%%%%%%%%%%%%%%%%%%%%%
~~{\rm {Level}}-8 \,~~~~~~ (16)\,\CMTB{\cdot}~\oplus~
(19)\,{\CMTB{\ydiagram{1}}}_{\rm IR} ~\oplus~
(21)\,{\CMTB{\ydiagram{1,1}}}_{\rm IR} ~\oplus~
(15)\,{\CMTB{\ydiagram{2}}}_{\rm IR} ~\oplus~
(25)\,{\CMTB{\ydiagram{1,1,1}}}_{\rm IR} ~\oplus~
(4)\,{\CMTB{\ydiagram{3}}}_{\rm IR} \\[15pt]
~~~~~~~~~~~~~~~~~~~~\oplus~
(17)\,{\CMTB{\ydiagram{2,1}}}_{\rm IR} ~\oplus~
(6)\,{\CMTB{\ydiagram{2,2}}}_{\rm IR} ~\oplus~
{\CMTB{\ydiagram{4}}}_{\rm IR} 
~\oplus~(15)\,{\CMTB{\ydiagram{2,1,1}}}_{\rm IR} ~\oplus~ 
{\CMTB{\ydiagram{2,2,2}}}_{\rm IR} \\[15pt]
~~~~~~~~~~~~~~~~~~~~\oplus~
(3)\,{\CMTB{\ydiagram{3,1}}}_{\rm IR} ~\oplus~
(3)\,{\CMTB{\ydiagram{2,2,1}}}_{\rm IR}~\oplus~
(4)\,{\CMTB{\ydiagram{3,1,1}}}_{\rm IR} 
~~~,  \\
{~~~~~~}  {~~~~} \vdots  {~~~~~~~~~\,~~~~~~} \vdots
\end{cases}
\label{equ:V_7D}
\ytableausetup{boxsize=1.2em}
\end{equation}
where Level-9 to Level-16 have exactly the same expressions as Level-7 to Level-0. 

\subsection{Index Structures and Irreducible Conditions of Component Fields in 7D Minimal Scalar Superfield}\label{sec:7D_Index}

In this section, we will translate the irreducible bosonic and mixed Young Tableaux into field variables. 
We follow the same $\{\}$-indices notation as well as ``$|$'' to separate indices in YT with different heights and ``,'' to separate
indices in YT with the same heights. 

The vector index $\un{a}$ runs from 0 to 6. 
The $\{\}$-indices, irreducible bosonic Young
Tableaux, and Dynkin Labels are equivalent and have the one-to-one correspondence.

The general expression is as below in Equations (\ref{equ:index-notation1_7D}) and (\ref{equ:index-notation2_7D}),
\begin{equation}
    \begin{split}
       & \{ \un{a}_1,\dots,\un{a}_p ~|~ \un{b}_1\un{c}_1,\dots,\un{b}_q \un{c}_q ~|~  \un{d}_1\un{e}_1\un{f}_1,\dots,\un{d}_r\un{e}_r\un{f}_r\}  \\[10pt]
      & ~{ \CMTB{{\ytableaushort{\aone}} } }~~~~~{ \CMTB{{\ytableaushort{\ap}} } }
      ~~~~{ \CMTB{{\ytableaushort{\bone,\cone}} } }~~~~~~~{ \CMTB{{\ytableaushort{\bq,\cunq}} } }
      ~~~~~~{ \CMTB{{\ytableaushort{\done,\eone,\fone}} } }~~~~~~~~~~{ \CMTB{{\ytableaushort{\dr,\er,\fr}} } }
      \\[10pt]
      & ~~~ \CMTB{[p,0,0]} ~~~~~~~~~ \CMTB{[0,q,0]} ~~~~~~~~~~~~~ \CMTB{[0,0,2r]}  
    \end{split}
    \label{equ:index-notation1_7D}
\end{equation}
where above we have ``disassembled'' the YT to show how each column is affiliated with each type of
subscript structure. Below we have assembled all the columns into a proper YT.
\begin{equation}
\begin{split}
    & { \CMTB{{\ytableaushort{\done\dots\dr\bone\dots\bq\aone\dots\ap,\eone\dots\er\cone\dots\cunq,\fone\dots\fr}} } }_{\rm IR}\\[10pt]
    &~~~~~~~~~~~~~ \CMTB{[p,q,2r]}
\end{split}
\label{equ:index-notation2_7D}
\end{equation}

As one moves from the YT's shown in Equation~(\ref{equ:index-notation1_7D}) to
Equation~(\ref{equ:index-notation2_7D}), it is clear the number of vertical boxes is tabulating the number of
1-forms, 2-forms, and 3-forms in the YT's.  These are the entries between
the vertical $|$ bars. These precisely correspond to the integers $p$, $q$, and $r$ appeared in Dynkin Labels.  
An example of the correspondence between the subscript conventions,
the affiliated YT, and Dynkin Label is shown in (\ref{equ:index-notation_ex1_7D}). 
\begin{equation}
    \{{\un a}_2 , {\un a}_3| {\un a}_1  {\un b}_1   {\un c}_1 \} ~~\equiv~~
    { \CMTB{{\ytableaushort{\aone \atwo \athree,\bone,\cone}} } }_{{\rm IR}}
    ~~\equiv~~ \CMTB{[2,0,2]} ~~~.
    \label{equ:index-notation_ex1_7D}
\end{equation}

The index structures as well as irreducible conditions of all bosonic and fermionic fields are identified below along 
with the level at which the fields occur in the adinkra of the scalar superfield. The spinor index $\alpha$ runs from 1 to 8.

\begin{itemize}
    \item Level-0: $\Phi(x)$~~~,
    %%%%%%%%%%%%%%%%%%%%%%%%%%%%%%%%%%%%%%%%
    \item Level-1: (2)\,$\Psi_{\alpha}(x)$~~~,
    %%%%%%%%%%%%%%%%%%%%%%%%%%%%%%%%%%%%%%%%
    \item Level-2: $\Phi(x)$~~~,~~~(3)\,$\Phi_{ \{\aone\} }(x)$ ~~~,~~~(3)\,$\Phi_{ \{\aone\bone\} }(x)$ ~~~,~~~ $\Phi_{ \{\aone\bone\cone\} }(x)$~~~,
    %%%%%%%%%%%%%%%%%%%%%%%%%%%%%%%%%%%%%%%%
    \item Level-3: (6)\,$\Psi_{\alpha}(x)$~~~,~~~
    (6)\,$\Psi_{\{\aone\}\alpha}(x):~ (\g^{\aone})_{\beta}{}^{\alpha}\Psi_{\{\aone\}\alpha}(x)~=~0 $~~~,~~~\\[10pt]
    (2)\,$\Psi_{\{\aone\bone\}\alpha}(x):~ (\g^{\aone})_{\beta}{}^{\alpha}\Psi_{\{\aone\bone\}\alpha}(x)~=~0 $~~~,
    %%%%%%%%%%%%%%%%%%%%%%%%%%%%%%%%%%%%%%%%
    \item Level-4: (6)\,$\Phi(x)$~~~,~~~(8)\,$\Phi_{ \{\aone\} }(x)$ ~~~,~~~(6)\,$\Phi_{ \{\aone\bone\} }(x)$ ~~~,~~~(6)\,$\Phi_{ \{\aone,\un{a}_2 \} }(x):~ \eta^{\aone\un{a}_2}\Phi_{ \{\aone,\un{a}_2 \} }(x)~=~0$~~~,~~~\\[10pt]
    (9)\,$\Phi_{ \{\aone\bone\cone\} }(x)$~~~,~~~
       (4)\,$\Phi_{ \{\un{a}_2|\aone\bone \} }(x):~ \eta^{\aone\un{a}_2}\Phi_{ \{\un{a}_2|\aone\bone \} }(x)~=~0$~~~,\\[10pt]
       $ \Phi{}_{\{{\un a}_1  {\un b}_1,   {\un a}_2
{\un b}_2 \}}(x)$~~:~~ $\begin{cases}
\eta^{{\un a}_1  {\un a}_2}\Phi{}_{\{{\un a}_1  {\un b}_1,   {\un a}_2
{\un b}_2 \}}(x)~=~0~~~,\\
\eta^{{\un a}_1  {\un a}_2}\eta^{{\un b}_1  {\un b}_2}\Phi{}_{\{{\un a}_1  {\un b}_1,  
{\un a}_2 {\un b}_2 \}}(x)~=~0~~~,
\end{cases}$\\[10pt]
(3)\,$\Phi_{ \{\un{a}_2|\aone\bone\cone\} }(x):~ \eta^{\aone\un{a}_2}\Phi_{ \{\un{a}_2|\aone\bone\cone\} }(x)~=~0$~~~,
%%%%%%%%%%%%%%%%%%%%%%%%%%%%%%%%%%%%%%%%
    \item Level-5: $(16)\,\Psi_{\alpha}(x)$~~~,~~~
    $(18)\,\Psi_{\{\aone\}\alpha}(x):~ (\g^{\aone})_{\beta}{}^{\alpha}\Psi_{\{\aone\}\alpha}(x)~=~0 $~~~,~~~\\[10pt]
    (8)\,$\Psi_{\{\aone\bone\}\alpha}(x):~ (\g^{\aone})_{\beta}{}^{\alpha}\Psi_{\{\aone\bone\}\alpha}(x)~=~0 $~~~,\\[10pt]
    $(4)\,\Psi_{\{\aone\bone\cone\}\alpha}(x):~ (\g^{\aone})_{\beta}{}^{\alpha}\Psi_{\{\aone\bone\cone\}\alpha}(x)~=~0 $~~~,\\[10pt]
    $(6)\,\Psi_{\{\aone,\un{a}_2\}\alpha}(x):~\begin{cases}
    (\g^{\aone})_{\beta}{}^{\alpha}\Psi_{\{\aone,\un{a}_2\}\alpha}(x)~\equiv~ \psi_{\{\un{a}_2\}\beta} ~=~0~~~, \\
    (\g^{\un{a}_2})_{\g}{}^{\beta}\psi_{\{\un{a}_2\}\beta} ~=~0~~~,
    \end{cases}$\\[10pt]
    $(2)\,\Psi_{\{\un{a}_2|\aone\bone\}\alpha}(x):~\begin{cases}
    (\g^{\bone})_{\beta}{}^{\alpha}\Psi_{\{\un{a}_2|\aone\bone\}\alpha}(x)~\equiv~ \psi_{\{\aone,\un{a}_2\}\beta} ~=~0 ~~~,\\
    (\g^{\un{a}_2})_{\beta}{}^{\alpha}\Psi_{\{\un{a}_2|\aone\bone\}\alpha}(x)~\equiv~ \psi_{\{\aone\bone\}\beta} ~=~0~~~, \\
    (\g^{\bone})_{\g}{}^{\beta}\psi_{\{\aone\bone\}\beta} ~=~0 ~~~,
    \end{cases}$
    %%%%%%%%%%%%%%%%%%%%%%%%%%%%%%%%%%%%%%%%
    \item Level-6: 
    (6)\,$\Phi(x)$~~~,~~~(18)\,$\Phi_{ \{\aone\} }(x)$ ~~~,~~~(21)\,$\Phi_{ \{\aone\bone\} }(x)$ ~~~,~~~(9)\,$\Phi_{ \{\aone,\un{a}_2 \} }(x):~ \eta^{\aone\un{a}_2}\Phi_{ \{\aone,\un{a}_2 \} }(x)~=~0$~~~,~~~\\[10pt]
    (15)\,$\Phi_{ \{\aone\bone\cone\} }(x)$~~~,~~~
       (3)\,$\Phi_{ \{\aone,\un{a}_2,\un{a}_3 \} }(x):~ \eta^{\aone\un{a}_2}\Phi_{ \{\aone,\un{a}_2,\un{a}_3 \} }(x)~=~0$~~~,\\[10pt]
       (12)\,$\Phi_{ \{\un{a}_2|\aone\bone \} }(x):~ \eta^{\aone\un{a}_2}\Phi_{ \{\un{a}_2|\aone\bone \} }(x)~=~0$~~~,\\[10pt]
       $ \Phi{}_{\{{\un a}_1  {\un b}_1,   {\un a}_2
{\un b}_2 \}}(x)$~~:~~ $\begin{cases}
\eta^{{\un a}_1  {\un a}_2}\Phi{}_{\{{\un a}_1  {\un b}_1,   {\un a}_2
{\un b}_2 \}}(x)~=~0~~~,\\
\eta^{{\un a}_1  {\un a}_2}\eta^{{\un b}_1  {\un b}_2}\Phi{}_{\{{\un a}_1  {\un b}_1,  
{\un a}_2 {\un b}_2 \}}(x)~=~0~~~,
\end{cases}$\\[10pt]
(12)\,$\Phi_{ \{\un{a}_2|\aone\bone\cone\} }(x):~ \eta^{\aone\un{a}_2}\Phi_{ \{\un{a}_2|\aone\bone\cone\} }(x)~=~0$~~~,\\[10pt]
(3)\,$\Phi_{ \{\un{a}_2,\un{a}_3|\aone\bone\} }(x):~ \begin{cases}
    \eta^{\un{a}_2\un{a}_3}\Phi_{ \{\un{a}_2,\un{a}_3|\aone\bone\} }(x)~=~0~~~,\\
    \eta^{\un{b}_1\un{a}_3}\Phi_{ \{\un{a}_2,\un{a}_3|\aone\bone\} }(x)~=~0~~~,
    \end{cases}$\\[10pt]
    (3)\,$\Phi_{ \{\un{a}_2\un{b}_2|\aone\bone\cone\} }(x):~ \begin{cases}
    \eta^{\aone\un{a}_2}\Phi_{ \{\un{a}_2\un{b}_2|\aone\bone\cone\} }(x)~=~0~~~,\\
    \eta^{\aone\un{a}_2}\eta^{{\un b}_1  {\un b}_2}\Phi_{ \{\un{a}_2\un{b}_2|\aone\bone\cone\} }(x)~=~0~~~,
    \end{cases}$\\[10pt]
    $\Phi_{ \{\un{a}_2,\un{a}_3|\aone\bone\cone\} }(x):~ \begin{cases}
    \eta^{\un{a}_2\un{a}_3}\Phi_{ \{\un{a}_2,\un{a}_3|\aone\bone\cone\} }(x)~=~0~~~,\\
    \eta^{\un{c}_1\un{a}_3}\Phi_{ \{\un{a}_2,\un{a}_3|\aone\bone\cone\} }(x)~=~0~~~,
    \end{cases}$
    %%%%%%%%%%%%%%%%%%%%%%%%%%%%%%%%%%%%%%%%
    \item Level-7: $(26)\,\Psi_{\alpha}(x)$~~~,~~~
    $(30)\,\Psi_{\{\aone\}\alpha}(x):~ (\g^{\aone})_{\beta}{}^{\alpha}\Psi_{\{\aone\}\alpha}(x)~=~0 $~~~,~~~\\[10pt]
    (18)\,$\Psi_{\{\aone\bone\}\alpha}(x):~ (\g^{\aone})_{\beta}{}^{\alpha}\Psi_{\{\aone\bone\}\alpha}(x)~=~0 $~~~,\\[10pt]
    $(6)\,\Psi_{\{\aone\bone\cone\}\alpha}(x):~ (\g^{\aone})_{\beta}{}^{\alpha}\Psi_{\{\aone\bone\cone\}\alpha}(x)~=~0 $~~~,\\[10pt]
    $(12)\,\Psi_{\{\aone,\un{a}_2\}\alpha}(x):~\begin{cases}
    (\g^{\aone})_{\beta}{}^{\alpha}\Psi_{\{\aone,\un{a}_2\}\alpha}(x)~\equiv~ \psi_{\{\un{a}_2\}\beta} ~=~0~~~, \\
    (\g^{\un{a}_2})_{\g}{}^{\beta}\psi_{\{\un{a}_2\}\beta} ~=~0~~~,
    \end{cases}$\\[10pt]
    (2)$\Psi_{\{\aone,\un{a}_2,\un{a}_3\}\alpha}(x):~\begin{cases}
    (\g^{\aone})_{\beta}{}^{\alpha}\Psi_{\{\aone,\un{a}_2,\un{a}_3\}\alpha}(x)~\equiv~ \psi_{\{\un{a}_2,\un{a}_3\}\b} ~=~0~~~, \\
    (\g^{\un{a}_2})_{\g}{}^{\b}\psi_{\{\un{a}_2,\un{a}_3\}\beta} ~\equiv~ \psi_{\{\un{a}_3\}\g}~=~0~~~,\\
    (\g^{\un{a}_3})_{\e}{}^{\g}\psi_{\{\un{a}_3\}\g} ~=~0~~~,
    \end{cases}$\\[10pt]
    $(6)\,\Psi_{\{\un{a}_2|\aone\bone\}\alpha}(x):~\begin{cases}
    (\g^{\bone})_{\beta}{}^{\alpha}\Psi_{\{\un{a}_2|\aone\bone\}\alpha}(x)~\equiv~ \psi_{\{\aone,\un{a}_2\}\beta} ~=~0 ~~~,\\
    (\g^{\un{a}_2})_{\beta}{}^{\alpha}\Psi_{\{\un{a}_2|\aone\bone\}\alpha}(x)~\equiv~ \psi_{\{\aone\bone\}\beta} ~=~0~~~, \\
    (\g^{\bone})_{\g}{}^{\beta}\psi_{\{\aone\bone\}\beta} ~=~0 ~~~,
    \end{cases}$\\[10pt]
    (2)\,$\Psi_{\{\un{a}_2|\aone\bone\cone\}\alpha}(x):~\begin{cases}
    (\g^{\cone})_{\beta}{}^{\alpha}\Psi_{\{\un{a}_2|\aone\bone\cone\}\alpha}(x)~\equiv~ \psi_{\{\un{a}_2|\aone\bone\}\beta} ~=~0~~~, \\
    (\g^{\un{a}_2})_{\beta}{}^{\alpha}\Psi_{\{\un{a}_2|\aone\bone\cone\}\alpha}(x)~\equiv~ \psi_{\{\aone\bone\cone\}\beta} ~=~0~~~, \\
    (\g^{\cone})_{\g}{}^{\beta}\psi_{\{\aone\bone\cone\}\beta} ~=~0~~~, 
    \end{cases}$
    %%%%%%%%%%%%%%%%%%%%%%%%%%%%%%%%%%%%%%%%
    \item Level-8:
    (16)\,$\Phi(x)$~~~,~~~(19)\,$\Phi_{ \{\aone\} }(x)$ ~~~,~~~(21)\,$\Phi_{ \{\aone\bone\} }(x)$ ~~~,~~~(15)\,$\Phi_{ \{\aone,\un{a}_2 \} }(x):~ \eta^{\aone\un{a}_2}\Phi_{ \{\aone,\un{a}_2 \} }(x)~=~0$~~~,~~~\\[10pt]
    (25)\,$\Phi_{ \{\aone\bone\cone\} }(x)$~~~,~~~
       (4)\,$\Phi_{ \{\aone,\un{a}_2,\un{a}_3 \} }(x):~ \eta^{\aone\un{a}_2}\Phi_{ \{\aone,\un{a}_2,\un{a}_3 \} }(x)~=~0$~~~,\\[10pt]
       (17)\,$\Phi_{ \{\un{a}_2|\aone\bone \} }(x):~ \eta^{\aone\un{a}_2}\Phi_{ \{\un{a}_2|\aone\bone \} }(x)~=~0$~~~,\\[10pt]
       (6)\,$ \Phi{}_{\{{\un a}_1  {\un b}_1,   {\un a}_2
{\un b}_2 \}}(x)$~~:~~ $\begin{cases}
\eta^{{\un a}_1  {\un a}_2}\Phi{}_{\{{\un a}_1  {\un b}_1,   {\un a}_2
{\un b}_2 \}}(x)~=~0~~~,\\
\eta^{{\un a}_1  {\un a}_2}\eta^{{\un b}_1  {\un b}_2}\Phi{}_{\{{\un a}_1  {\un b}_1,  
{\un a}_2 {\un b}_2 \}}(x)~=~0~~~,
\end{cases}$\\[10pt]
$\Phi_{ \{\aone,\un{a}_2,\un{a}_3,\un{a}_4\} }(x):~ \begin{cases}
    \eta^{\un{a}_1\un{a}_2}\Phi_{ \{\aone,\un{a}_2,\un{a}_3,\un{a}_4\} }(x)~=~0~~~,\\
     \eta^{\un{a}_1\un{a}_2}\eta^{\un{a}_3\un{a}_4}\Phi_{ \{\aone,\un{a}_2,\un{a}_3,\un{a}_4\} }(x)~=~0~~~,
    \end{cases}$\\[10pt]
(15)\,$\Phi_{ \{\un{a}_2|\aone\bone\cone\} }(x):~ \eta^{\aone\un{a}_2}\Phi_{ \{\un{a}_2|\aone\bone\cone\} }(x)~=~0$~~~,\\[10pt]
$\Phi_{ \{\aone\bone\cone,\un{a}_2\un{b}_2\un{c}_2\} }(x):~ \begin{cases}
    \eta^{\aone\un{a}_2}\Phi_{ \{\aone\bone\cone,\un{a}_2\un{b}_2\un{c}_2\} }(x)~=~0~~~,\\
    \eta^{\aone\un{a}_2}\eta^{{\un b}_1  {\un b}_2}\Phi_{ \{\aone\bone\cone,\un{a}_2\un{b}_2\un{c}_2\} }(x)~=~0~~~,\\
    \eta^{\aone\un{a}_2}\eta^{{\un b}_1  {\un b}_2}\eta^{{\un c}_1  {\un c}_2}\Phi_{ \{\aone\bone\cone,\un{a}_2\un{b}_2\un{c}_2\} }(x)~=~0~~~,
    \end{cases}$\\[10pt]
(3)\,$\Phi_{ \{\un{a}_2,\un{a}_3|\aone\bone\} }(x):~ \begin{cases}
    \eta^{\un{a}_2\un{a}_3}\Phi_{ \{\un{a}_2,\un{a}_3|\aone\bone\} }(x)~=~0~~~,\\
    \eta^{\un{b}_1\un{a}_3}\Phi_{ \{\un{a}_2,\un{a}_3|\aone\bone\} }(x)~=~0~~~,
    \end{cases}$\\[10pt]
    (3)\,$\Phi_{ \{\un{a}_2\un{b}_2|\aone\bone\cone\} }(x):~ \begin{cases}
    \eta^{\aone\un{a}_2}\Phi_{ \{\un{a}_2\un{b}_2|\aone\bone\cone\} }(x)~=~0~~~,\\
    \eta^{\aone\un{a}_2}\eta^{{\un b}_1  {\un b}_2}\Phi_{ \{\un{a}_2\un{b}_2|\aone\bone\cone\} }(x)~=~0~~~,
    \end{cases}$\\[10pt]
    (4)\,$\Phi_{ \{\un{a}_2,\un{a}_3|\aone\bone\cone\} }(x):~ \begin{cases}
    \eta^{\un{a}_2\un{a}_3}\Phi_{ \{\un{a}_2,\un{a}_3|\aone\bone\cone\} }(x)~=~0~~~,\\
    \eta^{\un{c}_1\un{a}_3}\Phi_{ \{\un{a}_2,\un{a}_3|\aone\bone\cone\} }(x)~=~0~~~,
    \end{cases}$
\end{itemize}

\subsection{$(1,0)$ Multiplet Decompositions}
In seven spacetime dimensions, although we don't have chiral spinors, instead we have two copies of $\CMTred{[0,0,1]}$ due to the SU(2)-Majorana condition. Starting from one copy of $\CMTred{[0,0,1]}$ spinor, we can construct a supermultiplet which is the subset of the one constructed from the minimal scalar superfield. We call it as $(1,0)$ multiplet decomposition.
Apply the Plethysm function and results are as below. Using branching rules for $\mathfrak{su(8)}\supset\mathfrak{so}(7)$ gives the same results. The projection matrix is presented in Equation (\ref{equ:Psu8toso7}). 

\begin{itemize}
    \item Level-0: $\CMTB{[0,0,0]}~(\CMTB{\{1\}})$
    \item Level-1: $\CMTred{[0,0,1]}~(\CMTred{\{8\}})$
    \item Level-2: $\CMTB{[0,1,0]}~(\CMTB{\{21\}})~\oplus~ \CMTB{[1,0,0]}~(\CMTB{\{7\}})$
    \item Level-3: $\CMTred{[1,0,1]}~(\CMTred{\{48\}})~\oplus~\CMTred{[0,0,1]}~(\CMTred{\{8\}})$
    \item Level-4: $\CMTB{[0,0,2]}~(\CMTB{\{35\}})~\oplus~\CMTB{[2,0,0]}~(\CMTB{\{27\}})~\oplus~\CMTB{[0,0,0]}~(\CMTB{\{1\}})~\oplus~\CMTB{[1,0,0]}~(\CMTB{\{7\}})$
    \item Level-5: $\CMTred{[1,0,1]}~(\CMTred{\{48\}})~\oplus~\CMTred{[0,0,1]}~(\CMTred{\{8\}})$
    \item Level-6: $\CMTB{[0,1,0]}~(\CMTB{\{21\}})~\oplus~ \CMTB{[1,0,0]}~(\CMTB{\{7\}})$
    \item Level-7: $\CMTred{[0,0,1]}~(\CMTred{\{8\}})$
    \item Level-8: $\CMTB{[0,0,0]}~(\CMTB{\{1\}})$
\end{itemize}

Starting from the above decompositions, we can reproduce the minimal scalar superfield decompositions using the similar idea as when we construct 10D Type IIB scalar superfield from 10D Type I superfield. Basically we can label spinors as $\theta^{\alpha}$ and $\CMTgrn{\theta^{\alpha}}$ both corresponding to the same irrep $\CMTred{[0,0,1]}$. Then expand the scalar superfield only with respect to $\CMTgrn{\theta^{\alpha}}$ first. Namely,
\begin{equation}
\begin{split} 
    \mathcal{V}(x,\theta^{\a},\CMTgrn{\theta^{\a}})~=&~ \mathcal{V}(x,\theta^{\a}) ~+~ \CMTgrn{\theta^{\a}}\mathcal{V}_{\CMTgrn{\a}}(x,\theta^{\a}) ~+~ \CMTgrn{\theta^{\a}}\CMTgrn{\theta^{\b}}\mathcal{V}_{\CMTgrn{\a\beta}}(x,\theta^{\a})~+~\cdots \\
    ~=&~ \mathcal{V}^{(0)}(x)~+~ \theta^{\a}\mathcal{V}^{(1)}_{\a}(x) ~+~ \theta^{\a}\theta^{\b}\mathcal{V}^{(2)}_{\a\b}(x) ~+~ \cdots \\
    &~+~ \CMTgrn{\theta^{\a}}\Big[\mathcal{V}^{(0)}_{\CMTgrn{\a}}(x)~+~ \theta^{\a}\mathcal{V}^{(1)}_{\CMTgrn{\a}\a}(x) ~+~ \theta^{\a}\theta^{\b}\mathcal{V}^{(2)}_{\CMTgrn{\a}\a\b}(x) ~+~ \cdots\Big] \\
    &~+~ \CMTgrn{\theta^{\a}\theta^{\b}}\Big[\mathcal{V}_{\CMTgrn{\a\b}}^{(0)}(x)~+~ \theta^{\a}\mathcal{V}^{(1)}_{\CMTgrn{\a\b}\a}(x) ~+~ \theta^{\a}\theta^{\b}\mathcal{V}^{(2)}_{\CMTgrn{\a\b}\a\b}(x) ~+~ \cdots\Big]\\
    &~+~\cdots
\end{split}  
\end{equation}
which implies that for example, Level-2 in the scalar superfield decomposition is nothing but $(2)\,\CMTred{\{8\}}\wedge\CMTred{\{8\}}~\oplus~\CMTred{\{8\}}\otimes\CMTred{\{8\}}~=~\CMTB{\{1\}} \oplus (3) \CMTB{\{7\}} \oplus (3) \CMTB{\{21\}} \oplus \CMTB{\{35\}}$.

\begin{figure}[htp!]
\centering
\begin{minipage}{0.46\textwidth}
    \centering
    \includegraphics[width=0.6\textwidth]{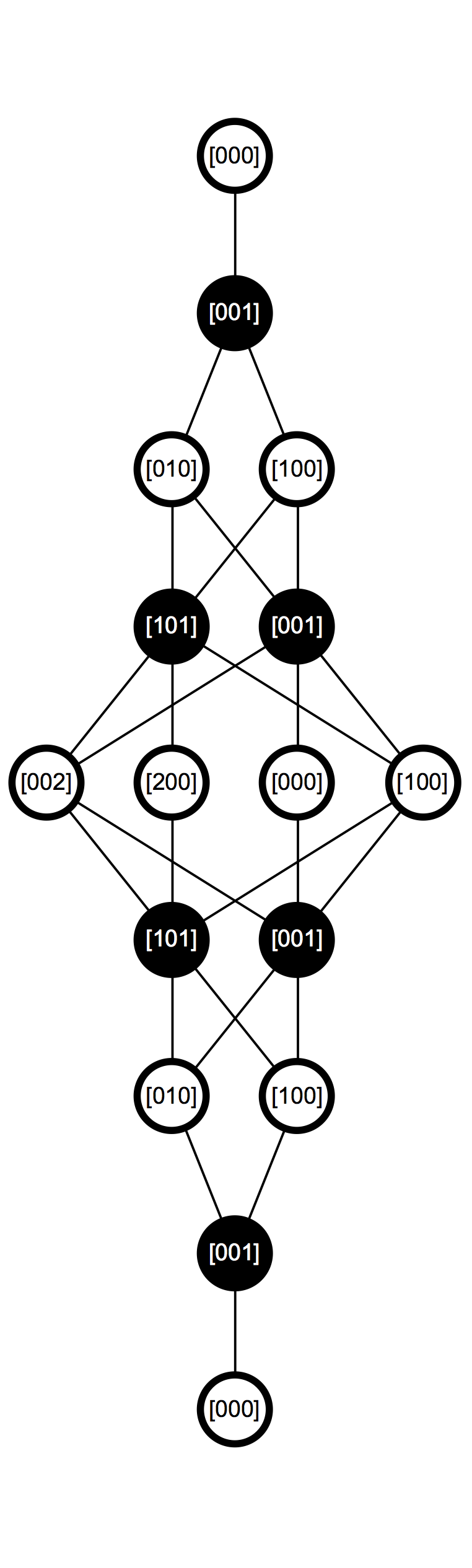}
    \caption{Adynkra Diagram for 7D $(1,0)$ Multiplet}
    \label{Fig:7Dchiral_Dynkin}
\end{minipage}
\begin{minipage}{0.46\textwidth}
    \centering
    \includegraphics[width=0.6\textwidth]{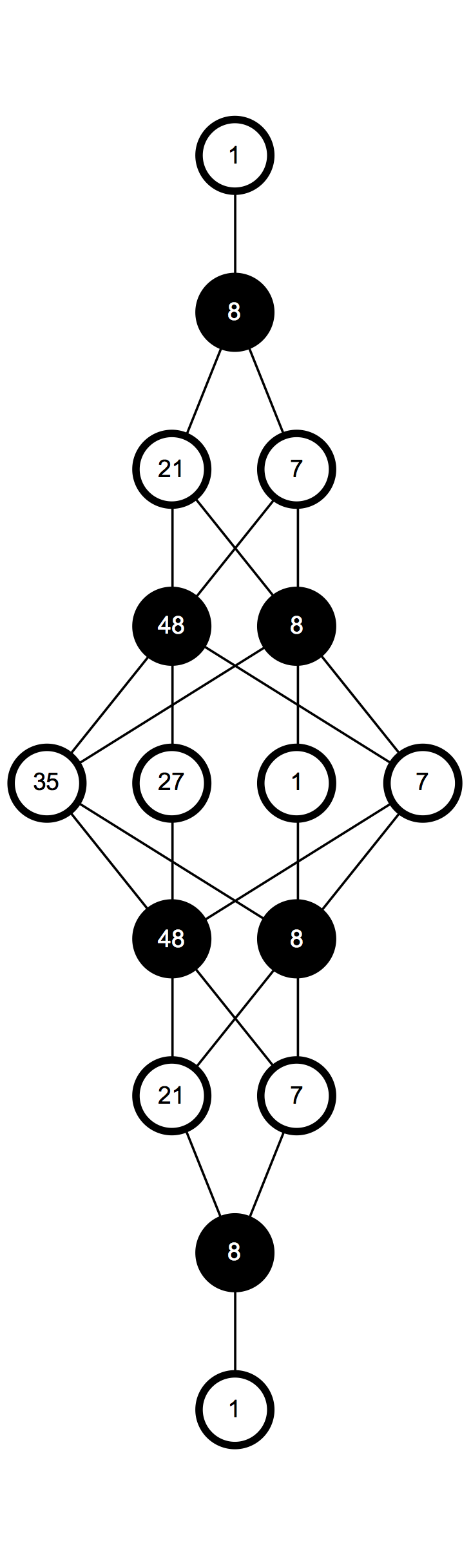}
    \caption{Adinkra Diagram for 7D $(1,0)$ Multiplet}
   \label{Fig:7Dchiral}
\end{minipage}
\end{figure}

Last but not least, we can draw the adynkra and adinkra diagrams corresponding to the $(1,0)$ multiplet, which are Figures \ref{Fig:7Dchiral_Dynkin} and \ref{Fig:7Dchiral}. 

\newpage
\section{6D Minimal Scalar Superfield Decomposition}

% The Lorentz group we are working on in 6D is SU(4) $\cong$ SO(6). 

\subsection{Component Decompostion Results}

The 6D minimal superfield component decompostion results by Dynkin Labels are listed below.
\begin{itemize}
\item Level-0: $\CMTB{[0,0,0]}$
\item Level-1: $ \CMTred{[0,1,0]} \oplus \CMTred{[0,0,1]} $
\item Level-2: $\CMTB{[0,0,0]} \oplus (2) \CMTB{[1,0,0]} \oplus \CMTB{[0,1,1]} $
\item Level-3: $(2) \CMTred{[0,1,0]} \oplus (2) \CMTred{[0,0,1]} \oplus \CMTred{[1,0,1]} \oplus \CMTred{[1,1,0]} $
\item Level-4: $(3) \CMTB{[0,0,0]} \oplus (2) \CMTB{[1,0,0]} \oplus \CMTB{[0,2,0]} \oplus \CMTB{[0,0,2]} \oplus \CMTB{[0,1,1]} \oplus \CMTB{[2,0,0]} $
\item Level-5: $(2) \CMTred{[0,1,0]} \oplus (2) \CMTred{[0,0,1]} \oplus \CMTred{[1,0,1]} \oplus \CMTred{[1,1,0]} $
\item Level-6: $\CMTB{[0,0,0]} \oplus (2) \CMTB{[1,0,0]} \oplus \CMTB{[0,1,1]} $
\item Level-7: $ \CMTred{[0,1,0]} \oplus \CMTred{[0,0,1]} $
\item Level-8: $\CMTB{[0,0,0]}$
\end{itemize}

And this is the component decompostion results by dimensions.
\begin{itemize}
    \item Level-0: $\CMTB{\{1\}}$
    \item Level-1: $ \CMTred{\{4\}} \oplus \CMTred{\{\overline{4}\}} $
    \item Level-2: $\CMTB{\{1\}} \oplus (2) \CMTB{\{6\}} \oplus \CMTB{\{15\}} $
    \item Level-3: $(2) \CMTred{\{4\}} \oplus (2) \CMTred{\{\overline{4}\}} \oplus \CMTred{\{20\}} \oplus \CMTred{\{\overline{20}\}} $
    \item Level-4: $(3) \CMTB{\{1\}} \oplus (2) \CMTB{\{6\}} \oplus \CMTB{\{10\}} \oplus \CMTB{\{\overline{10}\}} \oplus \CMTB{\{15\}} \oplus \CMTB{ \{20'\}} $
    \item Level-5: $(2) \CMTred{\{4\}} \oplus (2) \CMTred{\{\overline{4}\}} \oplus \CMTred{\{20\}} \oplus \CMTred{\{\overline{20}\}} $
    \item Level-6: $\CMTB{\{1\}} \oplus (2) \CMTB{\{6\}} \oplus \CMTB{\{15\}} $
    \item Level-7: $ \CMTred{\{4\}} \oplus \CMTred{\{\overline{4}\}} $
    \item Level-8: $\CMTB{\{1\}}$
\end{itemize}

\subsection{6D Minimal Adinkra Diagram}

\begin{figure}[htp!]
\centering
\includegraphics[width=0.7\textwidth]{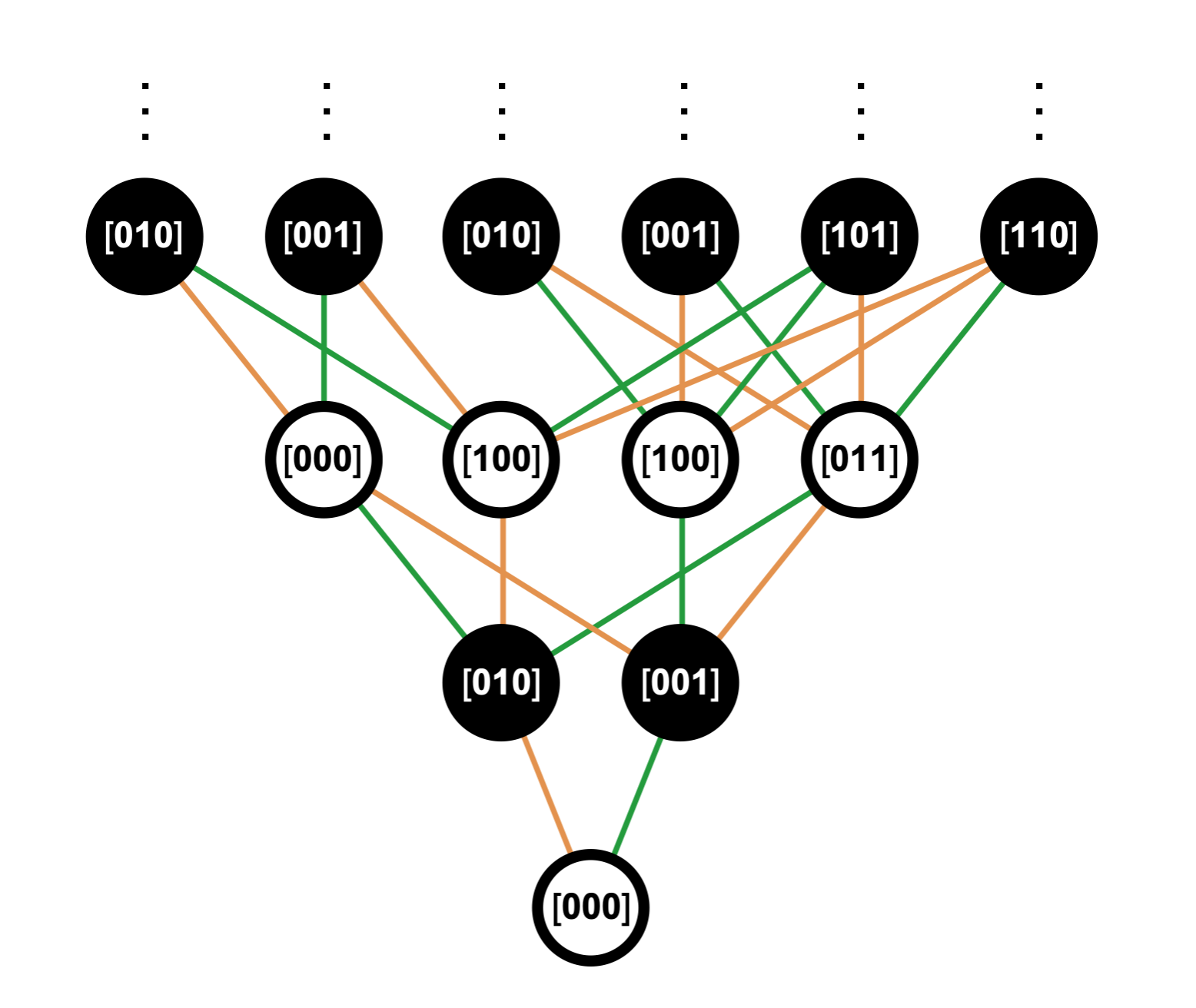}
\caption{Adynkra Diagram for 6D minimal scalar superfield}
\label{Fig:6D_Dynkin}
\end{figure}

\begin{figure}[htp!]
\centering
\includegraphics[width=0.7\textwidth]{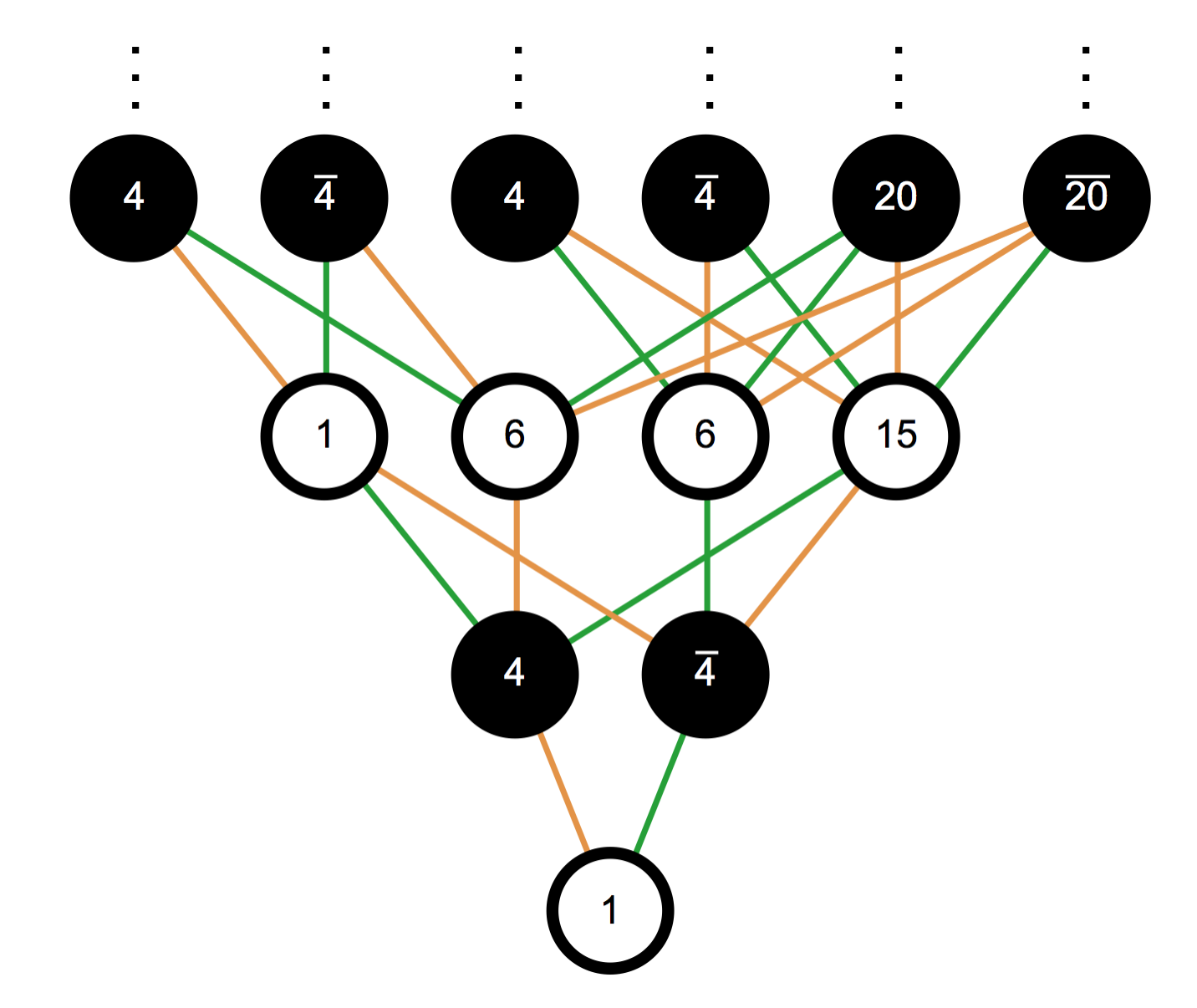}
\caption{Adinkra Diagram for 6D minimal scalar superfield}
\label{Fig:6D}
\end{figure}

The Adynkra and Adinkra diagrams for 6D minimal scalar superfield (up to Level-3) are Figures \ref{Fig:6D_Dynkin} and \ref{Fig:6D}. 
In six dimensions, we have two types of spinors having opposite chiralities. Thus we have two types of spinorial derivatives $\CMTorg{{\rm D}_{\Dot\alpha}}$ and $\CMTG{{\rm D}_{\alpha}}$. In Figure \ref{Fig:6D_Dynkin} and \ref{Fig:6D} we use orange links to denote the supersymmetry transformations carried by $\CMTorg{{\rm D}_{\Dot\alpha}}$ and green links for $\CMTG{{\rm D}_{\alpha}}$. 

\subsection{Young Tableaux Descriptions of Component Fields in 6D Minimal Scalar Superfield}
\label{sec:6D_component}

Consider the projection matrix for $\mathfrak{su}(6)\supset \mathfrak{so}(6)$~\cite{yamatsu2015},
\begin{equation}
P_{\mathfrak{su}(6)\supset \mathfrak{so}(6)} ~=~
\begin{pmatrix}
1 & 0 & 0 & 0 & 1 \\
0 & 1 & 0 & 1 & 0 \\
0 & 1 & 2 & 1 & 0 \\
\end{pmatrix}~~~.
\end{equation}
The highest weight of a specified irrep of $\mathfrak{su}(6)$ is a row vector $[p_1,p_2,p_3,p_4,p_5]$,
where $p_1$ to $p_5$ are non-negative integers. Since the $\mathfrak{su}(6)$ YT with $n$ vertical 
boxes is the conjugate of the one with $6-n$ vertical boxes, we only need to consider the $p_4=p_5=0$ case.

Starting from the weight vector $[p_1,p_2,p_3,0,0]$ in $\mathfrak{su}(6)$, we define its projected
weight vector $[p_1,p_2,p_2+2p_3]$ in $\mathfrak{so}(6)$ as the Dynkin Label of the corresponding
irreducible bosonic Young Tableau.
\begin{equation}
    [p_1,p_2,p_2+2p_3] ~=~ [p_1,p_2,p_3,0,0] \, P^T_{\mathfrak{su}(6)\supset\mathfrak{so}(6)} ~~~.
\end{equation}
Note that in 6D, duality condition needs to be considered for three-forms. Similar as the situation in 8D and 10D discussed in \ref{sec:8D_component} and \cite{nDx} respectively, the Dynkin Label $[p_1,p_2+2p_3,p_2]$ carries the same dimensionality and corresponds to the same YT shape as $[p_1,p_2,p_2+2p_3]$. 
Like in 10D, the three-form with Dynkin Label $[0,0,2]$ satisfies self-dual condition and the one with Dynkin Label $[0,2,0]$ satisfies anti-self-dual condition. Generally, the field/BYT corresponding to Dynkin Label $[p_1,p_2,p_2+2p_3]$ satisfies self-dual condition and the one with Dynkin Label $[p_1,p_2+2p_3,p_2]$ satisfies anti-self-dual condition.

Consider the congruence class of a representation with 
Dynkin Label $[a,b,c]$ in $\mathfrak{so}(6)$,
\begin{equation}
    \begin{split}
        C_{c1}(R) ~:=&~ b + c ~~ ({\rm mod} ~ 2) ~~~,\\
        C_{c2}(R) ~:=&~ 2a + b + 3c ~~ ({\rm mod} ~ 4) ~~~.
    \end{split}
\end{equation}
Based on the above equation, there are totally four congruence classes in $\mathfrak{so}(6)$,
\begin{equation}
[C_{c1}, C_{c2} ](R) ~=~
\begin{cases}
     [ 0,0 ]\\
     [ 0,2 ]\\
     [ 1,1 ]\\
     [ 1,3 ]\\
\end{cases}     
\end{equation}
$C_{c1}(R)$ actually classifies the bosonic irreps and spinorial
irreps:  $C_{c1}(R) = 0$ is bosonic and  $C_{c1}(R)=1$ is spinorial.  Consequently, a
bosonic irrep satisfies  $|c-b|
= 0 ~~ ({\rm mod} ~ 2)$.

Given an irreducible bosonic Young Tableau with $p_1$ columns of one box,
$p_2$ columns of two vertical boxes, and $p_3$ columns of three vertical boxes, the Dynkin Label of its
corresponding bosonic irrep is $[p_1,p_2,p_2+2p_3]$ (or $[p_1,p_2+2p_3,p_2]$). Reversely, given a bosonic irrep with Dynkin Label $[a,b,c]$, its corresponding irreducible 
bosonic Young Tableau is composed of $a$ columns of one box, $\min\{b,c\}$ columns of two 
vertical boxes, and $|c-b|/2$ columns of three vertical boxes. 

The simplest examples,
also the fundamental building blocks of an irreducible BYT, are given below.
\begin{equation}
\begin{gathered}
    {\CMTB{\ydiagram{1}}}_{{\rm IR}} ~\equiv~ \CMTB{[1,0,0]}  ~~~,~~~
    {\CMTB{\ydiagram{1,1}}}_{{\rm IR}} ~\equiv~ \CMTB{[0,1,1]} ~~~, \\ 
    {\CMTB{\ydiagram{1,1,1}}}_{{\rm IR},+} ~\equiv~ \CMTB{[0,0,2]} ~~~,~~~  
    {\CMTB{\ydiagram{1,1,1}}}_{{\rm IR},-} ~\equiv~ \CMTB{[0,2,0]} ~~~.
\end{gathered} \label{equ:BYTbasic_6D}
\end{equation}
For the YT satisfying self-dual condition, we use the subscript $\{{\rm IR}, +\}$ to label it; and for the YT satisfying anti-self-dual condition, we use the subscript $\{{\rm IR}, -\}$ to label it.

For spinorial irreps, the basic SYTs are defined by
\begin{equation}
\begin{split}
    \CMTred{\ytableaushort{\tinyfour}} ~\equiv&~~ \CMTred{[0,1,0]} ~~~, \\
    \CMTred{\ytableaushort{\tinyfourbar}} ~\equiv&~~ \CMTred{[0,0,1]} ~~~.
\end{split}
\label{equ:SYTbasic_6D}
\end{equation}
We could translate the Dynkin Label of any spinorial irrep to a mixed YT (which contains a BYT part and one of the basic SYTs above) with irreducible conditions by applying the same idea discussed in Chapter five in \cite{nDx}. 

Putting together the columns in (\ref{equ:BYTbasic_6D}) and (\ref{equ:SYTbasic_6D}) corresponds to adding their Dynkin Labels.

In summary, the irreducible Young Tableau descriptions of the 6D minimal scalar superfield decomposition is presented below.
\begin{equation}
\ytableausetup{boxsize=0.8em}
{\cal V} ~=~ \begin{cases}
~~{\rm {Level}}-0 \,~~~~~~~~~ \CMTB{\cdot} ~~~,  \\
%%%%%%%%%%%%%%%%%%%%%%%%%%%%%%%%%%%%%%%%%%%%%%%%
~~{\rm {Level}}-1 \,~~~~~~~~~ \,\CMTred{\ytableaushort{\tinyfour}}~\oplus~ \CMTred{\ytableaushort{\tinyfourbar}}~~~,  \\
%%%%%%%%%%%%%%%%%%%%%%%%%%%%%%%%%%%%%%%%%%%%%%%%
~~{\rm {Level}}-2 \,~~~~~~~~~ \CMTB{\cdot}~\oplus~
(2)\,{\CMTB{\ydiagram{1}}}_{\rm IR}~\oplus~
{\CMTB{\ydiagram{1,1}}}_{\rm IR} ~~~,  \\[10pt]
%%%%%%%%%%%%%%%%%%%%%%%%%%%%%%%%%%%%%%%%%%%%%%%%
~~{\rm {Level}}-3 \,~~~~~~~~~
(2)\,\CMTred{\ytableaushort{\tinyfour}}~\oplus~
(2)\,\CMTred{\ytableaushort{\tinyfourbar}}~\oplus~
{\CMTB{\ydiagram{1}}\CMTred{\ytableaushort{\tinyfour}}}_{\rm IR}~\oplus~
{\CMTB{\ydiagram{1}}\CMTred{\ytableaushort{\tinyfourbar}}}_{\rm IR}  ~~~,  \\[10pt]
%%%%%%%%%%%%%%%%%%%%%%%%%%%%%%%%%%%%%%%%%%%%%%%%
~~{\rm {Level}}-4 \,~~~~~~~~~ (3)\,\CMTB{\cdot}~\oplus~
(2)\,{\CMTB{\ydiagram{1}}}_{\rm IR}~\oplus~
{\CMTB{\ydiagram{1,1,1}}}_{{\rm IR},-}~\oplus~
{\CMTB{\ydiagram{1,1,1}}}_{{\rm IR},+}~\oplus~
{\CMTB{\ydiagram{1,1}}}_{\rm IR} ~\oplus~{\CMTB{\ydiagram{2}}}_{\rm IR}
 ~~~,  \\[10pt]
 %%%%%%%%%%%%%%%%%%%%%%%%%%%%%%%%%%%%%%%%%%%%%%%%
~~{\rm {Level}}-5 \,~~~~~~~~~
(2)\,\CMTred{\ytableaushort{\tinyfour}}~\oplus~
(2)\,\CMTred{\ytableaushort{\tinyfourbar}}~\oplus~
{\CMTB{\ydiagram{1}}\CMTred{\ytableaushort{\tinyfour}}}_{\rm IR}~\oplus~
{\CMTB{\ydiagram{1}}\CMTred{\ytableaushort{\tinyfourbar}}}_{\rm IR}
~~~,  \\[10pt]
%%%%%%%%%%%%%%%%%%%%%%%%%%%%%%%%%%%%%%%%%%%%%%%%
~~{\rm {Level}}-6 \,~~~~~~~~~ \CMTB{\cdot}~\oplus~
(2)\,{\CMTB{\ydiagram{1}}}_{\rm IR}~\oplus~
{\CMTB{\ydiagram{1,1}}}_{\rm IR}
~~~,  \\
%%%%%%%%%%%%%%%%%%%%%%%%%%%%%%%%%%%%%%%%%%%%%%%%
~~{\rm {Level}}-7 \,~~~~~~~~~
\CMTred{\ytableaushort{\tinyfour}}~\oplus~ \CMTred{\ytableaushort{\tinyfourbar}}
~~~,  \\
%%%%%%%%%%%%%%%%%%%%%%%%%%%%%%%%%%%%%%%%%%%%%%%%
~~{\rm {Level}}-8 \,~~~~~~~~~~ \CMTB{\cdot} 
~~~.  
\end{cases}
\label{equ:V_6D}
\ytableausetup{boxsize=1.2em}
\end{equation}

\subsection{Index Structures and Irreducible Conditions of Component Fields in 6D Minimal Scalar Superfield}\label{sec:6D_Index}

In this section, we will translate the irreducible bosonic and mixed Young Tableaux into field variables. 
We follow the same $\{\}$-indices notation as well as ``$|$'' to separate indices in YT with different heights and ``,'' to separate
indices in YT with the same heights. 

The vector index $\un{a}$ runs from 0 to 5. 
The $\{\}$-indices, irreducible bosonic Young
Tableaux, and Dynkin Labels are equivalent and have the one-to-one correspondence.

The general expression is as below in Equations (\ref{equ:index-notation1_6D}) and (\ref{equ:index-notation2_6D}),
\begin{equation}
\begin{split}
    & \{ \un{a}_1,\dots,\un{a}_p ~|~ \un{b}_1\un{c}_1,\dots,\un{b}_q \un{c}_q ~|~  \un{d}_1\un{e}_1\un{f}_1,\dots,\un{d}_r\un{e}_r\un{f}_r\}^{\pm}  \\[10pt]
    & ~{ \CMTB{{\ytableaushort{\aone}} } }~~~~~{ \CMTB{{\ytableaushort{\ap}} } }
    ~~~~{ \CMTB{{\ytableaushort{\bone,\cone}} } }~~~~~~~{ \CMTB{{\ytableaushort{\bq,\cunq}} } }
    ~~~~~~{ \CMTB{{\ytableaushort{\done,\eone,\fone}} } }~~~~~~~~~~{ \CMTB{{\ytableaushort{\dr,\er,\fr}} } }
    \\[10pt]
    & ~~~~ \CMTB{[p,0,0]} ~~~~~~~~ \CMTB{[0,q,q]} ~~~~~~~ \begin{matrix}\CMTB{[0,0,2r]} ~~ \text{with superscript } + \\
    \CMTB{[0,2r,0]} ~~ \text{with superscript } -
    \end{matrix}  
\end{split}
\label{equ:index-notation1_6D}
\end{equation}
where above we have ``disassembled'' the YT to show how each column is affiliated with each type of
subscript structure. Below we have assembled all the column into a proper YT.
\begin{equation}
\begin{split}
    &~ { \CMTB{{\ytableaushort{\done\dots\dr\bone\dots\bq\aone\dots\ap,\eone\dots\er\cone\dots\cunq,\fone\dots\fr}} } }_{{\rm IR},\pm}\\[10pt]
    & \begin{matrix}\CMTB{[p,q,q+2r]}~~\text{with subscript }+\\
    \CMTB{[p,q+2r,q]}~~\text{with subscript }-\end{matrix}
\end{split}
\label{equ:index-notation2_6D}
\end{equation}

As one moves from the YT's shown in Equation~(\ref{equ:index-notation1_6D}) to
Equation~(\ref{equ:index-notation2_6D}), it is clear the number of vertical boxes is tabulating the number of
1-forms, 2-forms, and 3-forms in the YT's.  These are the entries between
the vertical $|$ bars. These precisely correspond to the integers $p$, $q$, and $r$ appeared in Dynkin Labels.  
An example of the correspondence between the subscript conventions, the affiliated YT, and Dynkin Label is shown in (\ref{equ:index-notation_ex1_6D}). 
\begin{equation}
    \{{\un a}_2 , {\un a}_3| {\un a}_1  {\un b}_1   \} ~~\equiv~~
    { \CMTB{{\ytableaushort{\aone \atwo \athree,\bone}} } }_{{\rm IR}}
    ~~\equiv~~ \CMTB{[2,1,1]} ~~~.
    \label{equ:index-notation_ex1_6D}
\end{equation}

In 6D, we have two types of spinor indices corresponding to $\CMTred{\ytableaushort{\tinyfour}}$ and $\CMTred{\ytableaushort{\tinyfourbar}}$ respectively.
We define the field $\Psi^{\alpha}$ or $\Psi_{\Dot \alpha}$ correpsonds to $\CMTred{\ytableaushort{\tinyfour}}$ and the field $\Psi_{\alpha}$ or $\Psi^{\Dot \alpha}$ correpsonds to $\CMTred{\ytableaushort{\tinyfourbar}}$
The spinor indices $\alpha$ and $\Dot{\alpha}$ run from 1 to 4. 

The index structures as well as irreducible conditions of all bosonic and fermionic fields are identified below along 
with the level at which the fields occur in the adinkra of the superfield. 

\begin{itemize}
    \item Level-0: $\Phi(x)$~~~,
    \item Level-1: $\Psi^{\a}$~~~,~~~$\Psi_{\a}$~~~,
    \item Level-2: $\Phi(x)$~~~,~~~(2)\,$\Phi_{\{\aone\}}(x)$~~~,~~~$\Phi_{\{\aone\bone\}}(x)$~~~,
    \item Level-3: (2)\,$\Psi^{\a}$~~~,~~~(2)\,$\Psi_{\a}$~~~, ~~~
    $\Psi_{\{\aone\}\a}:~ (\s^{\aone})^{\a\b}\Psi_{\{\aone\}\a}~=~0$~~~,
    ~~~$\Psi_{\{\aone\}}{}^{\a}:~ (\s^{\aone})_{\a\b}\Psi_{\{\aone\}}{}^{\a}~=~0$~~~,
    \item Level-4: (3)\,$\Phi(x)$~~~,~~~(2)\,$\Phi_{\{\aone\}}(x)$~~~,~~~
    $\Phi_{\{\aone\bone\cone\}^{-}}(x)$~~~,~~~$\Phi_{\{\aone\bone\cone\}^{+}}(x)$~~~,\\[10pt]
    $\Phi_{\{\aone\bone\}}(x)$~~~,~~~
    $\Phi_{\{\aone,\un{a}_2\}}(x): \eta^{\aone\un{a}_2}\Phi_{\{\aone,\un{a}_2\}}(x)~=~0$~~~.
\end{itemize}
Level-5 to Level-8 have exactly the same expressions as Level-3 to Level-0.

\subsection{$(1,0)$ Multiplet Decompositions}
Similarly as in 8D, in 6D spinors also split into two pieces: $\CMTred{[0,1,0]}~\oplus~\CMTred{[0,0,1]}$ ($\CMTred{\{4\}}~\oplus~\CMTred{\{\overline{4}\}}$). Therefore, we can study the subset of decomposition results which generated solely by $\CMTred{[0,1,0]}$ or $\CMTred{[0,0,1]}$. The results obtained by only considering $\CMTred{[0,1,0]} ~ (\CMTred{\{4\}})$ piece are as below, which are called $(1,0)$ multiplet decomposition. 
\begin{itemize}
    \item Level-0: $\CMTB{[0,0,0]}~(\CMTB{\{1\}})$
    \item Level-1: $\CMTred{[0,1,0]}~(\CMTred{\{4\}})$
    \item Level-2: $\CMTB{[1,0,0]}~(\CMTB{\{6\}})$
    \item Level-3: $\CMTred{[0,0,1]}~(\CMTred{\{\overline{4}\}})$
    \item Level-4: $\CMTB{[0,0,0]}~(\CMTB{\{1\}})$
\end{itemize}

Starting from the $(1,0)$ multiplet decomposition, we can reproduce the scalar superfield decomposition following the same procedures discussed in 8D. Basically we can label spinors as $\CMTorg{\theta^{\alpha}}$ and $\CMTgrn{\theta^{\Dot{\alpha}}}$ corresponding to $\CMTred{[0,1,0]}$ and $\CMTred{[0,0,1]}$ respectively. Then expand the scalar superfield only with respect to $\CMTgrn{\theta^{\Dot{\alpha}}}$ first where the accompanying component fields are actually $(1,0)$ superfields. For example, Level-2 in the scalar superfield decomposition is nothing but $\CMTred{\{4\}}\wedge\CMTred{\{4\}}\oplus\CMTred{\{\overline{4}\}}\wedge\CMTred{\{\overline{4}\}}\oplus\CMTred{\{4\}}\otimes\CMTred{\{\overline{4}\}}~=~\CMTB{\{1\}} \oplus (2) \CMTB{\{6\}} \oplus \CMTB{\{15\}}$.

\begin{figure}[htp!]
\centering
\begin{minipage}{0.46\textwidth}
    \centering
    \includegraphics[width=0.4\textwidth]{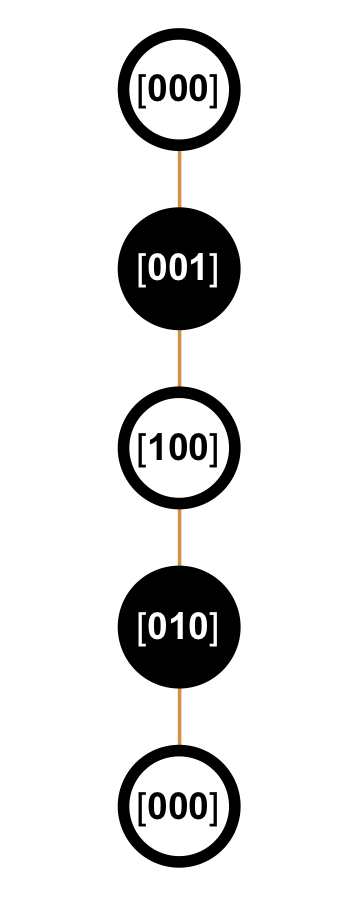}
    \caption{Adynkra Diagram for 6D $(1,0)$ Multiplet}
    \label{Fig:6Dchiral_Dynkin}
\end{minipage}
\begin{minipage}{0.46\textwidth}
    \centering
    \includegraphics[width=0.4\textwidth]{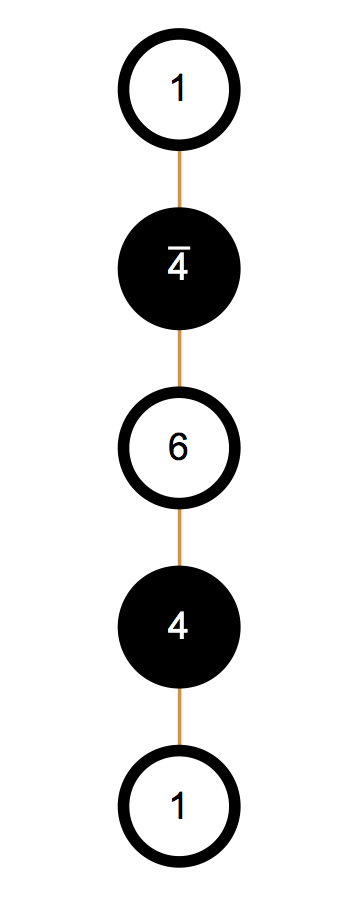}
    \caption{Adinkra Diagram for 6D $(1,0)$ Multiplet}
   \label{Fig:6Dchiral}
\end{minipage}
\end{figure}

Last but not least, we can draw the adynkra and adinkra diagrams corresponding to the $(1,0)$ multiplet, which are Figures \ref{Fig:6Dchiral_Dynkin} and \ref{Fig:6Dchiral}. 

\newpage
\section{5D Minimal Scalar Superfield Decomposition}

% The Lorentz group we are working on in 5D is Usp(4) $\cong$ SO(5). 

\subsection{Component Decompostion Results}

The 5D minimal scalar superfield component decompostion results by Dynkin Labels are given below.
\begin{itemize}
\item Level-0: $\CMTB{[0,0]}$
\item Level-1: $(2)\CMTred{[0,1]} $
\item Level-2: $(3)\CMTB{[0,0]}\oplus(3)\CMTB{[1,0]}\oplus\CMTB{[0,2]} $
\item Level-3: $(6)\CMTred{[0,1]}\oplus(2)\CMTred{[1,1]} $
\item Level-4: $(6)\CMTB{[0,0]}\oplus(4)\CMTB{[1,0]}\oplus(3)\CMTB{[0,2]}\oplus\CMTB{[2,0]} $
\item Level-5: $(6)\CMTred{[0,1]}\oplus(2)\CMTred{[1,1]} $
\item Level-6: $(3)\CMTB{[0,0]}\oplus(3)\CMTB{[1,0]}\oplus\CMTB{[0,2]} $
\item Level-7: $(2)\CMTred{[0,1]} $
\item Level-8: $\CMTB{[0,0]}$
\end{itemize}

The corresponding component decompostion results by dimensions are as follows.
\begin{itemize}
    \item Level-0: $\CMTB{\{1\}}$
    \item Level-1: $(2)\CMTred{\{4\}} $
    \item Level-2: $(3)\CMTB{\{1\}}\oplus(3)\CMTB{\{5\}}\oplus\CMTB{\{10\}} $
    \item Level-3: $(6)\CMTred{\{4\}}\oplus(2)\CMTred{\{16\}} $
    \item Level-4: $(6)\CMTB{\{1\}}\oplus(4)\CMTB{\{5\}}\oplus(3)\CMTB{\{10\}}\oplus \CMTB{ \{14\}} $
    \item Level-5: $(6)\CMTred{\{4\}}\oplus(2)\CMTred{\{16\}} $
    \item Level-6: $(3)\CMTB{\{1\}}\oplus(3)\CMTB{\{5\}}\oplus\CMTB{\{10\}} $
    \item Level-7: $(2)\CMTred{\{4\}} $
    \item Level-8: $\CMTB{\{1\}}$
\end{itemize}

\newpage
\subsection{5D Minimal Adinkra Diagram}

\begin{figure}[htp!]
\centering
\includegraphics[width=0.8\textwidth]{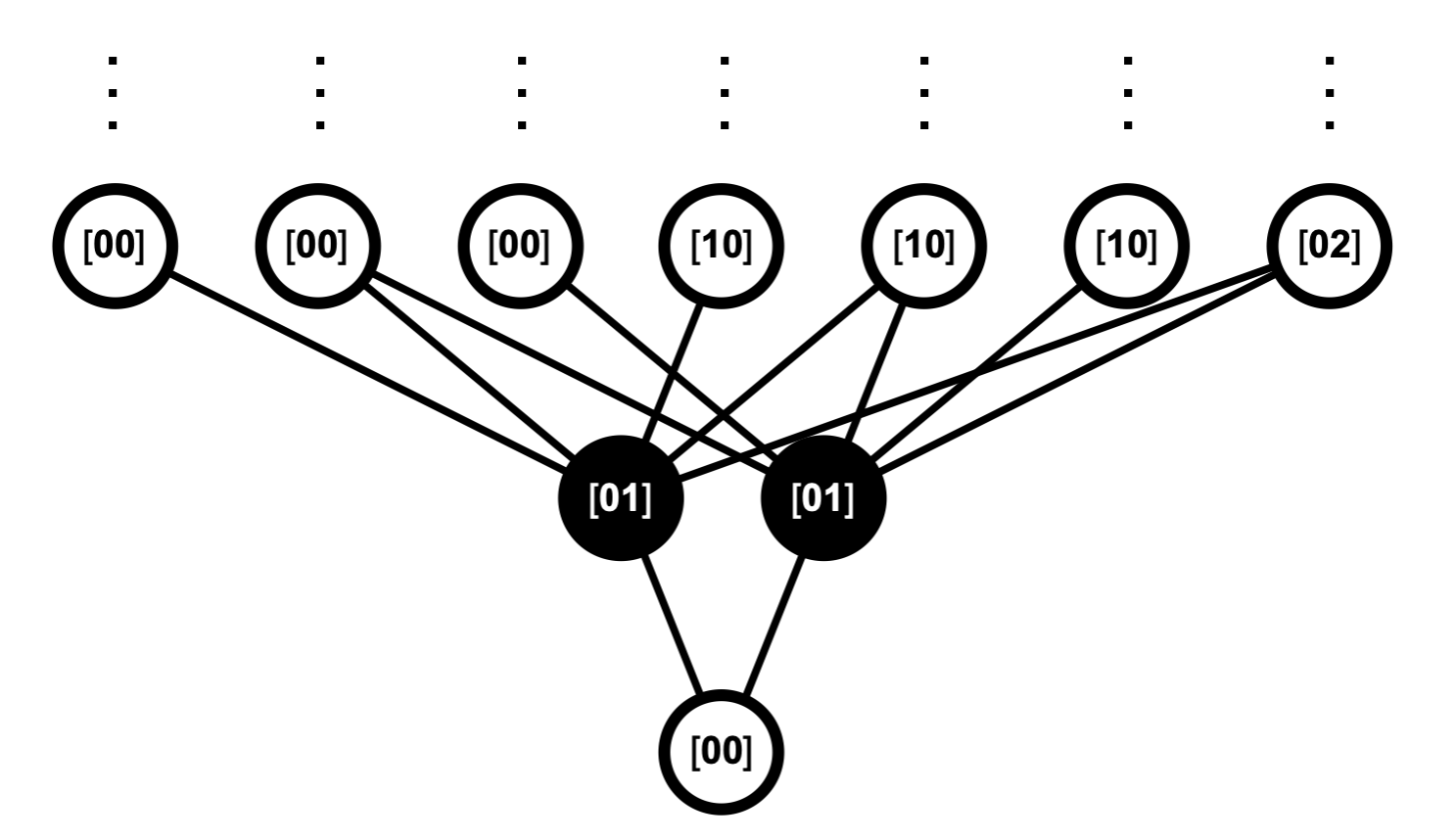}
\caption{Adynkra Diagram for 5D minimal scalar superfield}
\label{Fig:5D_Dynkin}
\end{figure}

\begin{figure}[htp!]
\centering
\includegraphics[width=0.8\textwidth]{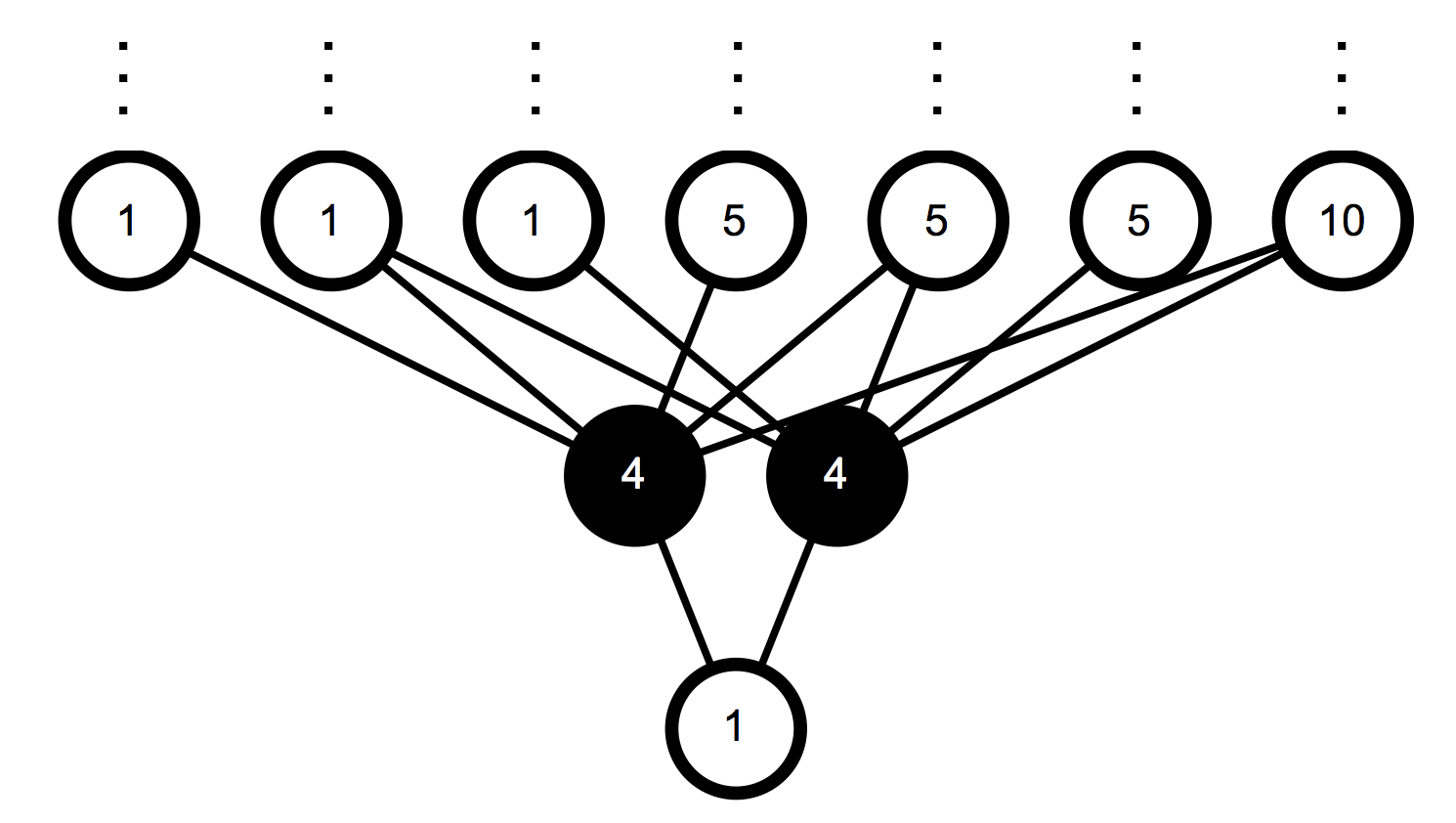}
\caption{Adinkra Diagram for 5D minimal scalar superfield}
\label{Fig:5D}
\end{figure}

The Adynkra and Adinkra diagrams for 5D minimal scalar superfield (up to Level-2) are Figures \ref{Fig:5D_Dynkin} and \ref{Fig:5D}. 

\subsection{Young Tableaux Descriptions of Component Fields in 5D Minimal Scalar Superfield}

Consider the projection matrix for $\mathfrak{su}(5)\supset \mathfrak{so}(5)$~\cite{yamatsu2015},
\begin{equation}
P_{\mathfrak{su}(5)\supset \mathfrak{so}(5)} ~=~
\begin{pmatrix}
1 & 0 & 0 & 1 \\
0 & 2 & 2 & 0 \\
\end{pmatrix}~~~.
\end{equation}
The highest weight of a specified irrep of $\mathfrak{su}(5)$ is a row vector $[p_1,p_2,p_3,p_4]$,
where $p_1$ to $p_4$ are non-negative integers. Since the $\mathfrak{su}(5)$ YT with $n$ vertical 
boxes is the conjugate of the one with $5-n$ vertical boxes, we need only consider the $p_3=p_4=0$ case.

Starting from the weight vector $[p_1,p_2,0,0]$ in $\mathfrak{su}(5)$, we define its projected
weight vector $[p_1,2p_2]$ in $\mathfrak{so}(5)$ as the Dynkin Label of the corresponding
irreducible bosonic Young Tableau.
\begin{equation}
    [p_1,2p_2] ~=~ [p_1,p_2,0,0] \, P^T_{\mathfrak{su}(5)\supset\mathfrak{so}(5)} ~~~.
\end{equation}
The congruence class of a representation with 
Dynkin Label $[a,b]$ in $\mathfrak{so}(5)$ is
\begin{equation}
    \begin{split}
        C_{c}(R) ~:=&~ b  ~~ ({\rm mod} ~ 2) ~~~,
    \end{split}
\end{equation}
which classifies the bosonic irreps and spinorial
irreps:  $C_{c}(R) = 0$ is bosonic and  $C_{c}(R)=1$ is spinorial.

Given an irreducible bosonic Young Tableau with $p_1$ columns of one box and
$p_2$ columns of two vertical boxes, the Dynkin Label of its
corresponding bosonic irrep is $[p_1,2p_2]$. Reversely, given a bosonic irrep with Dynkin Label $[a,b]$, its corresponding irreducible 
bosonic Young Tableau is composed of $a$ columns of one box and $b/2$ columns of two 
vertical boxes. 

The simplest examples,
also the fundamental building blocks of an irreducible BYT, are given below.
\begin{equation}
\begin{gathered}
    {\CMTB{\ydiagram{1}}}_{{\rm IR}} ~\equiv~ \CMTB{[1,0]}  ~~~,~~~
    {\CMTB{\ydiagram{1,1}}}_{{\rm IR}} ~\equiv~ \CMTB{[0,2]} ~~~.
\end{gathered} \label{equ:BYTbasic_5D}
\end{equation}

For spinorial irreps, the basic SYT is given by
\begin{equation}
    \CMTred{\ytableaushort{\tinyfour}} ~\equiv~~ \CMTred{[0,1]} ~~~.
\label{equ:SYTbasic_5D}
\end{equation}
We could translate the Dynkin Label of any spinorial irrep to a mixed YT (which contains a BYT part and a basic SYT above) with irreducible conditions by applying the same idea discussed in Chapter five in \cite{nDx}. 

Putting together the columns in (\ref{equ:BYTbasic_5D}) and (\ref{equ:SYTbasic_5D}) corresponds to adding their Dynkin Labels.

In summary, the irreducible Young Tableau descriptions of the 5D minimal scalar superfield decomposition is presented below.
\begin{equation}
\ytableausetup{boxsize=0.8em}
{\cal V} ~=~ \begin{cases}
~~{\rm {Level}}-0 \,~~~~~~~~~ \CMTB{\cdot} ~~~,  \\
%%%%%%%%%%%%%%%%%%%%%%%%%%%%%%%%%%%%%%%%%%%%%%%%
~~{\rm {Level}}-1 \,~~~~~~~~~ (2)\,\CMTred{\ytableaushort{\tinyfour}}~~~,  \\
%%%%%%%%%%%%%%%%%%%%%%%%%%%%%%%%%%%%%%%%%%%%%%%%
~~{\rm {Level}}-2 \,~~~~~~~~~ (3)\,\CMTB{\cdot}~\oplus~
(3)\,{\CMTB{\ydiagram{1}}}_{\rm IR}~\oplus~
{\CMTB{\ydiagram{1,1}}}_{\rm IR} ~~~,  \\[10pt]
%%%%%%%%%%%%%%%%%%%%%%%%%%%%%%%%%%%%%%%%%%%%%%%%
~~{\rm {Level}}-3 \,~~~~~~~~~
(6)\,\CMTred{\ytableaushort{\tinyfour}}~\oplus~
(2)\,{\CMTB{\ydiagram{1}}\CMTred{\ytableaushort{\tinyfour}}}_{\rm IR}  ~~~,  \\[10pt]
%%%%%%%%%%%%%%%%%%%%%%%%%%%%%%%%%%%%%%%%%%%%%%%%
~~{\rm {Level}}-4 \,~~~~~~~~~ (6)\,\CMTB{\cdot}~\oplus~
(4)\,{\CMTB{\ydiagram{1}}}_{\rm IR}~\oplus~
(3)\,{\CMTB{\ydiagram{1,1}}}_{\rm IR} ~\oplus~{\CMTB{\ydiagram{2}}}_{\rm IR}
 ~~~,  \\[10pt]
 %%%%%%%%%%%%%%%%%%%%%%%%%%%%%%%%%%%%%%%%%%%%%%%%
~~{\rm {Level}}-5 \,~~~~~~~~~
(6)\,\CMTred{\ytableaushort{\tinyfour}}~\oplus~
(2)\,{\CMTB{\ydiagram{1}}\CMTred{\ytableaushort{\tinyfour}}}_{\rm IR}
~~~,  \\[10pt]
%%%%%%%%%%%%%%%%%%%%%%%%%%%%%%%%%%%%%%%%%%%%%%%%
~~{\rm {Level}}-6 \,~~~~~~~~~ (3)\,\CMTB{\cdot}~\oplus~
(3)\,{\CMTB{\ydiagram{1}}}_{\rm IR}~\oplus~
{\CMTB{\ydiagram{1,1}}}_{\rm IR}
~~~,  \\
%%%%%%%%%%%%%%%%%%%%%%%%%%%%%%%%%%%%%%%%%%%%%%%%
~~{\rm {Level}}-7 \,~~~~~~~~~
(2)\,\CMTred{\ytableaushort{\tinyfour}}
~~~,  \\
%%%%%%%%%%%%%%%%%%%%%%%%%%%%%%%%%%%%%%%%%%%%%%%%
~~{\rm {Level}}-8 \,~~~~~~~~~~ \CMTB{\cdot} 
~~~.  
\end{cases}
\label{equ:V_5D}
\ytableausetup{boxsize=1.2em}
\end{equation}

\subsection{Index Structures and Irreducible Conditions of Component Fields in 5D Minimal Scalar Superfield}\label{sec:5D_Index}

In this section, we will translate the irreducible bosonic and mixed Young Tableaux into field representation. 
We follow the same $\{\}$-indices notation as well as ``$|$'' to separate indices in YT with different heights and ``,'' to separate
indices in YT with the same heights. 

The vector index $\un{a}$ runs from 0 to 4. 
The $\{\}$-indices, irreducible bosonic Young
Tableaux, and Dynkin Labels are equivalent and have the one-to-one correspondence.

The general expression is as below in Equations (\ref{equ:index-notation1_5D}) and (\ref{equ:index-notation2_5D}),
\begin{equation}
\begin{split}
    & \{ \un{a}_1,\dots,\un{a}_p ~|~ \un{b}_1\un{c}_1,\dots,\un{b}_q \un{c}_q \}  \\[10pt]
    & ~ { \CMTB{{\ytableaushort{\aone}} } }~~~~~{ \CMTB{{\ytableaushort{\ap}} } }
    ~~~~{ \CMTB{{\ytableaushort{\bone,\cone}} } }~~~~~~~{ \CMTB{{\ytableaushort{\bq,\cunq}} } }
    \\[10pt]
    & \,~~~~ \CMTB{[p,0]} \,~~~~~~~~~~ \CMTB{[0,2q]} 
\end{split}
\label{equ:index-notation1_5D}
\end{equation}
where above we have ``disassembled'' the YT to show how each column is affiliated with each type of
subscript structure. Below we have assembled all the column into a proper YT.
\begin{equation}
\begin{split}
    & { \CMTB{{\ytableaushort{\bone\dots\bq\aone\dots\ap,\cone\dots\cunq}} } }_{\rm IR}\\[10pt]
    &~~~~~~~~ \CMTB{[p,2q]}
\end{split}
\label{equ:index-notation2_5D}
\end{equation}

As one moves from the YT's shown in Equation~(\ref{equ:index-notation1_5D}) to
Equation~(\ref{equ:index-notation2_5D}), it is clear the number of vertical boxes is tabulating the number of
1-forms and 2-forms in the YT's.  These are the entries between
the vertical $|$ bars. These precisely correspond to the integers $p$ and $q$ appeared in Dynkin Labels.  
An example of the correspondence between the subscript conventions,
the affiliated YT, and Dynkin Label is shown in (\ref{equ:index-notation_ex1_5D}). 
\begin{equation}
    \{{\un a}_2 , {\un a}_3| {\un a}_1  {\un b}_1    \} ~~\equiv~~
    { \CMTB{{\ytableaushort{\aone \atwo \athree,\bone}} } }_{{\rm IR}}
    ~~\equiv~~ \CMTB{[2,2]} ~~~.
    \label{equ:index-notation_ex1_5D}
\end{equation}

The index structures as well as irreducible conditions of all bosonic and fermionic fields are identified below along 
with the level at which the fields occur in the adinkra of the superfield. The spinor index $\alpha$ runs from 1 to 4. 

\begin{itemize}
    \item Level-0: $\Phi(x)$~~~,
    \item Level-1: (2)\,$\Psi_{\a}$~~~,
    \item Level-2: (3)\,$\Phi(x)$~~~,~~~(3)\,$\Phi_{\{\aone\}}(x)$~~~,~~~$\Phi_{\{\aone\bone\}}(x)$~~~,
    \item Level-3: (6)\,$\Psi_{\a}$~~~,~~~ 
    (2)\,$\Psi_{\{\aone\}\a}:~ (\g^{\aone})^{\a\b}\Psi_{\{\aone\}\a}~=~0$~~~,
    \item Level-4: (6)\,$\Phi(x)$~~~,~~~(4)\,$\Phi_{\{\aone\}}(x)$~~~,~~~
    (3)\,$\Phi_{\{\aone\bone\}}(x)$~~~,\\[10pt]
    $\Phi_{\{\aone,\un{a}_2\}}(x): \eta^{\aone\un{a}_2}\Phi_{\{\aone,\un{a}_2\}}(x)~=~0$~~~.
\end{itemize}
Level-5 to Level-8 have exactly the same expressions as Level-3 to Level-0.

\subsection{$(1,0)$ Multiplet Decompositions}

In five spacetime dimensions, the story is pretty similar as in seven dimensions since in both cases SU(2)-Majorana condition plays an important role. Consequently we have two copies of $\CMTred{[0,1]}$. Start from one copy of $\CMTred{[0,1]}$ spinor, we can also construct a supermultiplet called $(1,0)$ multiplet which is the subset of the one constructed from the scalar superfield. Apply the Plethysm function and results are as below. Using branching rules for $\mathfrak{su(4)}\supset\mathfrak{so}(5)$ gives the same results. The projection matrix is presented in Equation (\ref{equ:Psu4toso5}). 

\begin{itemize}
    \item Level-0: $\CMTB{[0,0]}~\CMTB{\{1\}}$)
    \item Level-1: $\CMTred{[0,1]}~(\CMTred{\{4\}})$
    \item Level-2: $\CMTB{[1,0]}~(\CMTB{\{5\}})~\oplus~ \CMTB{[0,0]}(\CMTB{\{1\}})$
    \item Level-3: $\CMTred{[0,1]}~(\CMTred{\{4\}})$
    \item Level-4: $\CMTB{[0,0]}~(\CMTB{\{1\}}$)
\end{itemize}

Starting from the above decomposition and we can reproduce the scalar superfield decomposition using the similar idea as in 7D. Basically we can label spinors as $\theta^{\alpha}$ and $\CMTgrn{\theta^{\alpha}}$ both corresponding to the same irrep $\CMTred{[0,1]}$. Then expand the scalar superfield only with respect to $\CMTgrn{\theta^{\alpha}}$ first. For example, Level-2 in the scalar superfield decomposition is nothing but $(2)\,\CMTred{\{4\}}\wedge\CMTred{\{4\}}~\oplus~\CMTred{\{4\}}\otimes\CMTred{\{4\}}~=~(3)\CMTB{\{1\}}\oplus(3)\CMTB{\{5\}}\oplus\CMTB{\{10\}}$.

\begin{figure}[htp!]
\centering
\begin{minipage}{0.46\textwidth}
    \centering
    \includegraphics[width=0.3\textwidth]{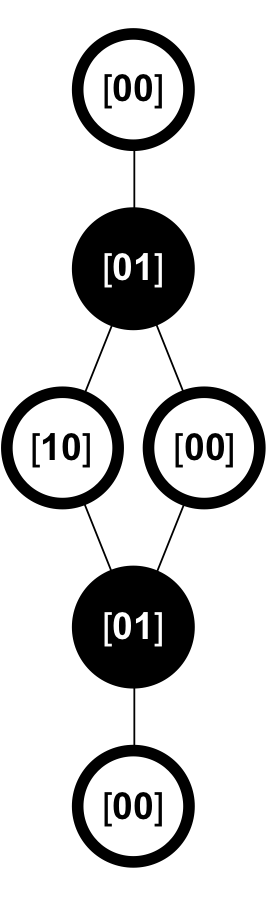}
    \caption{Adynkra Diagram for 5D $(1,0)$ Multiplet}
    \label{Fig:5Dchiral_Dynkin}
\end{minipage}
\begin{minipage}{0.46\textwidth}
    \centering
    \includegraphics[width=0.3\textwidth]{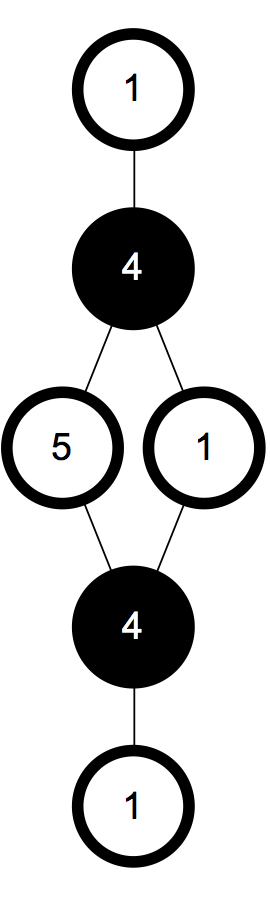}
    \caption{Adinkra Diagram for 5D $(1,0)$ Multiplet}
   \label{Fig:5Dchiral}
\end{minipage}
\end{figure}

Last but not least, we can draw the adynkra and adinkra diagrams corresponding to the $(1,0)$ multiplet, which are Figures \ref{Fig:5Dchiral_Dynkin} and \ref{Fig:5Dchiral}. 

\newpage

\newpage
\section{4D Minimal Scalar Superfield Decomposition}

% The Lorentz group we are working on in 4D is SU(2) $\times$ SU(2) $\cong$ SO(4). 

\subsection{Component Decompostion Results}

The 4D minimal scalar superfield component decompostion results by Dynkin Labels is given as follows.
\begin{itemize}
\item Level-0: $\CMTB{[0,0]}$
\item Level-1: $\CMTred{[1,0]}\oplus\CMTred{[0,1]}$
\item Level-2: $(2)\CMTB{[0,0]}\oplus\CMTB{[1,1]}$
\item Level-3: $\CMTred{[1,0]}\oplus\CMTred{[0,1]}$
\item Level-4: $\CMTB{[0,0]}$
\end{itemize}

Here we also list the component decompostion results by dimensions.
\begin{itemize}
\item Level-0: $\CMTB{\{1\}}$
\item Level-1: $\CMTred{\{2\}}\oplus\CMTred{\{\overline{2}\}}$
\item Level-2: $(2)\CMTB{\{1\}}\oplus\CMTB{\{4\}}$
\item Level-3: $\CMTred{\{2\}}\oplus\CMTred{\{\overline{2}\}}$
\item Level-4: $\CMTB{\{1\}}$
\end{itemize}

In Table \ref{Tab:Clifford}, we see that 4D is the only dimension there are two choices of spinor conventions, 
2-component Weyl spinors or 4-component Majorana spinors. 
Here $\CMTred{\{2\}}\oplus\CMTred{\{\overline{2}\}}$ corresponds to two-component Weyl spinors $\psi_{\a}$ and $\overline{\psi}_{\Dot{\a}}$, 
where $\a, \Dot{\a} = 1, 2$ and the superspace takes the form $(x^{\un{a}},\theta^{\a},\overline{\theta}^{\Dot{\a}})$.
However, we could also adopt the Majorana spinor convention, 
where we put together the two 2-component Weyl spinors into one 4-component Majorana spinor.
In that case, we have $\psi_{\a}$ that corresponds to $\CMTred{\{4\}} = \CMTred{\{2\}}\oplus\CMTred{\{\overline{2}\}}$,
and our superspace looks like $(x^{\un{a}},\theta^{\a})$ with $\a = 1, \dots, 4$. 
If we choose the Majorana notation, the 4D results here agree with what appeared in our previous work \cite{CNT10d}. In the rest of this chapter, we will stick to the Majorana spinor convention.

\newpage
\subsection{4D Minimal Adinkra Diagram}

\begin{figure}[htp!]
\centering
\begin{minipage}{0.46\textwidth}
    \centering
    \includegraphics[width=0.5\textwidth]{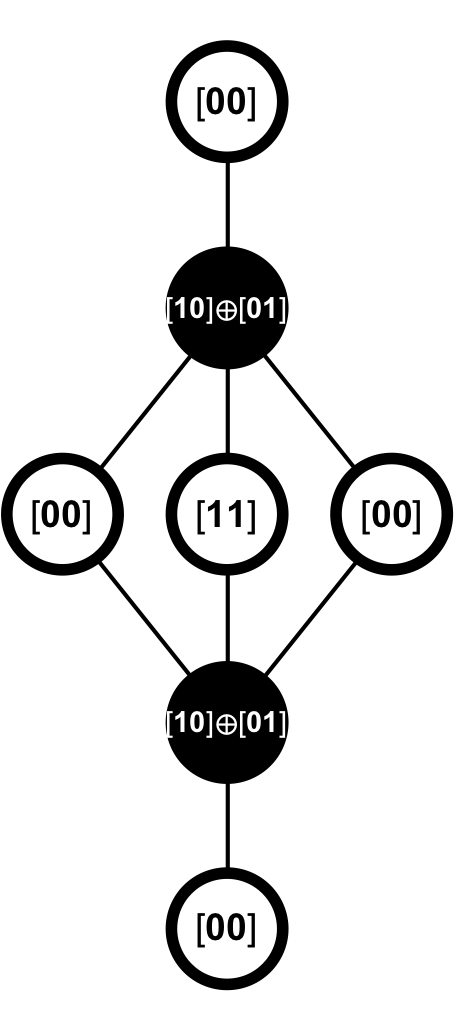}
    \caption{Adynkra Diagram for 4D minimal scalar superfield}
    \label{Fig:4D_Dynkin}
\end{minipage}
\begin{minipage}{0.46\textwidth}
    \centering
    \includegraphics[width=0.5\textwidth]{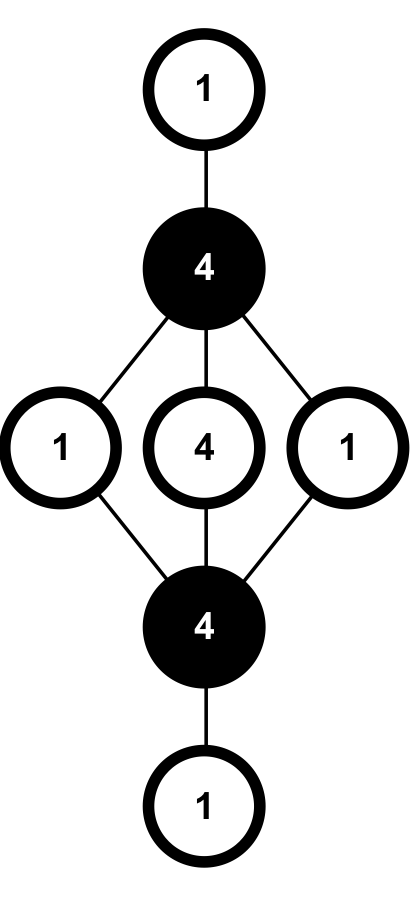}
    \caption{Adinkra Diagram for 4D minimal scalar superfield}
   \label{Fig:4D}
\end{minipage}
\end{figure}

The Adynkra and Adinkra diagrams for 4D minimal scalar superfield are Figures \ref{Fig:4D_Dynkin} and \ref{Fig:4D}. 

\subsection{Young Tableaux Descriptions of Component Fields in 4D Minimal Scalar Superfield}
\label{sec:4D_components}

Consider the projection matrix for $\mathfrak{su}(4)\supset \mathfrak{so}(4)$~\cite{yamatsu2015},
\begin{equation}
P_{\mathfrak{su}(4)\supset \mathfrak{so}(4)} ~=~
\begin{pmatrix}
1 & 0 & 1 \\
1 & 2 & 1 \\
\end{pmatrix}~~~.
\label{eqn:Psu4so4vector}
\end{equation}
The highest weight of a specified irrep of $\mathfrak{su}(4)$ is a row vector $[p_1,p_2,p_3]$,
where $p_1$ to $p_3$ are non-negative integers. Since the $\mathfrak{su}(4)$ YT with $n$ vertical 
boxes is the conjugate of the one with $4-n$ vertical boxes, we only need to consider the $p_3=0$ case.

Starting from the weight vector $[p_1,p_2,0]$ in $\mathfrak{su}(4)$, we define its projected
weight vector $[p_1,p_1+2p_2]$ in $\mathfrak{so}(4)$ as the Dynkin Label of the corresponding
irreducible bosonic Young Tableau.
\begin{equation}
    [p_1,p_1+2p_2] ~=~ [p_1,p_2,0] \, P^T_{\mathfrak{su}(4)\supset\mathfrak{so}(4)} ~~~.
\end{equation}
Note that in 4D, duality condition needs to be considered for two-form. Similar as the situations in 6D, 8D and 10D discussed in \ref{sec:6D_component}, \ref{sec:8D_component} and \cite{nDx} respectively, the Dynkin Label $[p_1+2p_2,p_1]$ carries the same dimensionality and corresponds to the same YT shape as $[p_1,p_1+2p_2]$. 
Like in 10D, the two-form with Dynkin Label $[0,2]$ satisfies self-dual condition and the one with Dynkin Label $[2,0]$ satisfies anti-self-dual condition. Generally, the field/BYT corresponding to Dynkin Label $[p_1,p_1+2p_2]$ satisfies self-dual condition and the one with Dynkin Label $[p_1+2p_2,p_1]$ satisfies anti-self-dual condition.

Given an irreducible bosonic Young Tableau with $p_1$ columns of one box and
$p_2$ columns of two vertical boxes, the Dynkin Label of its
corresponding bosonic irrep is $[p_1,p_1+2p_2]$ (or $[p_1+2p_2,p_1]$). Reversely, given a bosonic irrep with Dynkin Label $[a,b]$, its corresponding irreducible 
bosonic Young Tableau is composed of $\min\{a,b\}$ columns of one box and $|b-a|/2$ columns of two 
vertical boxes. 

The simplest examples,
also the fundamental building blocks of an irreducible BYT, are given below.
\begin{equation}
\begin{gathered}
    {\CMTB{\ydiagram{1}}}_{{\rm IR}} ~\equiv~ \CMTB{[1,1]}  ~~~,~~~
    {\CMTB{\ydiagram{1,1}}}_{{\rm IR},+} ~\equiv~ \CMTB{[0,2]} ~~~,~~~
    {\CMTB{\ydiagram{1,1}}}_{{\rm IR},-} ~\equiv~ \CMTB{[2,0]} ~~~.
\end{gathered} \label{equ:BYTbasic_4D}
\end{equation}

For spinorial irreps, by adopting the Majorana convention, the basic SYT is
\begin{equation}
    \CMTred{\ytableaushort{\tinyfour}} ~\equiv~~ \CMTred{[1,0]} ~\oplus~ \CMTred{[0,1]} ~~~.
\label{equ:SYTbasic_4D}
\end{equation}

Putting together the columns in (\ref{equ:BYTbasic_4D}) and (\ref{equ:SYTbasic_4D}) corresponds to adding their Dynkin Labels.

In summary, if we adopt the 4-component Majorana notation, the irreducible Young Tableau descriptions of the 4D minimal scalar superfield decomposition can be presented as follows.
\begin{equation}
\ytableausetup{boxsize=0.8em}
{\cal V} ~=~ \begin{cases}
~~{\rm {Level}}-0 \,~~~~~~~~~~ \CMTB{\cdot} ~~~,  \\
%%%%%%%%%%%%%%%%%%%%%%%%%%%%%%%%%%%%%%%%%%%%%%%%
~~{\rm {Level}}-1 \,~~~~~~~~~~ \CMTred{\ytableaushort{\tinyfour}} ~~~,  \\
%%%%%%%%%%%%%%%%%%%%%%%%%%%%%%%%%%%%%%%%%%%%%%%%
~~{\rm {Level}}-2 \,~~~~~~~~~~ (2)~\CMTB{\cdot}~\oplus~
{\CMTB{\ydiagram{1}}}_{\rm IR} ~~~,  \\
%%%%%%%%%%%%%%%%%%%%%%%%%%%%%%%%%%%%%%%%%%%%%%%%
~~{\rm {Level}}-3 \,~~~~~~~~~~
\CMTred{\ytableaushort{\tinyfour}} ~~~,  \\
%%%%%%%%%%%%%%%%%%%%%%%%%%%%%%%%%%%%%%%%%%%%%%%%
~~{\rm {Level}}-4 \,~~~~~~~~~~ \CMTB{\cdot} ~~~.
\end{cases}
\label{equ:V_4D}
\ytableausetup{boxsize=1.2em}
\end{equation}

\subsection{Index Structures and Irreducible Conditions of Component Fields in 4D Minimal Scalar Superfield}\label{sec:4D_Index}

In this section, we will translate the irreducible bosonic Young Tableaux into field representation. 
We follow the same $\{\}$-indices notation as well as ``$|$'' to separate indices in YT with different heights and ``,'' to separate
indices in YT with the same heights. 

The vector index $\un{a}$ runs from 0 to 3. 
The $\{\}$-indices, irreducible bosonic Young
Tableaux, and Dynkin Labels are equivalent and have the one-to-one correspondence.

The general expression is as below in Equations (\ref{equ:index-notation1_4D}) and (\ref{equ:index-notation2_4D}),
\begin{equation}
\begin{split}
    & \{ \un{a}_1,\dots,\un{a}_p ~|~ \un{b}_1\un{c}_1,\dots,\un{b}_q \un{c}_q \}^{\pm}  \\[10pt]
    & ~ { \CMTB{{\ytableaushort{\aone}} } }~~~~~{ \CMTB{{\ytableaushort{\ap}} } }
    ~~~~{ \CMTB{{\ytableaushort{\bone,\cone}} } }~~~~~~~{ \CMTB{{\ytableaushort{\bq,\cunq}} } }
    \\[10pt]
    & \,~~~~ \CMTB{[p,p]} \,~~~~~~~~~~  
    \begin{matrix}\CMTB{[0,2q]} ~~ \text{with superscript } + \\
    \CMTB{[2q,0]} ~~ \text{with superscript } -
    \end{matrix}  
\end{split}
\label{equ:index-notation1_4D}
\end{equation}
where above we have ``disassembled'' the YT to show how each column is affiliated with each type of
subscript structure. Below we have assembled all the columns into a proper YT.
\begin{equation}
\begin{gathered}
    { \CMTB{{\ytableaushort{\bone\dots\bq\aone\dots\ap,\cone\dots\cunq}} } }_{{\rm IR},\pm} \\[10pt]
    \begin{matrix}\CMTB{[p,p+2q]} ~~ \text{with superscript } + \\
    \CMTB{[p+2q,p]} ~~ \text{with superscript } -
    \end{matrix}  
\end{gathered}
\label{equ:index-notation2_4D}
\end{equation}

As one moves from the YT's shown in Equation~(\ref{equ:index-notation1_4D}) to
Equation~(\ref{equ:index-notation2_4D}), it is clear the number of vertical boxes is tabulating the number of
1-forms and 2-forms in the YT's.  These are the entries between
the vertical $|$ bars. These precisely correspond to the Dynkin Labels $p$ and $q$.  
An example of the correspondence between the subscript conventions,
the affiliated YT, and Dynkin Label is shown in (\ref{equ:index-notation_ex1_4D}). 
\begin{equation}
    \{{\un a}_2 , {\un a}_3| {\un a}_1  {\un b}_1    \} ~~\equiv~~
    { \CMTB{{\ytableaushort{\aone \atwo \athree,\bone}} } }_{{\rm IR},+}
    ~~\equiv~~ \CMTB{[2,4]} ~~~.
    \label{equ:index-notation_ex1_4D}
\end{equation}

The index structures as well as irreducible conditions of all bosonic and fermionic fields are identified below along 
with the level at which the fields occur in the adinkra of the superfield. The Majorana spinor index $\alpha$ runs from 1 to 4. 

\begin{itemize}
    \item Level-0: $\Phi(x)$ ~~~,
    \item Level-1: $\Psi_{\a}(x)$ ~~~,
    \item Level-2: (2)\,$\Phi(x)$ ~~~,~~~ $\Phi_{\{\aone\}}(x)$ ~~~,
    \item Level-3: $\Psi^{\a}(x)$ ~~~,
    \item Level-4: $\Phi(x)$ ~~~.
\end{itemize}

\newpage
\section{Conclusion}

In this work, we have established the basic libraries of adynkras that can be
used to explore problems of embedding component fields into superfields
in the context of spacetimes with Lorentzian signature and D $-$ 1
spatial dimensions where 4 $\le$ D $\le$ 9.
In addition to minimal scalar superfields, we also explored $(1,0)$ multiplets in several dimensions.

How to write the complete set of supersymmetry transformations for these component fields is an important question. 
In every adynkra diagram we draw, the supersymmetry transformations are determined by a sufficient but not neccesary condition as stated in Section 8.3 of \cite{nDx}.
A fully satisfactory answer would require the construction of supercovariant derivative operators acting on adynkrafields. We will discuss the 4D case in a paper in preparation and its generalizations deserve further study.

The importance of the adynkra libraries is that for the first time in the
history of the subject of supersymmetry, these will support the creation
of algorithms that solve the embedding problem.  One can now start with {\it {any}}
spectrum of on-shell component fields in these various dimensions and algorithmically
{\it{derive}} the minimal dimension superfield representation (as well as
alternatives) that contains this specified component field spectrum.  The technique 
for this has been christened as an``adynkra digital analysis" (ADA) scan as described in the 
the work of \cite{10dCon}.

\vspace{.05in}
 \begin{center}
\parbox{4in}{{\it ``I am thinking about something much more important\\ $~\,$ than bombs. 
I am thinking about computers.'' \\ ${~}$ 
 %\\ ${~}$ 
\\ ${~}$ }\,\,-\,\, John von Neumann $~~~~~~~~~$}
 \parbox{4in}{
 $~~$}  
 \end{center}
 \noindent
{\bf Acknowledgements}\\[.1in] \indent

The research of S.\ J.\ G., Y.\ Hu, and S.-N.\ Mak is supported in part by the endowment
of the Ford Foundation Professorship of Physics at Brown University and they gratefully
acknowledge the support of the Brown Theoretical Physics Center. The authors highly appreciate the reviewer's insightful and helpful feedback on our manuscript.

\newpage
\appendix
\section{Irreducible Representations\label{appen:irrep}}

%\subsection{SO(9) Irreducible Representations}
Here we list parts of irreducible representations by both Dynkin labels and dimensionalities of Lorentz algebras in 9D to 4D \cite{yamatsu2015}.

\begin{table}[h!]
\centering
\begin{tabular}{|c|c|}
\hline
 Dynkin label & Dimension \\ \hline 
 [0,0,0,0] & $\CMTB{1}$ \\ \hline 
 [1,0,0,0] & $\CMTB{9}$ \\ \hline 
 [0,0,0,1] & $\CMTred{16}$ \\ \hline 
 [0,1,0,0] & $\CMTB{36}$ \\ \hline 
 [2,0,0,0] & $\CMTB{44}$ \\ \hline 
 [0,0,1,0] & $\CMTB{84}$ \\ \hline 
 [0,0,0,2] & $\CMTB{126}$ \\ \hline 
 [1,0,0,1] & $\CMTred{128}$ \\ \hline 
 [3,0,0,0] & $\CMTB{156}$ \\ \hline 
 [1,1,0,0] & $\CMTB{231}$ \\ \hline 
 [0,1,0,1] & $\CMTred{432}$ \\ \hline 
 [4,0,0,0] & $\CMTB{450}$ \\ \hline 
 [0,2,0,0] & $\CMTB{495}$ \\ \hline 
 [2,0,0,1] & $\CMTred{576}$ \\ \hline 
 [1,0,1,0] & $\CMTB{594}$ \\ \hline 
 [0,0,0,3] & $\CMTred{672}$ \\ \hline 
 [0,0,1,1] & $\CMTred{768}$ \\ \hline 
 [2,1,0,0] & $\CMTB{910}$ \\ \hline 
 [1,0,0,2] & $\CMTB{924}$ \\ \hline 
\end{tabular}
\caption{$\mathfrak{so}(9)$ irreducible representations \#1 \protect\cite{yamatsu2015}}
\end{table}

\newpage
\begin{table}[h!]
\centering
\begin{tabular}{|c|c|}
\hline
Dynkin label & Dimension \\ \hline 
 [5,0,0,0] & $\CMTB{1122}$ \\ \hline 
 [0,1,1,0] & $\CMTB{1650}$ \\ \hline 
 [3,0,0,1] & $\CMTred{1920}$ \\ \hline 
 [0,0,2,0] & $\CMTB{1980}$ \\ \hline 
 [2,0,1,0] & $\CMTB{2457}$ \\ \hline 
 [6,0,0,0] & $\CMTB{2508}$ \\ \hline 
 [1,1,0,1] & $\CMTred{2560}$ \\ \hline 
 [1,2,0,0] & $\CMTB{2574}$ \\ \hline 
 [0,1,0,2] & $\CMTB{2772}$ \\ \hline 
 [0,0,0,4] & $\CMTB{2772'}$ \\ \hline 
 [3,1,0,0] & $\CMTB{2772''}$ \\ \hline 
 [2,0,0,2] & $\CMTB{3900}$ \\ \hline 
 [0,3,0,0] & $\CMTB{4004}$ \\ \hline 
 [0,0,1,2] & $\CMTB{4158}$ \\ \hline 
 [1,0,0,3] & $\CMTred{4608}$ \\ \hline 
 [0,2,0,1] & $\CMTred{4928}$ \\ \hline 
 [1,0,1,1] & $\CMTred{5040}$ \\ \hline 
\end{tabular}
\caption{$\mathfrak{so}(9)$ irreducible representations \#2 \protect\cite{yamatsu2015}}
\end{table}

\newpage
%\subsection{SO(8) Irreducible Representations}

%Here we list the SO(8) irreducible representations by Dynkin labels and dimensions \cite{yamatsu2015}.

\begin{table}[h!]
\centering
\begin{tabular}{|c|c|}
\hline
 Dynkin label & Dimension \\ \hline 
 [0,0,0,0] & $\CMTB{1}$ \\ \hline 
 [1,0,0,0] & $\CMTB{8_{v}}$ \\ \hline 
 [0,0,0,1] & $\CMTred{8_{s}}$ \\ \hline 
 [0,0,1,0] & $\CMTred{8_{c}}$ \\ \hline 
 [0,1,0,0] & $\CMTB{28}$ \\ \hline 
 [2,0,0,0] & $\CMTB{35_{v}}$ \\ \hline 
 [0,0,2,0] & $\CMTB{35_{c}}$ \\ \hline 
 [0,0,0,2] & $\CMTB{35_{s}}$ \\ \hline 
 [0,0,1,1] & $\CMTB{56_{v}}$ \\ \hline 
 [1,0,1,0] & $\CMTred{56_{s}}$ \\ \hline 
 [1,0,0,1] & $\CMTred{56_{c}}$ \\ \hline 
 [3,0,0,0] & $\CMTB{112_{v}}$ \\ \hline 
 [0,0,0,3] & $\CMTred{112_{s}}$ \\ \hline 
 [0,0,3,0] & $\CMTred{112_{c}}$ \\ \hline 
 [1,1,0,0] & $\CMTB{160_{v}}$ \\ \hline 
 [0,1,0,1] & $\CMTred{160_{s}}$ \\ \hline 
 [0,1,1,0] & $\CMTred{160_{c}}$ \\ \hline 
 [1,0,2,0] & $\CMTB{224_{cv}}$ \\ \hline 
 [1,0,0,2] & $\CMTB{224_{sv}}$ \\ \hline 
 [2,0,0,1] & $\CMTred{224_{vs}}$ \\ \hline 
 [0,0,2,1] & $\CMTred{224_{cs}}$ \\ \hline 
 [2,0,1,0] & $\CMTred{224_{vc}}$ \\ \hline 
 [0,0,1,2] & $\CMTred{224_{sc}}$ \\ \hline 
 [4,0,0,0] & $\CMTB{294_{v}}$ \\ \hline 
 [0,0,4,0] & $\CMTB{294_{c}}$ \\ \hline 
 [0,0,0,4] & $\CMTB{294_{s}}$ \\ \hline 
\end{tabular}
\caption{$\mathfrak{so}(8)$ irreducible representations \#1 \protect\cite{yamatsu2015}}
\end{table}

\newpage
\begin{table}[h!]
\centering
\begin{tabular}{|c|c|}
\hline
 Dynkin label & Dimension \\ \hline 
 [0,2,0,0] & $\CMTB{300}$ \\ \hline 
 [1,0,1,1] & $\CMTB{350}$ \\ \hline 
 [2,1,0,0] & $\CMTB{567_{v}}$ \\ \hline 
 [0,1,2,0] & $\CMTB{567_{c}}$ \\ \hline 
 [0,1,0,2] & $\CMTB{567_{s}}$ \\ \hline 
 [0,0,3,1] & $\CMTB{672_{cs}}$ \\ \hline 
 [0,0,1,3] & $\CMTB{672_{sc}}$ \\ \hline 
 [3,0,1,0] & $\CMTred{672_{vc}}$ \\ \hline 
 [1,0,3,0] & $\CMTred{672_{cv}}$ \\ \hline 
 [3,0,0,1] & $\CMTred{672_{vs}}$ \\ \hline 
 [1,0,0,3] & $\CMTred{672_{sv}}$ \\ \hline 
 [5,0,0,0] & $\CMTB{672'_{v}}$ \\ \hline 
 [0,0,0,5] & $\CMTred{672'_{s}}$ \\ \hline 
 [0,0,5,0] & $\CMTred{672'_{c}}$ \\ \hline 
 [0,1,1,1] & $\CMTB{840_{v}}$ \\ \hline 
 [1,1,1,0] & $\CMTred{840_{s}}$ \\ \hline 
 [1,1,0,1] & $\CMTred{840_{c}}$ \\ \hline 
 [2,0,2,0] & $\CMTB{840'_{s}}$ \\ \hline 
 [2,0,0,2] & $\CMTB{840'_{c}}$ \\ \hline 
 [0,0,2,2] & $\CMTB{840'_{v}}$ \\ \hline 
 [2,0,1,1] & $\CMTB{1296_{v}}$ \\ \hline 
 [1,0,1,2] & $\CMTred{1296_{s}}$ \\ \hline 
 [1,0,2,1] & $\CMTred{1296_{c}}$ \\ \hline 
\end{tabular}
\caption{$\mathfrak{so}(8)$ irreducible representations \#2 \protect\cite{yamatsu2015}}
\end{table}

%\subsection{SO(7) Irreducible Representations}

%Here we list the SO(7) irreducible representations by Dynkin labels and dimensions \cite{yamatsu2015}.

\newpage

\begin{table}[h!]
\centering
\begin{tabular}{|c|c|}
\hline
 Dynkin label & Dimension \\ \hline 
 [0,0,0] & $\CMTB{1}$ \\ \hline 
 [1,0,0] & $\CMTB{7}$ \\ \hline 
 [0,0,1] & $\CMTred{8}$ \\ \hline 
 [0,1,0] & $\CMTB{21}$ \\ \hline 
 [2,0,0] & $\CMTB{27}$ \\ \hline 
 [0,0,2] & $\CMTB{35}$ \\ \hline 
 [1,0,1] & $\CMTred{48}$ \\ \hline 
 [3,0,0] & $\CMTB{77}$ \\ \hline 
 [1,1,0] & $\CMTB{105}$ \\ \hline 
 [0,1,1] & $\CMTred{112}$ \\ \hline 
 [0,0,3] & $\CMTred{112'}$ \\ \hline 
 [2,0,1] & $\CMTred{168}$ \\ \hline 
 [0,2,0] & $\CMTB{168'}$ \\ \hline 
 [4,0,0] & $\CMTB{182}$ \\ \hline 
 [1,0,2] & $\CMTB{189}$ \\ \hline 
 [0,0,4] & $\CMTB{294}$ \\ \hline 
 [2,1,0] & $\CMTB{330}$ \\ \hline 
 [0,1,2] & $\CMTB{378}$ \\ \hline 
 [5,0,0] & $\CMTB{378'}$ \\ \hline 
 [3,0,1] & $\CMTred{448}$ \\ \hline 
 [1,1,1] & $\CMTred{512}$ \\ \hline 
 [1,0,3] & $\CMTred{560}$ \\ \hline 
 [2,0,2] & $\CMTB{616}$ \\ \hline 
\end{tabular}
\caption{$\mathfrak{so}(7)$ irreducible representations \protect\cite{yamatsu2015}}
\end{table}

%\subsection{SO(6) Irreducible Representations}

%Here we list the SO(6) irreducible representations by Dynkin labels and dimensions \cite{yamatsu2015}.

\newpage

\begin{table}[h!]
\centering
\begin{tabular}{|c|c|}
\hline
 Dynkin label & Dimension \\ \hline 
 [0,0,0] & $\CMTB{1}$ \\ \hline 
 [0,1,0] & $\CMTred{4}$ \\ \hline 
 [0,0,1] & $\CMTred{\overline{4}}$ \\ \hline 
 [1,0,0] & $\CMTB{6}$ \\ \hline 
 [0,2,0] & $\CMTB{10}$ \\ \hline 
 [0,0,2] & $\CMTB{\overline{10}}$ \\ \hline 
 [0,1,1] & $\CMTB{15}$ \\ \hline 
 [1,0,1] & $\CMTred{20}$ \\ \hline 
 [1,1,0] & $\CMTred{\overline{20}}$ \\ \hline 
 [2,0,0] & $\CMTB{20'}$ \\ \hline 
 [0,0,3] & $\CMTred{20''}$ \\ \hline 
 [0,3,0] & $\CMTred{\overline{20''}}$ \\ \hline 
\end{tabular}
\caption{$\mathfrak{so}(6)$ irreducible representations \protect\cite{yamatsu2015}}
\end{table}

%\subsection{SO(5) Irreducible Representations}

%Here we list the SO(5) irreducible representations by Dynkin labels and dimensions \cite{yamatsu2015}.

\begin{table}[h!]
\centering
\begin{tabular}{|c|c|}
\hline
 Dynkin label & Dimension \\ \hline 
 [0,0] & $\CMTB{1}$ \\ \hline 
 [0,1] & $\CMTred{4}$ \\ \hline 
 [1,0] & $\CMTB{5}$ \\ \hline 
 [0,2] & $\CMTB{10}$ \\ \hline 
 [2,0] & $\CMTB{14}$ \\ \hline 
 [1,1] & $\CMTred{16}$ \\ \hline 
\end{tabular}
\caption{$\mathfrak{so}(5)$ irreducible representations \protect\cite{yamatsu2015}}
\end{table}

\begin{table}[h!]
\centering
\begin{tabular}{|c|c|}
\hline
 Dynkin label & Dimension \\ \hline 
 [0,0] & $\CMTB{1}$ \\ \hline 
 [1,0] & $\CMTred{2}$ \\ \hline 
 [0,1] & $\CMTred{\overline{2}}$ \\ \hline 
 [2,0] & $\CMTB{3}$ \\ \hline 
 [0,2] & $\CMTB{\overline{3}}$ \\ \hline 
 [1,1] & $\CMTB{4}$ \\ \hline 
 [4,0] & $\CMTB{5}$ \\ \hline 
 [0,4] & $\CMTB{\overline{5}}$ \\ \hline 
 [1,2] & $\CMTred{6}$ \\ \hline 
 [2,1] & $\CMTred{\overline{6}}$ \\ \hline 
\end{tabular}
\caption{$\mathfrak{so}(4)$ irreducible representations \protect\cite{yamatsu2015}}
\end{table}

\newpage
\section{Projection Matrices for Branching Rules}
\label{appen:proj}

Here we list the projection matrices for the branching rules we used to obtain both the Lorentz component contents of minimal scalar superfields and $(1,0)$ superfields from 9D to 5D. 

\subsection{Minimal Scalar Superfield Decompositions}

The relevant branching rules for decomposing minimal scalar superfields are $\mathfrak{su}(d)\supset\mathfrak{so}(\rD)$, where $d$ is the number of real components of Grassman coordinates and $\rD$ is the spacetime dimension. 

\noindent
For 9D, 
\begin{equation}
\begin{split}
P_{\mathfrak{su}(16)\supset\mathfrak{so}(9)} 
~=&~ P_{\mathfrak{so}(10)\supset\mathfrak{so}(9)} \, P_{\mathfrak{su}(16)\supset\mathfrak{so}(10)} \\
~=&~ 
\left(
\begin{array}{ccccccccccccccc}
 0 & 0 & 0 & 0 & 1 & 0 & 1 & 0 & 1 & 0 & 1 & 0 & 0 & 0 & 0 \\
 0 & 0 & 1 & 2 & 1 & 1 & 0 & 0 & 0 & 1 & 1 & 2 & 1 & 0 & 0 \\
 0 & 1 & 0 & 0 & 0 & 0 & 1 & 2 & 1 & 0 & 0 & 0 & 0 & 1 & 0 \\
 1 & 0 & 1 & 0 & 1 & 2 & 1 & 0 & 1 & 2 & 1 & 0 & 1 & 0 & 1 \\
\end{array}
\right) ~~~.
\end{split}
\end{equation}
For 8D,
\begin{equation}
\begin{split}
P_{\mathfrak{su}(16)\supset\mathfrak{so}(8)} 
~=&~ P_{\mathfrak{so}(9)\supset\mathfrak{so}(8)} \, P_{\mathfrak{so}(10)\supset\mathfrak{so}(9)} \, P_{\mathfrak{su}(16)\supset\mathfrak{so}(10)} \\
~=&~
\left(
\begin{array}{ccccccccccccccc}
 0 & 1 & 0 & 0 & 0 & 0 & 1 & 2 & 1 & 0 & 0 & 0 & 0 & 1 & 0 \\
 0 & 0 & 1 & 2 & 1 & 1 & 0 & 0 & 0 & 1 & 1 & 2 & 1 & 0 & 0 \\
 0 & 0 & 0 & 0 & 1 & 0 & 1 & 0 & 1 & 0 & 1 & 0 & 0 & 0 & 0 \\
 -1 & -2 & -3 & -4 & -4 & -4 & -4 & -4 & -4 & -4 & -4 & -4 & -3 & -2 & -1 \\
\end{array}
\right) ~~~.
\end{split}
\end{equation}
For 7D,
\begin{equation}
\begin{split}
P_{\mathfrak{su}(16)\supset\mathfrak{so}(7)} 
~=&~ P_{\mathfrak{so}(8)\supset\mathfrak{so}(7)} \, P_{\mathfrak{so}(9)\supset\mathfrak{so}(8)} \, P_{\mathfrak{so}(10)\supset\mathfrak{so}(9)} \, P_{\mathfrak{su}(16)\supset\mathfrak{so}(10)} \\
~=&~
\left(
\begin{array}{ccccccccccccccc}
 0 & 1 & 0 & 0 & 0 & 0 & 1 & 2 & 1 & 0 & 0 & 0 & 0 & 1 & 0 \\
 0 & 0 & 1 & 2 & 1 & 1 & 0 & 0 & 0 & 1 & 1 & 2 & 1 & 0 & 0 \\
 -1 & -2 & -3 & -4 & -3 & -4 & -3 & -4 & -3 & -4 & -3 & -4 & -3 & -2 & -1 \\
\end{array}
\right) ~~~.
\end{split}
\end{equation}
For 6D,
\begin{equation}
\begin{split}
P_{\mathfrak{su}(8)\supset\mathfrak{so}(6)} 
~=&~ P_{\mathfrak{so}(7)\supset\mathfrak{so}(6)} \, P_{\mathfrak{so}(8)\supset\mathfrak{so}(7)} \, P_{\mathfrak{su}(8)\supset\mathfrak{so}(8)}^{(\text{spinor})} \\
~=&~
\left(
\begin{array}{ccccccc}
 0 & 1 & 0 & 0 & 0 & 1 & 0 \\
 0 & 0 & 1 & 0 & 1 & 0 & 0 \\
 -1 & -2 & -2 & -2 & -2 & -2 & -1 \\
\end{array}
\right) ~~~.
\end{split}
\end{equation}
For 5D, 
\begin{equation}
\begin{split}
P_{\mathfrak{su}(8)\supset\mathfrak{so}(5)} 
~=&~ P_{\mathfrak{so}(6)\supset\mathfrak{so}(5))} \, P_{\mathfrak{so}(7)\supset\mathfrak{so}(6)} \, P_{\mathfrak{so}(8)\supset\mathfrak{so}(7)} \, P_{\mathfrak{su}(8)\supset\mathfrak{so}(8)}^{(\text{spinor})} \\
~=&~ 
\left(
\begin{array}{ccccccc}
 0 & 1 & 0 & 0 & 0 & 1 & 0 \\
 -1 & -2 & -1 & -2 & -1 & -2 & -1 \\
\end{array}
\right) ~~~.
\end{split}
\end{equation}
For 4D, 
\begin{equation}
P_{\mathfrak{su}(4)\supset\mathfrak{so}(4)}^{(\text{spinor})} 
~=~ 
\left(
\begin{array}{ccc}
 0 & 1 & 0 \\
 1 & 1 & 1 \\
\end{array}
\right) ~~~.
\label{eqn:Psu4so4spinor}
\end{equation}
We will discuss $P_{\mathfrak{su}(8)\supset\mathfrak{so}(8)}^{(\text{spinor})}$ and $P_{\mathfrak{su}(4)\supset\mathfrak{so}(4)}^{(\text{spinor})}$ in the section \ref{appen:Proj_Mat_SU8}.

\subsection{$(1,0)$ Superfield Decompositions}

For 8D, 7D, 6D and 5D, the dimensions of the spinor representations in $\mathfrak{so}(\rD)$ are one-halves of the numbers of real components of Grassman coordinates $d$. Therefore, we could write down the component contents of the $(1,0)$ superfields by utilizing the branching rules $\mathfrak{su}(\fracm{d}{2})\supset\mathfrak{so}(\rD)$.

\noindent
For 8D, 
\begin{equation}
P_{\mathfrak{su}(8)\supset\mathfrak{so}(8)}^{(\text{spinor})}
~=~
\left(
\begin{array}{ccccccc}
 0 & 0 & 1 & 0 & 1 & 0 & 0 \\
 0 & 1 & 0 & 0 & 0 & 1 & 0 \\
 0 & 0 & 1 & 2 & 1 & 0 & 0 \\
 1 & 0 & 0 & 0 & 0 & 0 & 1 \\
\end{array}
\right) ~~~.
\label{eqn:Psu8so8spinor}
\end{equation}
For 7D, 
\begin{equation}\label{equ:Psu8toso7}
\begin{split}
P_{\mathfrak{su}(8)\supset\mathfrak{so}(7)}
~=&~ P_{\mathfrak{so}(8)\supset\mathfrak{so}(7)} \, P_{\mathfrak{su}(8)\supset\mathfrak{so}(8)}^{(\text{spinor})} \\
~=&~ 
\left(
\begin{array}{ccccccc}
 0 & 0 & 1 & 0 & 1 & 0 & 0 \\
 0 & 1 & 0 & 0 & 0 & 1 & 0 \\
 1 & 0 & 1 & 2 & 1 & 0 & 1 \\
\end{array}
\right) ~~~.
\end{split}
\end{equation}
For 6D,
\begin{equation}\label{equ:Psu4toso6}
P_{\mathfrak{su}(4)\supset\mathfrak{so}(6)} ~=~ 
\left(
\begin{array}{ccc}
 0 & 1 & 0 \\
 1 & 0 & 0 \\
 0 & 0 & 1 \\
\end{array}
\right) ~~~.
\end{equation}
For 5D,
\begin{equation}\label{equ:Psu4toso5}
P_{\mathfrak{su}(4)\supset\mathfrak{so}(5)} ~=~ 
\left(
\begin{array}{ccc}
 0 & 1 & 0 \\
 1 & 0 & 1 \\
\end{array}
\right) ~~~.
\end{equation}

\subsection{A Note on the Projection Matrices of $\mathfrak{su}(8)\supset\mathfrak{so}(8)$ and $\mathfrak{su}(4)\supset\mathfrak{so}(4)$}
\label{appen:Proj_Mat_SU8}

Speaking of the projection matrix for $\mathfrak{su}(8)\supset\mathfrak{so}(8)$, one may recall our discussions on translating $\mathfrak{so}(8)$ Dynkin labels to BYTs in Section \ref{sec:8D_component}.
In Equation (\ref{eqn:Psu8so8vector}), we have such an expression for the projection matrix,
\begin{equation}
P_{\mathfrak{su}(8)\supset\mathfrak{so}(8)}^{(\text{vector})} ~=~
\left(
\begin{array}{ccccccc}
1 & 0 & 0 & 0 & 0 & 0 & 1\\
0 & 1 & 0 & 0 & 0 & 1 & 0\\
0 & 0 & 1 & 0 & 1 & 0 & 0\\
0 & 0 & 1 & 2 & 1 & 0 & 0\\
\end{array}
\right) ~~~,
\end{equation}
which is different than that in Equation (\ref{eqn:Psu8so8spinor}) above. 
Similarly, the projection matrix for $\mathfrak{su}(4)\supset\mathfrak{so}(4)$ for BYTs in Equation (\ref{eqn:Psu4so4vector}) in Section \ref{sec:4D_components} is
\begin{equation}
P_{\mathfrak{su}(4)\supset\mathfrak{so}(4)}^{(\text{vector})} ~=~
\left(
\begin{array}{ccc}
1 & 0 & 1 \\
1 & 2 & 1 \\
\end{array}
\right) ~~~,
\end{equation}
which also differs with that in Equation (\ref{eqn:Psu4so4spinor}).
Note that we put the superscripts (vector) and (spinor) for these two projection matrices. The reasons are as follows.

Recall the definition of a projection matrix $P_{\mathfrak{g}\supset\mathfrak{h}}$,
\begin{equation}
\label{equ:def_prjmat}
    v_{\mathfrak{h}}^T ~=~ P_{\mathfrak{g}\supset\mathfrak{h}} \, w_{\mathfrak{g}}^T ~~~,
\end{equation}
where $w_{\mathfrak{g}}$ and $v_{\mathfrak{h}}$ are weight vectors in $\mathfrak{g}$ and $\mathfrak{h}$ respectively.
In the discussion of $\mathfrak{so}(8)$ BYTs, we want the fundamental representation \{8\} in $\mathfrak{su}(8)$ projected to the vector in $\mathfrak{so}(8)$, i.e. 
\begin{equation}
    \{8\} ~=~ [ 1, 0, 0, 0, 0, 0, 0 ] ~\longrightarrow~ \CMTB{\{8_v\}} ~=~ \CMTB{[1,0,0,0]} ~~~,
\end{equation}
hence we call it $P_{\mathfrak{su}(8)\supset \mathfrak{so}(8)}^{(\text{vector})}$. 
In the discussion of the 8D $(1,0)$ superfield decompositions though, we want the \{8\} in $\mathfrak{su}(8)$ to correspond to the $\CMTred{\{8_s\}}$ spinor in $\mathfrak{so}(8)$, i.e. 
\begin{equation}
    \{8\} ~=~ [ 1, 0, 0, 0, 0, 0, 0 ] ~\longrightarrow~ \CMTred{\{8_s\}} ~=~ \CMTred{[0,0,0,1]} ~~~,
\end{equation}
hence we call it $P_{\mathfrak{su}(8)\supset \mathfrak{so}(8)}^{(\text{spinor})}$. One quick way to distinguish between two projection matrices is to look at their first column - if it is $\CMTB{[1,0,0,0]}^{T}$ then it is (vector), and if it is $\CMTred{[0,0,0,1]}^{T}$ then it is (spinor). 

Similarly, in the discussion of $\mathfrak{so}(4)$ BYTs, we project the fundamental representation $\{4\}$ in $\mathfrak{su}(4)$ to the vector in $\mathfrak{so}(4)$, i.e. 
\begin{equation}
    \{4\} ~=~ [ 1, 0, 0 ] ~\longrightarrow~ \CMTB{\{4\}} ~=~ \CMTB{[1,1]} ~~~,
\end{equation}
then we obtain $P_{\mathfrak{su}(4)\supset \mathfrak{so}(4)}^{(\text{vector})}$.
In decomposing the superfield components though, we project the $\{4\}$ to the spinor representation $\CMTred{\{4\}} = \CMTred{\{2\}} \oplus \CMTred{\{\overline{2}\}}$ in $\mathfrak{so}(4)$, i.e.
\begin{equation}
    \{4\} ~=~ [ 1, 0, 0 ] ~\longrightarrow~ \CMTred{\{4\}} ~=~ \CMTred{\{2\}} \oplus \CMTred{\{\overline{2}\}} ~=~ \CMTred{[1,0]} \oplus \CMTred{[0,1]} ~~~,
\end{equation}
then we get $P_{\mathfrak{su}(4)\supset \mathfrak{so}(4)}^{(\text{spinor})}$.

We conclude by saying that the coincidence of both vector and spinor having d.o.f. = 8 in 8D, or the triality of $\mathfrak{so}(8)$, creates the opportunity of writing two $P_{\mathfrak{su}(8)\supset \mathfrak{so}(8)}$ matrices while we think of them as descending from the same fundamental irrep \{8\} in $\mathfrak{su}(8)$, since there is no vectorial / spinorial distinction of irreps in $\mathfrak{su}(8)$.
The same is true for $\mathfrak{su}(4)\supset\mathfrak{so}(4)$, as both vector and spinor in 4D have a total number of 4 components.
Therefore, two versions of $P_{\mathfrak{su}(8)\supset\mathfrak{so}(8)}$ serve two different purposes, and so do the two versions of $P_{\mathfrak{su}(4)\supset\mathfrak{so}(4)}$, and there is no contradiction.

%%%%%%%%% add su4 to so4 discussions

\subsection{A Note on Isomorphisms}

In \cite{yamatsu2015} and \cite{Susyno}, the discussions or the programming tools for calculations of some particular algebras such as $\mathfrak{so}(6)$, $\mathfrak{so}(5)$ and $\mathfrak{so}(4)$ are absent. This is because they are isomorphic to some other simple lie algebras, 
\begin{align}
    \mathfrak{so}(6) ~\cong&~~ \mathfrak{su}(4) ~~~, \\
    \mathfrak{so}(5) ~\cong&~~ \mathfrak{usp}(4) ~~~, \\
    \mathfrak{so}(4) ~\cong&~~ \mathfrak{su}(2) \times \mathfrak{su}(2) ~~~.
\end{align}
In particular, the representations of two isomorphic algebras should have 1-1 correspondences. 
\begin{align}
    \begin{matrix} \mathfrak{so}(6) \\ [a,b,c] \end{matrix}
    ~~~~\cong&~~~~~
    \begin{matrix} \mathfrak{su}(4) \\ [b,a,c] \end{matrix} ~~~~~~, \\[10pt]
    \begin{matrix} \mathfrak{so}(5) \\ [a,b] \end{matrix}
    ~~~~\cong&~~~~~ 
    \begin{matrix} \mathfrak{usp}(4) \\ [b,a] \end{matrix} ~~~~~~, \\[10pt]
    \begin{matrix} \mathfrak{so}(4) \\ [a,b] \end{matrix}
    ~~~~\cong&~~~~~ 
    \begin{matrix} \mathfrak{su}(2) \times \mathfrak{su}(2) \\ [a] \times [b] \end{matrix} ~~~~~~.
\end{align}
Their Weyl dimension formulas, weight systems for each pair of isomorphic representations, and etc. are exactly the same with the appropriate changes in Dynkin labels. 
Two isomorphic algebras are subalgebras of each other. Therefore, it does not hurt to define the following projection matrices, 
\begin{align}
P_{\mathfrak{so}(6)\supset\mathfrak{su}(4)} ~=~ P_{\mathfrak{su}(4)\supset\mathfrak{so}(6)} ~=&~ 
\left(
\begin{array}{ccc}
0 & 1 & 0 \\
1 & 0 & 0 \\
0 & 0 & 1 \\
\end{array}
\right) ~~~, \\
P_{\mathfrak{so}(5)\supset\mathfrak{usp}(4)} ~=~ P_{\mathfrak{usp}(4)\supset\mathfrak{so}(5)} ~=&~ 
\left(
\begin{array}{cc}
0 & 1 \\
1 & 0 \\
\end{array}
\right) ~~~, \\
P_{\mathfrak{so}(4)\supset\mathfrak{su}(2)\times\mathfrak{su}(2)} ~=~ P_{\mathfrak{su}(2)\times\mathfrak{su}(2)\supset\mathfrak{so}(4)} ~=&~ 
\left(
\begin{array}{cc}
1 & 0 \\
0 & 1 \\
\end{array}
\right) ~~~,
\end{align}
which helps us translate between Dynkin labels of isomorphic algebras.

\newpage
$$~~$$

\end{document}